\documentclass[12pt]{article}
 
\usepackage{amssymb,amsmath,amsfonts,eurosym,geometry,ulem,graphicx,caption,color,setspace,sectsty,comment,footmisc,caption,pdflscape,subfigure,array,placeins,dcolumn,xcolor}
\usepackage[colorlinks=true,linkcolor=blue,citecolor=blue]{hyperref}
\usepackage[utf8]{inputenc}
\usepackage{enumitem}
\usepackage{tabularx}
\usepackage{caption}
\usepackage{booktabs}
\usepackage[super]{nth}
\usepackage{titlesec}
\usepackage[flushleft]{threeparttable}
 
\usepackage[
backend=bibtex,%
maxcitenames=2,%
mincitenames=1,%
style=authoryear,%
citestyle=authoryear,%
natbib=true,%
url=false,%
uniquename=false,%
uniquelist=false,%
doi=true,%
eprint=false%
]{biblatex}
\appto{\citesetup}{\color{blue}}

\newcolumntype{L}[1]{>{\raggedright\let\newline\\arraybackslash\hspace{0pt}}m{#1}}
\newcolumntype{C}[1]{>{\centering\let\newline\\arraybackslash\hspace{0pt}}m{#1}}
\newcolumntype{R}[1]{>{\raggedleft\let\newline\\arraybackslash\hspace{0pt}}m{#1}}

\newcommand*\dd{\text{d\kern 0.03em}}
\newcommand*\dz{\text{d\kern 0.05em}z_t}
\newcommand*\dq{\text{d\kern 0.05em}z_t^{\mathbb{Q}}}

\definecolor{ao(english)}{rgb}{0.0, 0.5, 0.0}
\newcommand{\hi}[1]{} 
\newcommand{\hide}[1]{}

\newcolumntype{d}{D{.}{.}{-3}}
\newcommand{\specialcell}[2][c]{\begin{tabular}[#1]{@{}c@{}}#2\end{tabular}} 

\titleformat{\section}{\normalfont\large\centering}{\bfseries {\bfseries\thesection.}}{0.5em}{\MakeUppercase}
\titlespacing*{\section}{0pt}{2em}{0.25em}

\titleformat{\subsection}{\normalfont\bfseries\centering}{\thesubsection.}{0.5em}{}
\titlespacing*{\subsection}{0pt}{1em}{0.3em}{}

\titleformat{\subsubsection}{\normalfont\itshape\normalsize\centering}{\thesubsubsection.}{0.5em}{}
\titlespacing*{\subsubsection}{0pt}{1em}{0.3em}{}

\begin{document}

\title{The Long-Term Effects of Early-Life Pollution Exposure: \\
\vspace{2mm} 
Evidence from the London Smog}
\author{Stephanie von Hinke\thanks{School of Economics, University of Bristol; Institute for Fiscal Studies. E-mail: \href{mailto:S.vonHinke@bristol.ac.uk}{S.vonHinke@bristol.ac.uk}} \and Emil N. S\o{}rensen\thanks{School of Economics, University of Bristol. E-mail: \href{mailto:E.Sorensen@bristol.ac.uk}{E.Sorensen@bristol.ac.uk}}}

\date{\today\thanks{We would like to thank the editor as well as two anonymous referees for valuable comments. We also thank Samuel Baker, Pietro Biroli, Jason Fletcher, Hans van Kippersluis, Niels Rietveld, Nicolai Vitt, and seminar participants from the University of Bologna, the University of Wisconsin-Madison, the 2022 ESSGN conference, and the 2022 workshop on ``Global Health, Environment and Labour'' (Royal Holloway)  for valuable comments. This work is based on data provided through www.visionofbritain.org.uk and uses historical material which is copyright of the Great Britain Historical GIS Project and the University of Portsmouth. We gratefully acknowledge financial support from NORFACE DIAL (Grant Reference 462-16-100) and the European Research Council (Starting Grant Reference 851725).}}
\maketitle

\begin{abstract}
\singlespacing
\noindent 
This paper uses a large UK cohort to investigate the impact of early-life pollution exposure on individuals' human capital and health outcomes in older age. We compare individuals who were exposed to the London smog in December 1952 whilst \textit{in utero} or in infancy to those born after the smog and those born at the same time but in unaffected areas. We find that those exposed to the smog have substantially lower fluid intelligence and worse respiratory health, with some evidence of a reduction in years of schooling.

\vspace{3mm}

\vspace{1mm}\newline
\end{abstract}

\clearpage
\doublespacing

\section{Introduction}

There is a growing literature on the contemporaneous effects of exposure to air pollution on individuals' human capital and health outcomes \citep[for a review see e.g.,][]{graff2013environment}. There is relatively little empirical evidence, however, on the much longer-term and cumulative effects of early-life pollution exposure, despite the fact that the early-life environment has been shown to be crucial in shaping individuals' health and economic outcomes in older age. The literature on the so-called ``Developmental Origins of Health and Disease'' (DOHaD) hypothesis -- proposing that circumstances early in life can have life-long, potentially irreversible impacts on individuals' health and well-being -- explores the importance of the prenatal as well as early childhood environment and has mainly focused on the longer-term effects of (adverse) nutritional, health and economic environments \citep[for reviews, see e.g.,][]{AlmondCurrie2011a, AlmondCurrie2011b, almond2018childhood, conti2019developmental}.\footnote{For example, research has explored the importance of maternal physical health \citep[][]{behrman2004returns, Almond2006, almond20051918}, maternal mental health \citep{von2022mental}, maternal health behaviours \citep[][]{nilsson2017alcohol, von2014alcohol}, maternal nutrition \citep[][]{vandenBerg2021}, the economic environment \citep[][]{van2006economic, banerjee2010long}, the early life disease environment \citep{bleakley2007disease, case2009early, cattan2021health}, or the home environment \citep{carneiro2015flying}.} 

The relative lack of studies on the very long-term effects of early life pollution exposure is likely to be at least partially driven by a general absence of high quality historical pollution data. Indeed, most studies that explore the effects of early-life pollution exposure investigate the \textit{immediate} effects on child birth outcomes \citep[see e.g.,][]{chay2003impact, currie2005air, almond2009chernobyl, Currie2009, jayachandran2009air, currie2011traffic, knittel2016caution, sanders2015have, arceo2016does, beach2018coal, hanlon2022london, jia2019china, rangel2019agricultural}, with only few exploring potential effects in childhood or early adulthood \citep[see e.g.,][]{reyes2007environmental, bharadwaj2017gray, almond2009chernobyl, sanders2012doesn, black2019only, isen2017every, persico2021effects, heissel2022does} and even fewer focusing on outcomes in older age \citep[][]{bharadwaj2016early, Ball2018, persico2020can}. As such, ignoring potential long-term effects of pollution may lead to a substantial underestimation of the total welfare effects caused by exposure to environmental toxins. 

We overcome the lack of historical pollution data by relying on reduced form analyses. More specifically, we examine the effect of early life exposure to the London smog: a severe pollution event that affected London residents between 5 and 9 December 1952. Although pollution levels in London are currently much lower than in the 1950s, the high levels recorded at the time are similar to the levels currently reported in industrialising economies such as India and China, so our study is relevant in particular to those settings. In early December 1952, pollution from residential and industrial chimneys, vehicle exhausts and coal burning became trapped under a layer of warm air due to a thermal inversion, which caused a thick smog to form over London. We investigate the long-term effects of exposure to this smog event by studying individuals' human capital and health outcomes in older age. The data we use is the UK Biobank: a large population-based cohort of approximately 500,000 individuals living in the United Kingdom. It includes rich data on individuals' later life health and economic outcomes, linked to administrative records. Using participants' eastings and northings of birth, our identification strategy exploits spatio-temporal variation in exposure to the London smog across birth dates and space using a difference-in-difference approach. In other words, we compare individuals who were exposed to the smog in the London area in early life to those living in other urban areas as well as to those conceived \textit{after} the smog, whilst controlling for local area-specific trends in the outcomes of interest across birth cohorts. Since we use those conceived \textit{after} the smog as the baseline (rather than those conceived \textit{before}), we refer to this as a `reverse' difference-in-difference approach.\footnote{We come back to the implications of this in Section~\ref{sec:methods}. Note that the term `reverse difference-in-difference' has also been used to refer to a difference-in-difference design where one group of units moves from control to treated, but another is treated throughout (i.e., the `control' group is always treated; see \citet{Kim2019}). They refer to our setting as a `time-reversed difference-in-difference'.} 

This paper has three main contributions. First, most of the literature that explores the effects of pollution exposure focuses on child \textit{birth} outcomes. Whilst it is important to better understand the effects on, e.g., infant mortality, it is one of the most extreme consequences of exposure to environmental toxins. Indeed, those who survive pollution exposure may be affected in other ways, with any `scarring' likely to reduce individuals' human capital and health potential. We investigate the effects on years of schooling, fluid intelligence, respiratory disease, cardiovascular disease, cancer and COVID-19 hospitalisations/mortality. Finding strong evidence of such longer-term effects would therefore indicate that any pollution impacts are much larger than what would be suggested by the literature focusing on the effects of pollution on birth outcomes. 

Second, we provide new empirical evidence in support of the DOHaD hypothesis. There is a large and growing literature in economics estimating the \textit{causal} developmental origins of later life economic and health outcomes. These have shown the consequences of many adverse circumstances, but generally lack evidence on the very long-term effects of pollution exposure.\footnote{The literature focusing on the contemporaneous effects of pollution exposure mainly shows large impacts on respiratory and cardiovascular disease as well as mortality, but also on brain health and cognitive decline \citep[see e.g.,][]{zhang2018impact, bishop2018hazed}. The literature suggests that high pollution concentrations affect lung function and cause irritation and inflammation of the respiratory, as well as cardiovascular system. Small pollution particles can penetrate into the lung tissue and interfere directly with the transfer of oxygen to the blood, as well as restrict blood flow to the heart. Both the elderly and the young are at increased risk of air pollution; the latter because their organs are still developing \citep{EPA2021} and because they inhale more air per body mass than adults \citep{Laskin2006}. In addition, because small particles can be passed through the placenta to the developing foetus, this can directly affect the oxygen available to the foetus, and with that, its development.} 
To examine the impacts of pollution exposure on outcomes in mid-adulthood, \citet{persico2020can} exploits the opening and closing of industrial plants to compare individuals who were exposed to pollution prenatally to its unexposed siblings. She finds that pollution reduces education and wages, and increases the likelihood of living in poverty as adults. 
We contribute to this literature by exploring the very long-term effects of early life exposure to pollution, investigating individuals' outcomes at age $\sim$60. Our identification is similar to \citet{bharadwaj2016early} and \citet{Ball2018}, who also focus on the long-term effects of the London smog on asthma and employment outcomes respectively. \citet{bharadwaj2016early} uses the English Longitudinal Study of Ageing (ELSA) to show that those exposed to the smog are more likely to develop asthma in later life, and \citet{Ball2018} uses the Office of National Statistics Longitudinal Study to show that those exposed to the smog were less likely to have a degree and worked fewer hours. An additional advantage of our data and setting is that it allows us to identify the gestational ages that are most sensitive to pollution. Indeed, with just 42 individuals exposed \textit{in utero} and 15 in infancy in \citet{bharadwaj2016early}, sample sizes do not allow for the analysis of trimester-specific effects. Although \citet{Ball2018} uses large samples, the analyses use individuals' \textit{year} of birth, implying it is not possible to look at gestation effects. An advantage of our data is that we have large samples as well as information on the year \textit{and month} of birth. This means we have more power to detect even relatively small effects on later life outcomes, and we are able to explore the importance of exposure at different gestational ages. With that, our research builds on studies examining the long-term effects of other relatively short-term events, such as the Ramadan \citep[see e.g.][]{almond2011health}, the Dutch Hunger Winter \citep[][]{lumey2011prenatal, bijwaard2021severe}, and temporary post-war confectionery derationing \citep[][]{berg2023prenatal}. 
In addition, our setting allows us to provide evidence on the human capital and health effects of pollution in a high pollution setting that is similar to current pollution levels in several industrialising countries, where evidence on the effects of pollution remains limited \citep{greenstone2015envirodevonomics}.

Our third contribution is that we explore heterogeneity of treatment effects with respect to three important sources of variation. First, we explore heterogeneity by gender. Since the literature suggests that male foetuses are frailer than female foetuses, we explore whether the long-term effects of pollution differ by gender due to either scarring or differential selection. Second, we investigate whether there is a social gradient in the effect of pollution exposure. For this, we characterise the local area that individuals are born in with respect to its social class and run our analysis separately for individuals in high and low social class areas. 

Finally, we build on a recent literature in social science genetics, directly modelling human capital and health outcomes as a function not only of individuals' environments (`nurture'), but also of their genetic predisposition to these outcomes (`nature') as well as the `nature-nurture' interaction. This acknowledges the major role that genetic variation has been shown to play in shaping individuals' life outcomes \citep[see e.g.][]{turkheimer2000three, polderman2015meta}, and allows nature and nurture to interact and \textit{jointly} contribute to individuals' human capital and health formation, as highlighted in the medical \citep[see e.g.,][]{Rutter2006} as well as economics literature \citep[see e.g.,][]{cunha2007technology}. 
Indeed, finding evidence of such `gene-environment interplay' provides a strong argument against ideas of genetic (or environmental) determinism. As such, we examine whether -- and to what extent -- one's genetic variation can protect against, or exacerbate, the effects of such adverse events. 
There is a large literature investigating the importance of gene-by-environment interactions ($G \times E$), with relatively recent contributions from economics and social science \citep[see, for example,][]{Biroli2015, bierut2018childhood, barth2020genetic, ronda2022family}. However, most existing studies tend to use endogenous environments, where it is not always clear how to interpret the main effects as well as the $G \times E$ interaction effect \citep[][]{biroli2022economics}.\footnote{Indeed, the coefficient on the genetic component may partially capture environmental circumstances due to `genetic nurture' (that is: parental genotypes can shape the offspring environment, and since the offspring's genetic variation is inherited from the parents, this may partially capture such environments; see e.g., \cite{belsky2018genetic,kong2018nature}), and the coefficient on the environmental circumstances may pick up variation driven by genetics due to gene-environment correlation (that is: the fact that individuals with a genetic predisposition to a specific trait can be more commonly found in certain environments).}  
We address this issue by exploiting the London smog as a natural experiment, ensuring that the environment is orthogonal to observed and unobserved individual characteristics. Hence, we add to only a small number of relatively recent studies that exploit \textit{exogenous} variation in the environment within a $G \times E$ setting, allowing us to identify the causal environmental impact within a $G \times E$ framework.\footnote{Other studies that exploit exogenous variation in the environment in a $G \times E$ setting include, e.g., \citet{Fletcher2012, schmitz2021impact, Schmitz2016, fletcher2019environmental, Barcellos2018, pereira2022interplay, biroli2021pubs, berg2023prenatal, berg2023early}, with \citet{muslimova2020complementarities} exploiting exogenous variation in \textit{both} genetic variation and environmental circumstances. See \citet{pereira2021} for a recent review of this literature.}

Our findings indicate large effects of both prenatal and childhood smog exposure on late-life fluid intelligence and -- to a lesser extent -- years of education. We also find a robust increase in the probability of being diagnosed with respiratory disease, but no differences in rates of cardiovascular disease, cancers, or COVID-19 hospitalisation/mortality. Furthermore, these effects are generally larger for individuals exposed in the first and second trimester of pregnancy, with overall reduced effect sizes for those exposed in the last trimester.

Although we cannot statistically distinguish the estimates for most of the subgroups in our data, the heterogeneity analysis suggests stronger effects on years of schooling for women, both for prenatal and early childhood exposure, whereas the estimates for fluid intelligence and respiratory disease are similar across the genders. Using the gender ratio as the outcome, we find no evidence of gender differences in survival, suggesting that the differential gender effects are driven by scarring rather than selection. Furthermore, our estimates suggest a social gradient in long-term pollution effects, with those born in higher social class areas (as proxied by a high proportion of the population being in professional, managerial or technical occupations) being less affected; similar to e.g., \citet{jans2018economic}. This in turn suggests either that the higher social classes were better able to avoid highly polluted areas, or that the health stock of lower class individuals is simply more vulnerable to adverse early life shocks. As Londoners at the time were not aware of the potential health risks of (severe) pollution, and there is little evidence of avoidance behaviour in the early 1950s, the former is perhaps less plausible, though we cannot say this with certainty.

Finally, we find that the negative effects of being exposed to the smog prenatally and in early childhood are generally stronger for those with a high genetic predisposition to the outcome. For respiratory disease, for example, this suggests that the respiratory health of individuals who are genetically predisposed is more vulnerable to severe pollution events compared to the health of individuals how are not genetically predisposed. 

The rest of the paper is structured as follows. Section~\ref{sec:background} provides the background to the London smog and Section~\ref{sec:data} describes the data used in our analysis. We set out the empirical strategy in Section~\ref{sec:methods}, and discuss the results in Section~\ref{sec:results}. We explore the sensitivity of our findings in Section~\ref{sec:robustness} and conclude in Section~\ref{sec:conclusion}.

\section{Background: The London smog}\label{sec:background}

On 4 December 1952, an anticyclone led to a temperature inversion over London, causing the cold air to be trapped under a layer of warm air. The resulting fog in combination with higher than usual coal smoke (due to the slightly colder temperature at the time) from residential and industrial chimneys, the pollution from vehicle exhausts (e.g. steam locomotives, diesel-fuelled buses) and other pollution (e.g. coal-fired power stations) formed a thick smog. 
With very little wind, it was not dispersed and led to an unprecedented accumulation of pollutants over the next five days, from 5--9 December 1952. 

\begin{figure}[!h]\caption{\label{fig:smoke_measurements_wilkins}Pollution and mortality during the London smog of December 1952.}\centering\includegraphics[width=\textwidth]{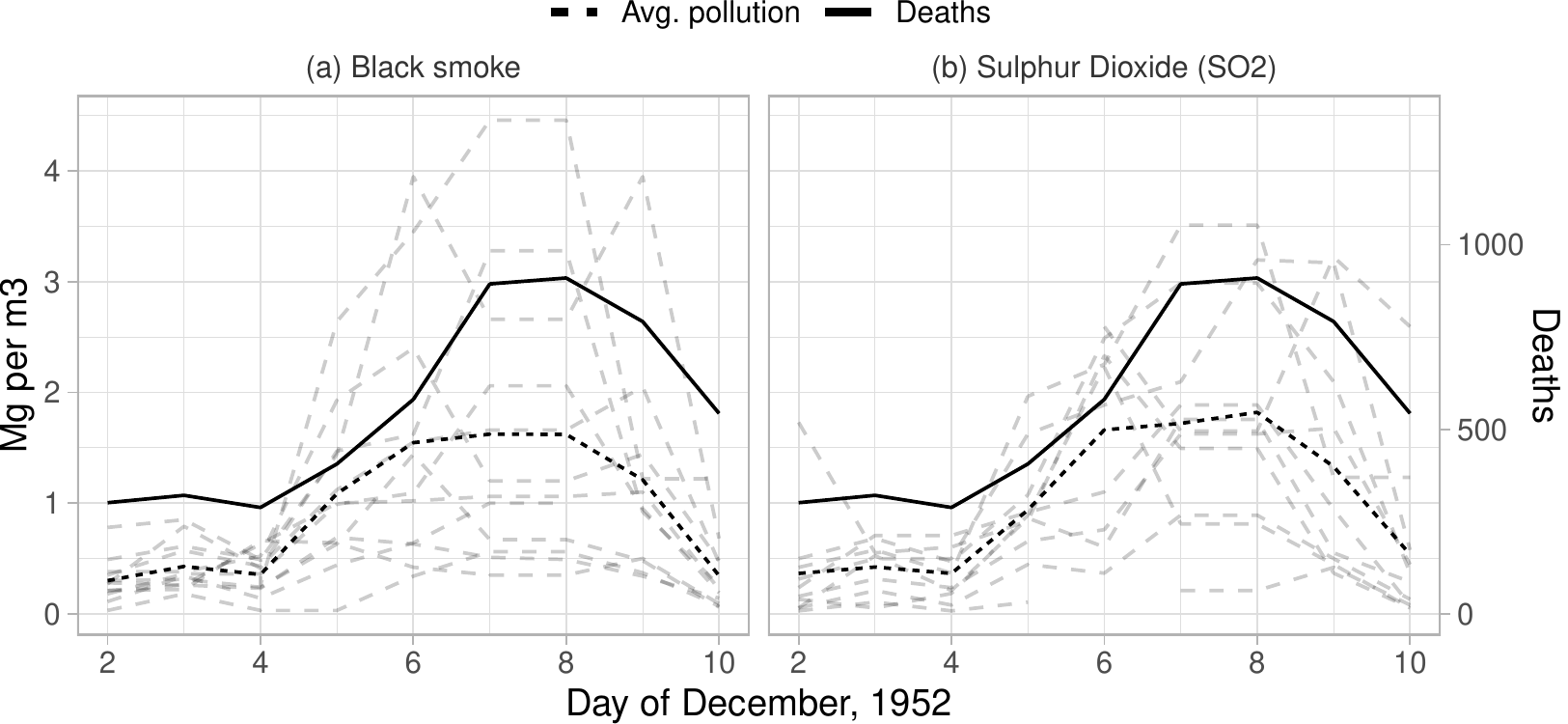}\vspace{0.5em}\caption*{\small{Historical measurements of pollution (black smoke and $\text{SO}_2$) from stations in London in December 1952. Each of the gray dashed lines represents the pollution measurements by a specific station. The dotted black line indicates the daily mean across all stations. The number of deaths in the Greater London area is overlaid with a solid black line. Pollution is digitised from Table~I in \citet{wilkins1954air} while deaths are digitised from Table~VIII in \citet{logan1953mortality}.}}\end{figure}

\citet{wilkins1954air} discusses the severity of the London smog in terms of changes in concentrations of black smoke and sulphur dioxide (SO$_2$), with the historical measurements for London from that paper presented in Figure~\ref{fig:smoke_measurements_wilkins}. This shows two interesting features. First, there is a rapid rise in both black smoke and SO$_2$ concentrations between 5 and 9 December, with average concentrations rising to multiple times their usual level, after which they returned to pre-smog levels.\footnote{Both black smoke and SO$_2$ are released into the atmosphere via fuel combustion, such as coal burning.
Consistent with the statistics shown in Figure~\ref{fig:smoke_measurements_wilkins}, \citet{wilkins1954air} reports that the average black smoke and SO$_2$ levels were approximately 4.7-4.75 times higher in London during the 1952 smog compared to December 1951, whilst they were 1.5-1.8 times higher in a group of `eight non-foggy towns’. Although this also suggests a small increase in pollution in other areas, the paper does not disclose the name or location of these `non-foggy' towns. This is important, since we cannot be certain that they do not include towns on the outskirts of London, which may have been partially affected. Indeed, \citet{Ball2018} shows trends and seasonality in pollution measurements for London as well as other urban (control) areas between 1950 and 1958 and shows a spike in pollution specifically in London for December 1952, with little change in the pollution for areas outside London. Either way, all the evidence points to a substantial increase in pollution levels in London relative to other areas.}
Second, there is substantial variation in pollution within London, indicated by the grey dashed lines, each representing a different measurement station. Note, however, that the black smoke concentrations shown here are likely to be underestimated. Indeed, the smoke filters that measured the pollution were so overloaded that concentrations were more likely to be around 7-8 $\text{mg}/\text{m}^3$ in the worst polluted areas of London \citep{WarrenSpringLaboratory1967}. 

Only levels of smoke and sulphur dioxide were measured at the time. However, as discussed in \citet{wilkins1954air}, it is likely that there were increases in tar, carbon monoxide (due to severe traffic congestion), carbon dioxide (due to a strong correlation with sulphur dioxide) and sulphuric acid (due to the oxidation of sulphuric dioxide). The nature of the smog also meant that it did not restrict itself to low-income areas, with the most affluent boroughs such as Kensington and Chelsea, as well as poorer boroughs such as Lambeth and Southwark being among the worst affected \citep{wilkins1954air}.

Although Londoners were used to smogs, the one in December 1952 was worse than any event Londoners had experienced before. Due to the dramatically reduced visibility, all public transport other than the London Underground was suspended, most flights to London were diverted, ambulance services stopped, and -- with its penetration into indoor areas -- concerts, theatres and cinema screenings were cancelled. Outdoor sporting events were also cancelled \citep[see, e.g.,][]{BBC1952}.\footnote{To the extent that this affected the health service, including potential pre- and post-natal care, this may have induced selection. We come back to the selection issue in more detail below. } 

Despite this, Londoners got on with everyday life, potentially since the health consequences of extreme pollution were unknown. However, medical statistics that were published in the following weeks showed a substantial increase in mortality, with an estimated 4,000 deaths caused by the smog. Indeed, the right vertical axis of Figure~\ref{fig:smoke_measurements_wilkins} presents the daily number of deaths over the period of the smog, depicted as the solid black line. This shows around 300 daily deaths before the smog, increasing to $\sim$900 at its peak, after which it reduced; a similar inverse U-shaped pattern as the pollution data \citep{logan1953mortality}.

Subsequent calculations showed that 90\% of the excess deaths were among those aged 45 and over \citep{MinistryOfHealth1954}. 
There was also an increase in mortality among newborns and infants, as well as foetal loss \citep{hanlon2022london, ball2018hidden}, but these capture a relatively small proportion of the total increase. We come back to this later. However, also in the months after the London smog, mortality exceeded normal levels.\footnote{Although an initial government report suggested these deaths were caused by influenza, there was no influenza outbreak in 1952, and \citet{Bell2004} find that only an extremely severe influenza epidemic could account for the excess deaths during this period. More recent analysis indeed suggests that the smog caused up to 12,000 deaths \citep{Bell2001}.} About half of these excess deaths were attributed to bronchitis or pneumonia, with other increases observed in respiratory tuberculosis, lung cancer, coronary disease, myocardial degeneration and other respiratory disease \citep{logan1953mortality}.

\section{Data}\label{sec:data}
Our primary dataset is the UK Biobank, a prospective, population-based cohort that contains detailed information on the health and well-being of approximately 500,000 individuals living in the United Kingdom.  Recruitment and collection of baseline information occurred between 2006 and 2010, when participants were 40--69 years old. The data include information on demographics, physical and mental health, health behaviours, cognition, and economic outcomes, obtained via questionnaires, interviews, and measurement taken by nurses. It has also been linked to hospital records, as well as the National Death Registry. Furthermore, samples of blood, urine and saliva have been collected, and all individuals have been genotyped. \citet{bycroft2018uk} give a detailed description of the sample. 

We are interested in the long term economic and health consequences of short term variation in the early-life pollution environment. We focus on a set of outcomes, each informed by previous literature on the effects of pollution.\footnote{The outcomes are selected based on their likely relationship with pollution according to the literature, taking into account data availability; we therefore do not use multiple hypothesis testing. In practice, this means that despite the existing literature showing an association between pollution and dementia (see e.g., \cite{bishop2018hazed}), we do not specify dementia as an outcome, since we only observe 64 hospitalisations linked to dementia in our sample of interest. This is driven by at least two factors with different implications. First, the UK Biobank is a sample of volunteers who have been shown to be healthier and wealthier than the general population \citep{fry2017comparison} and therefore are less likely to have dementia. The implication is that our sample is not necessarily representative of the population, something we return to in the conclusion. 
Second, the cohorts used in our analyses are still relatively young for dementia incidence to be significant. As shown in \autoref{fig:dementia_by_year_of_birth} in Appendix~\ref{sec:AppendixB}, dementia hospitalisations are substantially more common among those born in the early to mid 1940s, who are excluded from our sample window. It does suggest, however, that it would be interesting to repeat our analyses in 5--10 years' time, when dementia incidence has increased among the relevant cohorts. 
Similarly, we do not investigate the impacts of exposure to the smog on asthma, since asthma is more likely to be diagnosed at the GP, with few individuals being hospitalised with asthma ($n=453$ in our sample); we discuss this in Section~\ref{sec:results}.}
First, we build on the literature that shows medium-to-long term effects of early life pollution exposure on economic outcomes \citep[see e.g.,][]{almond2009chernobyl, Ball2018}, investigating the effects on educational attainment and fluid intelligence. Educational attainment is defined based on individuals' qualifications\footnote{Table~\ref{tab:years_educ_definition} in Appendix~\ref{sec:AppendixB} shows the mapping between qualifications and years of education, using a similar definition as in, e.g., \citet{Rietveld2013, Okbay2016, LeeEtAl2018, okbay2022polygenic}.}
 and fluid intelligence is a score based on problem solving questions that require logic and reasoning ability, independent of acquired knowledge. Unfortunately, there is no systematic data collection on individuals' labour market outcomes and hence we cannot explore this. 
 
Next, we build on the literature showing pollution effects on individuals' health \citep[see e.g.,][]{currie2011traffic} and explore the effects on respiratory disease, cardiovascular disease and cancer. We observe whether individuals are diagnosed with these conditions using the linkage to hospital records. As these records go back to 1997 for England and 1999 for Wales, we do not observe any earlier hospitalisations and hence, to the extent that individuals were diagnosed before these years but did not receive any treatment in the 10--15 years since, we may underestimate disease occurrence. 
The contemporaneous effects of air pollution on respiratory and cardiovascular disease are well-known. Less is known, however, about the potential long term effects of early life exposure. Indeed, since air pollution disproportionally affects individuals with compromised lung function, and much of the burden in adulthood is believed to be due to poor development of the lungs (rather than its accelerated decline; see e.g. \citet{lancet2019air}), early life pollution is a natural exposure to consider in the development of respiratory disease. 
Similarly, long-term exposure to air pollution has been shown to prematurely age blood vessels and lead to increased build up of calcium in the coronary artery, restricting blood flow to the heart \citep{kaufman2016association}. Both of these channels can in turn lead to cancer, for example via the build up of nanoparticles in the lungs affecting the replication of cells.

We create dummy variables to indicate whether the individual has been diagnosed with respiratory disease, cardiovascular disease, or cancer from the administrative hospitalisation data that have been merged into the UK Biobank. Furthermore, due to the links between respiratory disease and severe COVID-19 \citep{aveyard2021association}, and between short-term air pollution and COVID-19 \citep{persico2020deregulation, austin2020covid, isphording2021pandemic}, we additionally use a binary indicator for being hospitalised with, or having died from, COVID-19.\footnote{The hospitalisation data include all diagnoses in ICD-10 coding. We use ICD-10 J00-J99 to identify respiratory disease as diagnosis. We identify cardiovascular disease by codes I00-99 and G45, and cancer by ICD-10 codes C00-99. We use ICD-10 emergency codes U071 and U072 to identify COVID-19 related hospitalisations and deaths. The data on COVID-19 was last updated on 12/12/2021, including just under two years of UK COVID-19 morbidity and mortality, but excluding most infections with the (less severe) Omicron variant.} 

Using participants' eastings and northings of birth, we assign each individual one of the 1472 Local Government Districts of birth across England and Wales.\footnote{UK Biobank participants who indicated they were born in England, Wales, or Scotland were asked ``What is the town or district you first lived in when you were born?'' The interviewer then selected the corresponding place from a detailed list of place names in the UK, which were converted to north and east coordinates (British National Grid (OSBS1936)) with a 1km resolution. Our districts are defined based on the 1951 shapefiles from \href{https://www.visionofbritain.org.uk/}{Vision of Britain} \citep{humphrey2009vob}.} This spatial information, in combination with temporal information on individuals' year-month of birth, allows us to identify individuals who were exposed to the smog at different time points during the intrauterine and early childhood period. We split our sample along the time dimension by considering whether the prenatal period precedes, overlaps, or follows the smog event on December 5-9th, 1952. This allows us to define three groups: (i) those exposed to the smog during childhood (i.e., those born before the smog), (ii) those exposed to the smog \textit{in utero}, and (iii) those conceived after the smog event and therefore not exposed.\footnote{Note that we do not observe gestational age at birth. Hence, we assume that the prenatal period covers the nine months before the year-month of birth. The exact birth date cutoffs are as follows. Exposed in childhood: 1950-Dec to 1952-Nov. Exposed \textit{in utero}: 1952-Dec to 1953-Aug. Not exposed: 1953-Sep to 1956-Dec. We drop those born after December 1956 for two reasons. First, depending on their month of birth in 1957, individuals may have been directly affected by an educational reform -- the raising of the school leaving age -- which has been shown to have affected individuals' longer-term education as well as health outcomes \citep[see e.g.][]{HarmonWalker1995, DaviesEtAl2018}, though note that the evidence on the health effects are more mixed \citep[see e.g.,][]{ClarkRoyer2013}. Second, the first Clean Air Act allowed local authorities to create Smoke Control Areas; areas that prohibited all smoke emissions. The first orders of such Smoke Control Areas were announced in 1957 \citep{fukushima2021uk}. By dropping all births in 1957 onwards from our analysis, we avoid our estimates potentially capturing reductions in pollution due to the Smoke Control Areas. Finally, we restrict our analysis to those potentially exposed at age 0 and age 1, since the literature suggests that the first 1,000 days of a child's life is a crucial period in shaping individuals' outcomes, with effects of adverse circumstances reducing as children age.}

\begin{figure}[!h]\caption{\label{fig:exposure_indicator_coverage}Visibility and pollution measurements during the London smog.}\centering\includegraphics[width=0.8\textwidth]{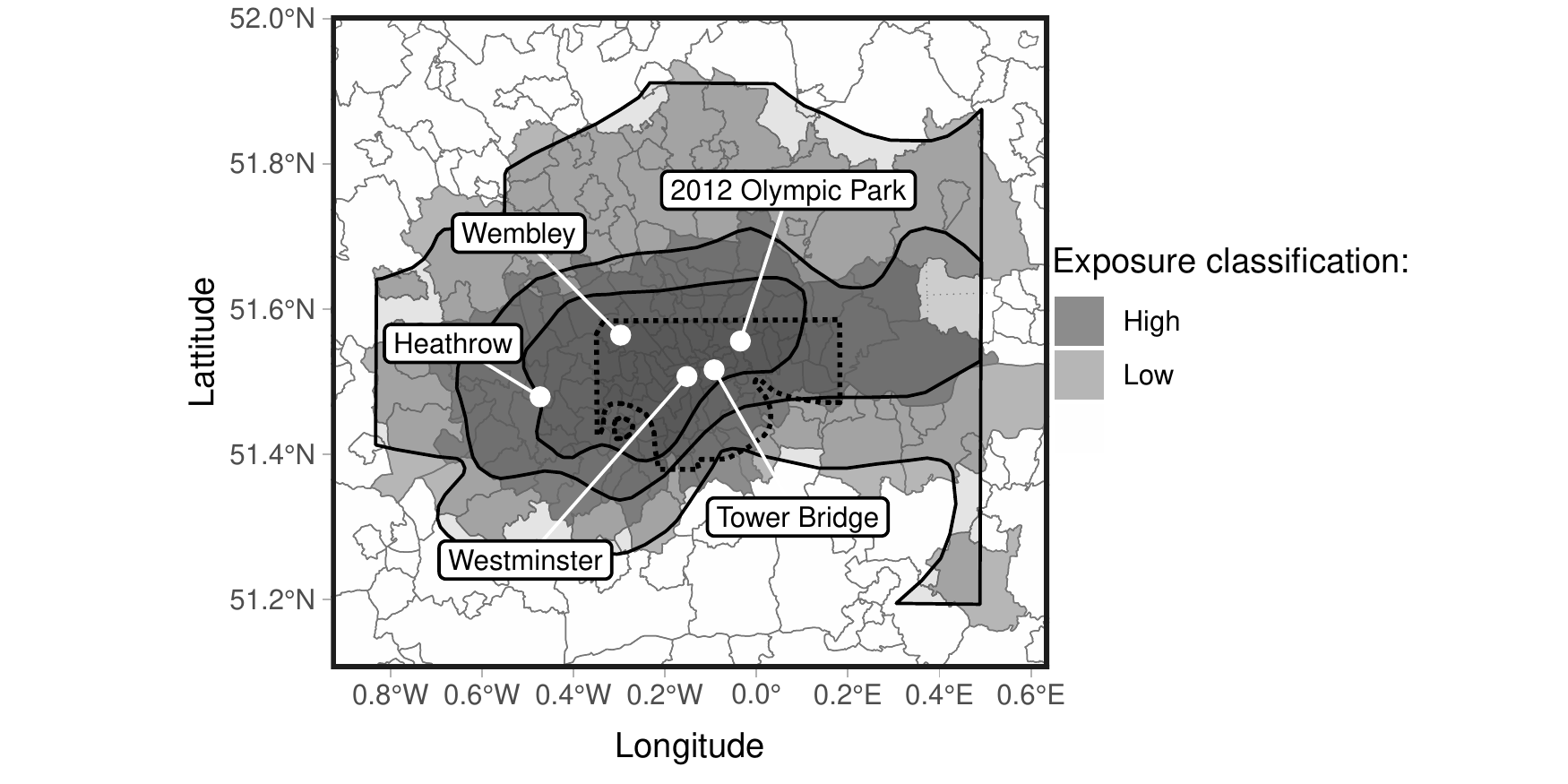}\caption*{The geographic boundaries of the London smog based on the maps in \citet{wilkins1954air}. The solid black outlines show the areas with reduced visibility. The inner boundaries experienced a more severe reduction in visibility. The dotted outline shows the area with high sulphur dioxide measurements. The map classifies the districts into `high exposure' (dark gray), `low exposure' (light gray), and `unexposed' (white) districts. The white dots mark London landmarks (i.e., Heathrow, Wembley, Westminster, Tower Bridge, and the 2012 Olympic Park) to help show the scale of the map.}\end{figure}

We split our sample along the spatial dimension by identifying the geographical areas in and around London that were exposed to high pollution during the smog. To do so, we overlay the reduced visibility and sulphur dioxide measurements from \citet{wilkins1954air} onto a district-level shapefile, where we highlight some well-known landmarks for a better sense of scale. This is shown in Figure~\ref{fig:exposure_indicator_coverage}, where the solid black outlines indicate the areas with low and severe reductions in visibility and the dotted outline indicates the area with high sulphur dioxide measurements.\footnote{The sulphur dioxide boundary, based on \citet{wilkins1954air}, shows measurements from different stations with limited geographical coverage, resulting in a boundary with a sharp border, while the visibility boundary is based on observations recorded by the Meteorological Office at 9am and 6pm throughout the smog event \citep{wilkins1954air}.} 
We define ``high exposure'' districts as those that experienced severe reductions in visibility (i.e., overlap with the two inner solid boundaries in Figure~\ref{fig:exposure_indicator_coverage}) and/or experienced high sulphur dioxide measurements (i.e., overlap with the dotted boundary in Figure~\ref{fig:exposure_indicator_coverage}). Districts that only overlap with the outer solid boundary, indicating the mildest reduction in visibility, are classified as ``low exposure''. In our main analysis, we do not distinguish between the high and low exposure districts but instead refer to them jointly as ``treated'' districts.
We compare these treated districts to a set of ``control'' districts that are defined as other urban districts in England and Wales with a population density exceeding $400$ individuals per $\text{km}^2$. In our robustness checks, we explore the sensitivity of our results to control districts with different population densities, to excluding the ``low exposure'' districts, to assigning exposure based on individuals' reported birth \textit{locations}, as well as by defining control districts as other major cities in England and Wales.\footnote{The control districts used in the main analysis are shown on a map in Figure~\ref{fig:districts_density}, Appendix~\ref{sec:AppendixB}, and colourised according to their population density. The data on population density is from Vision of Britain \citep{southall2011rebuilding}. 
\citet{Ball2018} argues that the only other city with unusually high pollution at the time of the London smog was Leeds. We therefore drop Leeds in all our analyses.}

Our sample selection process is as follows: we only consider the subsample of 332,393 individuals born in either treated or control districts, and restrict the sample to those born from December 1950 to December 1956, resulting in 66,139 individuals. Furthermore, we follow the (genetics) literature and restrict our sample to those of white European ancestry, dropping 1,058 individuals.\footnote{Because genetic variation differs by ancestry, this accounts for population stratification; a form of genetic confounding. We discuss the genetic data as well as its interpretation in more detail in Appendix~\ref{sec:AppendixA}. While we drop individuals of non-European ancestry in our main analysis, our results are robust to including all genetic ancestries; we discuss this in Section \ref{sec:results}.} This leaves us with between 26,877--65,060 participants for the main analysis, depending on the outcome of interest. 

Given the potential importance of the weather for exposure to smog as well as impacts of the weather on individuals' health \citep[see e.g.][]{hanlon2021temperature}, we merge in two auxiliary datasets that give an interpolated grid of measurements on ambient temperature, sunshine, wind speed, and rainfall at the time of the smog. For temperature, rain and sunshine, these data are available from the MET Office \citep{METoffice}, while, for wind, we use the ERA5 reanalysis data \citep{era5}. The datasets have an approximate grid resolution of 25km and we assign measurements to individuals by linking their location of birth, as measured in eastings and northings, to its nearest grid point. We merge in the weather data at individuals' birth location for the period of the smog. For ambient temperature, we assign the minimum temperature measured during the smog, while for sunshine, wind speed and rainfall, we use the average. Because the weather measurements link to individuals' eastings and northings of birth (rather than their birth \textit{districts}), these measurements vary \textit{within} districts and capture local atmospheric conditions experienced at individuals' birth locations at the time of the smog. Most of the variation in weather conditions, however, comes from between-district variation, with the within-district variation being relatively small.\footnote{Figure~\ref{fig:weather} in Appendix~\ref{sec:AppendixB} shows the monthly time series of weather conditions for our sample, distinguishing between treated and control districts. As mentioned in the introduction, this shows a slightly lower temperature in exposed districts at the time of the smog. However, the difference between treated and control districts is minimal (less than 0.5$^\circ$C). Hence, these do not suggest any notable differences in weather conditions between the district types.}

\autoref{tab:descriptives} presents the descriptive statistics on the full sample (column 1) as well as distinguishing between the treated (column 2) and control (column 3) samples. Panel~A shows the statistics for individuals, while Panel~B uses data from the Registrar General reports \citep{Southall2004} and the 1951 UK Census to calculate district-level descriptives. Panel~A shows that 44\% of the sample is male and individuals, on average, have 13.3 years of education. Individuals' fluid intelligence is only available for a smaller sample of individuals, which we standardise to have mean zero and unit variance on the full sample. 14.2\% of the individuals have been hospitalised with cancer, while it is 19.2\% for cardiovascular disease and 9.2\% for respiratory disease. Finally, 0.7\% of our sample has been either hospitalised or has died with COVID-19 as primary or secondary cause. For the districts, Panel~B shows an average of 15 births and 13 deaths per 1,000 population, and an infant ($<1$ year) and neonatal ($<4$ weeks) mortality rate of 30 and 19 respectively per 1,000 live births. The main occupation share is from skilled occupations (53\%).
For most of the variables, treated districts perform slightly `better' than control districts, with individuals in treated districts having higher education and fluid intelligence, as well as a lower probability of being hospitalised. Comparing the shares of workers in the five occupation classes of the 1951 census, control districts have a larger share of partly skilled workers, while treated districts have a higher share of workers in professional and intermediate occupations. The shares of skilled and unskilled workers are similar.

\begin{table}

\caption{\label{tab:descriptives}Descriptive statistics for the main outcomes and variables.}
\centering
\begin{threeparttable}
\fontsize{9}{11}\selectfont
\begin{tabular}[t]{ldrrdrrdrr}
\toprule
\multicolumn{1}{c}{\em{}} & \multicolumn{3}{c}{\em{Estimation sample}} & \multicolumn{3}{c}{\em{Treated}} & \multicolumn{3}{c}{\em{Control}} \\
\cmidrule(l{3pt}r{3pt}){2-4} \cmidrule(l{3pt}r{3pt}){5-7} \cmidrule(l{3pt}r{3pt}){8-10}
\multicolumn{1}{c}{} & \multicolumn{1}{c}{Mean} & \multicolumn{1}{c}{SD} & \multicolumn{1}{c}{Obs.} & \multicolumn{1}{c}{Mean} & \multicolumn{1}{c}{SD} & \multicolumn{1}{c}{Obs.} & \multicolumn{1}{c}{Mean} & \multicolumn{1}{c}{SD} & \multicolumn{1}{c}{Obs.}\\
\midrule
\addlinespace[0.3em]
\multicolumn{10}{l}{\textbf{Panel (a) -- Individuals}}\\
\hspace{1em}Male & 0.443 & 0.497 & 65,060 & 0.447 & 0.497 & 11,011 & 0.443 & 0.497 & 54,049\\
\hspace{1em}Educational attainment & 13.317 & 2.274 & 64,681 & 13.791 & 2.242 & 10,953 & 13.221 & 2.268 & 53,728\\
\hspace{1em}Fluid intelligence & -0.000 & 1.000 & 26,877 & 0.078 & 1.003 & 5,666 & -0.021 & 0.998 & 21,211\\
\hspace{1em}Cancer & 0.142 & 0.349 & 64,926 & 0.140 & 0.347 & 10,991 & 0.142 & 0.349 & 53,935\\
\hspace{1em}Cardio-vascular disease & 0.192 & 0.394 & 64,933 & 0.167 & 0.373 & 10,991 & 0.197 & 0.398 & 53,942\\
\hspace{1em}Respiratory disease & 0.092 & 0.290 & 64,923 & 0.080 & 0.272 & 10,988 & 0.095 & 0.293 & 53,935\\
\hspace{1em}COVID-19 & 0.007 & 0.081 & 65,060 & 0.005 & 0.071 & 11,011 & 0.007 & 0.083 & 54,049\\
\addlinespace[0.3em]
\multicolumn{10}{l}{\textbf{Panel (b) -- Districts}}\\
\hspace{1em}Shares of occupations &  &  &  &  &  &  &  &  & \\
\hspace{1em}\hspace{1em} -- Professional & 0.036 & 0.026 & 805 & 0.053 & 0.036 & 121 & 0.034 & 0.023 & 684\\
\hspace{1em}\hspace{1em} -- Intermediate & 0.147 & 0.058 & 805 & 0.169 & 0.059 & 121 & 0.143 & 0.057 & 684\\
\hspace{1em}\hspace{1em} -- Skilled & 0.531 & 0.055 & 805 & 0.537 & 0.056 & 121 & 0.530 & 0.055 & 684\\
\hspace{1em}\hspace{1em} -- Partly skilled & 0.157 & 0.067 & 805 & 0.115 & 0.033 & 121 & 0.164 & 0.069 & 684\\
\hspace{1em}\hspace{1em} -- Unskilled & 0.129 & 0.047 & 805 & 0.126 & 0.056 & 121 & 0.129 & 0.045 & 684\\
\hspace{1em}Birth rate & 0.015 & 0.002 & 809 & 0.015 & 0.002 & 124 & 0.015 & 0.003 & 685\\
\hspace{1em}Death rate & 0.013 & 0.003 & 809 & 0.011 & 0.002 & 124 & 0.014 & 0.003 & 685\\
\hspace{1em}Death rate, $<$ 1 year & 0.030 & 0.014 & 809 & 0.024 & 0.008 & 124 & 0.031 & 0.015 & 685\\
\hspace{1em}Death rate, $<$ 4 weeks & 0.019 & 0.013 & 809 & 0.017 & 0.007 & 124 & 0.020 & 0.014 & 685\\
\bottomrule
\end{tabular}
\begin{tablenotes}
\item Panels: (a) descriptives calculated on individual-level data, (b) descriptives calculated at the district-level using data from the Registrar General reports and the 1951 census. The columns report sample means, sample standard deviations, and number of observations, in, respectively, the full sample, the treated districts and the control districts. Birth and death rates are measured per 1,000 births/population. 
\end{tablenotes}
\end{threeparttable}
\end{table}

\FloatBarrier

\section{Empirical strategy}\label{sec:methods}
To investigate the long term effects on human capital and health outcomes of early life pollution exposure, we exploit spatio-temporal variation in exposure to the London smog across birth dates and residential locations using a `reverse difference-in-difference' approach. We observe the eastings and northings of individuals' residential location at birth, with a 1km resolution. We distinguish between those born inside and outside the exposed London area (spatial variation) while also considering the timing of birth relative to the smog event (temporal variation). Our main specification is:
\begin{align}\label{eq:E}
Y_{ijt} &= \alpha_{j} + \gamma_t + \tau_{k} t + \beta_{IU} E^{IU}_i \times L_j + \beta_{CH} E^{CH}_i \times L_j + \theta_{IU} E^{IU}_i + \theta_{CH} E^{CH}_i + \delta \mathbf{X}_{i} + \epsilon_{ijt},
\end{align}
where $Y_{ijt}$ denotes the outcome of interest for individual $i$, born in district $j$ in year $t$. Thus, $\alpha_j$ denote district fixed effects, and $\gamma_t$ are year of birth fixed effects. We additionally include administrative-county-specific time (year-month) trends, denoted by $\tau_k t$.\footnote{In our sensitivity analysis in Appendix~\ref{sec:trend_robustness}, we show that our results are robust to including administrative county-specific year (as opposed to year-month) trends, district-specific year or year-month trends, and not including any additional trends.} The vector $\mathbf{X}_i$ includes weather conditions during the smog, gender, and month-of-birth dummies to account for weather effects, gender differences, and seasonality in the outcome. Recall that the weather conditions are linked to individuals' eastings and northings, meaning that we can identify their impacts also in a specification that includes district fixed effects. 
The indicators $E^{IU}_i$ and $E^{CH}_i$ are dummy variables that are equal to one for individuals who are in utero and in childhood (i.e., $<$age 2) respectively during the time of the smog, while $L_j$ is a binary variable that indicates whether the individual is born in an area of London that was exposed to the smog (i.e., in treated districts). Hence, our identification strategy compares individuals' outcomes $Y_{ijt}$ for those exposed at different ages (i.e., \textit{in utero} and in early childhood) to those conceived after the London smog in treated districts, relative to others born at the same time, but in control districts.\footnote{We here assume no residential mobility around birth. Although we cannot test this directly, since we do not observe early life house moves, the evidence suggests that residential mobility was low. Indeed, \citet{falkingham2016residential} show that 7–10\% of 0–19 year old children moved between 1938–1947. Note that this is likely to be an overestimate of the true residential mobility at this time, as the period for which is was calculated includes the Second World War, where millions of children were evacuated out of urban areas. Indeed, exploiting the birth location data of the smaller sibling sample of the UK Biobank, \citet{vonHinkeVitt2023} estimate the probability of moving during this period and distinguish it from reporting error. They find that house moves are relatively rare: each additional year between the birth of siblings increases the probability of moving by approximately one percentage point. Hence, we find no evidence of substantial migration during our period of interest.}
The parameters of interest are therefore $\beta_{IU}$ and $\beta_{CH}$ which parameterise, respectively, the long-term effects of being exposed to the smog \emph{in utero} and in early childhood, relative to non-exposed cohorts, accounting for any administrative county-specific trends in $Y_{ijt}$, time and district fixed effects. We use robust standard errors, clustered by district throughout.

Our specification implicitly assumes that individuals who were born in districts that were affected by the London smog would have had similar trends in their outcomes of interest in the absence of the smog compared to those born in districts that were not affected. We explore this common trend assumption empirically in Appendix~\ref{sec:AppendixC}. Since those born \textit{prior to} the smog may have been exposed in early childhood, we should find common trends among those conceived \textit{after} the smog in treated and control districts.\footnote{We purposely do not consider those born \textit{before} the smog as an (alternative) control group, since these cohorts were likely to be affected by an earlier smog that affected the same London areas as the one in 1952. More specifically, London experienced another major smog on 26 November 1948 that lasted for six days. Although it was less severe than the 1952 smog, it similarly caused severe traffic disruptions, accidents and cancellation of events. It also led to a strong increase in mortality \citep{wilkins1954air}. Empirically, this implies that individuals born before the 1952 smog cannot be considered a suitable control group. Indeed, those born before 1948 include individuals exposed to the 1948 smog in early childhood, whilst the 1948--1949 cohorts were exposed to the 1948 smog prenatally and to the 1952 smog at ages 2--3. In other words, each of these alternative control groups are in effect also 'treated'. Running the analysis with these alternative control groups shows no impacts on our outcome variables. Specifying the controls to be born \textit{after} the 1952 smog, when there were no other major smogs until 1962, ensures a `clean' control group not affected by other major pollution events.} We find no evidence to suggest that those born in treated districts have differential trends in the outcomes of interest compared to those born control districts. 

Because we use a `reverse' difference-in-difference approach, comparing individuals conceived before/during the smog (i.e., affected \textit{in utero} or in childhood) to those conceived \textit{afterwards}, the validity of our approach also relies on the assumption that the outcomes of those conceived after the smog are the same as what would have been observed in the absence of the smog. This implicitly assumes away differential pollution avoidance behaviour post-smog in treated and control districts. Indeed, if individuals in both treated and control districts engaged in similar avoidance behaviour (e.g., closing windows in polluted areas), our difference-in-difference approach cancels these out and we would still obtain an unbiased estimate of the effect of exposure to the London smog. However, if those in treated districts engaged in more pollution avoidance behaviour compared to those in control districts, we may overestimate the impact of the London smog. In this case, our estimate captures (i) the actual effect of increased smog exposure as well as (ii) a larger decrease in exposure post-smog among those in treated districts compared to those in control districts, overestimating the difference. 

Similarly, we assume that individuals' residence choices did not change systematically after the smog. Although we cannot test this directly, we explore the validity of this assumption indirectly in two ways. First, we examine changes in the socio-economic composition of treated and control districts pre- and post-smog. In the absence of annual systematic data collection on each districts' socio-economic composition, we use the 1951, 1961 and 1971 censuses and plot the average district-level share of high and low social class households (defined using individuals' occupation) in each of the census years. The results, presented in \autoref{fig:census_ses_shares} in Appendix~\ref{sec:AppendixB}, shows that the average share of high and low social class households in treated and control districts remains relatively stable over time. We find a slight increase in the share of low social class households in treated districts between 1951 and 1961, but the change is negligible, both in absolute terms (0.013 percentage points) and in relative terms (relative to a standard deviation of 0.08). 

In the absence of data on household moves or parental residential location choices, the second way in which we explore the validity of the assumption that the London smog did not affect individuals' choice of residence is by exploiting the location of birth for a smaller set of sibling pairs (identified using the genetic kinship data). 
We specify a binary variable $move_{st}$ to equal one if sibling pair $s$ with (firstborn's) year of birth $t$ reports different locations of birth. Within our sample of sibling pairs, some are \textit{both} born before the smog, whilst for others, one sibling is born before and one after the smog. We restrict this sample to sibling pairs where the first sibling was born between 1946-1952 in the same districts as those used in our main analysis, and where the siblings are at most five years apart, yielding $n=3,079$ sibling pairs. We denote the group of sibling-pairs where the youngest sibling was born after the smog as $Post$ and run the following regression:
\begin{align}\label{eq:move}
move_{st} &= \eta_0 + \eta_1 Post_t + \eta_2 Treated_s + \eta_3 Treated_s \times Post_t + \eta_4 age_{s} + u_{st},
\end{align} 
where $Treated_s$ is a binary indicator that equals one if the older sibling is born in treated districts, and $age_s$ captures the age of the oldest sibling when at the birth of the younger sibling. We run this regression both with and without (the older child's) district, year-of-birth and month-of-birth fixed effects, capturing whether -- for cohorts where one sibling is born after the smog -- they are more likely to move if the first child was born in treated compared to control areas. Note that $Treated_s$ drops out when we include district fixed effects. We report the estimates in \autoref{tab:moving_probability} in Appendix~\ref{sec:AppendixB}. The parameter $\eta_1$ essentially captures a cohort effect, indicating whether sibling pairs with a younger sibling born after the smog are systematically different in terms of moving probabilities compared to those where both siblings were born before the smog. We find no such evidence. In the specification without fixed effects, the parameter $\eta_2$ however, shows that sibling pairs where the first sibling was born in a treated district are 25 percentage points more likely to move in between the births of two siblings compared to sibling pairs where the first sibling was born in a control district. We find no evidence though that this probability is different for sibling pairs where the youngest was born after the smog, as captured by the parameter on the interaction ($\eta_3$). The estimates, both with and without fixed effects, are small (2 and 4 percentage points, respectively) and statistically insignificant, suggesting that the London smog did not significantly increase the probability of moving for those with an older sibling born in treated districts. 

Another important issue in the above specification is potential (foetal) selection. Indeed, there is evidence of increased mortality among newborns and infants \citep{MinistryOfHealth1954}, as well as foetal loss \citep{hanlon2022london, ball2018hidden}. Although these effects were relatively small as the smog mainly affected deaths among the elderly \citep{MinistryOfHealth1954}, they do affect our analysis and interpretation. More specifically, assuming that the smog increased mortality among relatively frail infants, leaving the stronger ones to survive, this may have led to an improvement in average cohort-level human capital and health outcomes for those exposed in the affected districts. This, in turn, suggests that our estimates may be underestimates of the effects of interest. We report our analyses on the impacts of the smog on the annual birth, death and infant mortality rate in more detail in \autoref{fig:impacts_on_rates} in Appendix~\ref{sec:AppendixB}, where we find strong evidence of an increase in total annual deaths in 1952 and 1953, some evidence of a reduction in births in 1953 -- potentially driven by foetal loss -- but no significant effects on infant mortality. 

Finally, the fact that the London smog only lasted five days additionally allows us to explore the gestational ages that are most sensitive to pollution exposure. This identification exploits the fact that the period of pollution exposure is substantially shorter than the length of gestation. To do this, we replace $E^{IU}_i$ in \autoref{eq:E} with three binary variables indicating the trimester in which the individual was exposed to the London smog. 

\section{Results}\label{sec:results}
\subsection{The London smog}

\noindent We start by examining the long-term impact of pollution exposure on human capital outcomes. Columns (1) and (2) of \autoref{tab:lon_vs_urb-economic_outcomes} show the estimates from \autoref{eq:E} for years of education and fluid intelligence respectively. We find no strong differences in the outcomes among individuals exposed to the London smog prenatally or in early childhood in control districts compared to those conceived afterwards. However, within treated districts, we find a reduction in fluid intelligence for those exposed \textit{in utero} as well as in childhood. The latter show the largest negative effects, with $0.16$ standard deviations lower fluid intelligence score compared to those born at the same time in control districts. Being exposed \emph{in utero} reduces fluid intelligence by $0.11$ standard deviations. Looking at the estimates for years of education, we find that they are of a similar magnitude, but they are not significantly different from zero at conventional levels. 
Table~\ref{tab:lon_vs_urb-qualifications} in Appendix~\ref{sec:AppendixB} presents the results where we distinguish between different qualifications, showing the effect of exposure to the smog on the probability of exiting the education system with a degree (column 1), upper secondary (column 2; i.e., A/AS levels, or professional/vocational training), lower secondary (column 3; i.e., CSEs, GCSEs, O-levels), and no qualifications (column 4). This shows that the negative estimates for pre- and post-natal pollution exposure are mostly driven by a reduction in the probability of obtaining a degree or upper secondary qualification, and -- correspondingly -- a higher probability of exiting the education system with lower secondary qualifications.

\begin{table}[!h]

\caption{\label{tab:lon_vs_urb-economic_outcomes}Difference-in-Difference estimates comparing treated to control districts defined as urban England and Wales.}
\centering
\begin{threeparttable}
\begin{tabular}[t]{ldd}
\toprule
\multicolumn{1}{c}{\em{}} & \multicolumn{2}{c}{\em{Dependent variable:}} \\
\cmidrule(l{3pt}r{3pt}){2-3}
\multicolumn{1}{c}{} & \multicolumn{1}{c}{(1)} & \multicolumn{1}{c}{(2)} \\
\multicolumn{1}{c}{ } & \multicolumn{1}{c}{{\specialcell[b]{Educational \\ attainment}}} & \multicolumn{1}{c}{{\specialcell[b]{Fluid \\ intelligence}}}\\
\midrule
Treated $\times$ In utero & -0.100^{  } & -0.112^{ ** }\\
 & (0.089) & (0.051)\\
Treated $\times$ Childhood & -0.135^{  } & -0.158^{ ** }\\
 & (0.088) & (0.068)\\
In utero & -0.003^{  } & 0.008^{  }\\
 & (0.054) & (0.036)\\
Childhood & -0.014^{  } & -0.051^{  }\\
 & (0.120) & (0.074)\\
\midrule
\multicolumn{1}{l}{Observations} & \multicolumn{1}{c}{64,681} & \multicolumn{1}{c}{26,877}\\
\multicolumn{1}{l}{Mean dep. var.} & \multicolumn{1}{c}{13.317} & \multicolumn{1}{c}{0}\\
\multicolumn{1}{l}{$R^2$} & \multicolumn{1}{c}{0.08} & \multicolumn{1}{c}{0.067}\\
\bottomrule
\end{tabular}
\begin{tablenotes}
\item Columns: (1) educational attainment in years, (2) standardised fluid intelligence score. Includes fixed-effects for district, month of birth, and year of birth. Also controls for gender, weather variation, and year-month linear time trends by administrative county. Urban England and Wales are defined as districts that had a population density above 400 individuals per km$^2$ in 1951. Standard errors are clustered by district. (*): $p < 0.1$, (**): $p<0.05$, (***): $p<0.01$.
\end{tablenotes}
\end{threeparttable}
\end{table}

We now turn to the long-term consequences of pollution on health outcomes. 
\autoref{tab:lon_vs_urb-health_outcomes}, Column (1) presents the estimates from \autoref{eq:E} for respiratory disease. This shows an increase in the prevalence of respiratory disease for those exposed to the smog prenatally. More specifically, those \textit{in utero} during the smog and born in the exposed areas are two percentage points more likely to be diagnosed with respiratory disease compared to those conceived after the smog. Considering that the overall incidence of respiratory disease is 9\% in the sample, this is a large effect, similar to a 22\% increase.\footnote{Although the existing literature suggests long-term impacts of the smog on the development of asthma \citep{bharadwaj2016early}, we do not include these analyses for two reasons. First, our focus is on inpatient hospital admissions, and we observe very few hospitalisations due to asthma ($n=453$). Hence, we do not have sufficient variation to identify such effects. Second, we do not use self-reported asthma due to the substantial discrepancies between objective and subjective health conditions, as well as the strong income gradient in false negative reporting \citep{johnston2009comparing}. Our analysis on these outcomes (not shown here, but available upon request) shows no impact on asthma hospitalisations, but a three percentage point reduction in self-reported asthma for those exposed in utero. In light of the evidence on the discrepancies between objective and subjective health measures, however, it is difficult to interpret these estimates.} To explore which respiratory conditions drive this increase in hospitalisation risk, we estimate separate regressions for each of the ICD-10 sub-categories that make up our respiratory outcome and report the results in \autoref{tab:lon_vs_urb-icd_j_subgroups} in Appendix~\ref{sec:AppendixB}. These estimates show that our results are driven by hospitalisations relating to influenza and pneumonia (J09-J18), with no impacts on any of the other ICD-10 sub-categories. To explore this further, we report this as a separate outcome in Column~(2) of \autoref{tab:lon_vs_urb-health_outcomes} as well as throughout our further analyses. 
Next, we examine whether the negative effect on respiratory disease translates into COVID-related deaths or hospitalisations in Column~(3). With coefficients that are tightly estimated around zero, we find no evidence of increased COVID-related morbidity or mortality, and similarly, in Column~(4), we do not find increased prevalence of hospitalisations due to cancer. Finally, in Column~(5), we see no statistically significant increase in risk of being hospitalised with cardiovascular disease, but we note that the estimates both for prenatal and childhood exposure are larger in (absolute) magnitude than the estimates seen for cancer and COVID-19, giving some suggestive evidence of a relationship between smog exposure and late-life cardiovascular disease.\footnote{\autoref{tab:lon_vs_urb-robustness_add_regressors} in Appendix~\ref{sec:AppendixB} presents the estimates accounting for different configurations of regressors and fixed effects, starting with no covariates. They show relatively stable effect estimates for each of the specifications. Taking into account potential treatment effect heterogeneity, \autoref{tab:lon_vs_urb-borusyak} shows that we obtain very similar estimates of the average treatment effect when using the estimation approach of \citet{borusyak2021revisiting}. Finally, \autoref{tab:lon_vs_urb-economic_outcomes_no-ancestry} and \autoref{tab:lon_vs_urb-health_outcomes_no-ancestry} show the results that do not restrict the sample to only include those of white European ancestry, giving very similar results.} 

\begin{table}[!h]

\caption{\label{tab:lon_vs_urb-health_outcomes}Difference-in-Difference estimates comparing treated to control districts defined as urban England and Wales.}
\centering
\begin{threeparttable}
\fontsize{10}{12}\selectfont
\begin{tabular}[t]{lddddd}
\toprule
\multicolumn{1}{c}{\em{}} & \multicolumn{5}{c}{\em{Dependent variable:}} \\
\cmidrule(l{3pt}r{3pt}){2-6}
\multicolumn{1}{c}{} & \multicolumn{1}{c}{(1)} & \multicolumn{1}{c}{(2)} & \multicolumn{1}{c}{(3)} & \multicolumn{1}{c}{(4)} & \multicolumn{1}{c}{(5)} \\
\multicolumn{1}{c}{ } & \multicolumn{1}{c}{{\specialcell[b]{Respiratory, \\ any}}} & \multicolumn{1}{c}{{\specialcell[b]{Influenza/ \\ pneumonia}}} & \multicolumn{1}{c}{{\specialcell[b]{COVID-19}}} & \multicolumn{1}{c}{{\specialcell[b]{Cancer}}} & \multicolumn{1}{c}{{\specialcell[b]{Cardio- \\ vascular}}}\\
\midrule
Treated $\times$ In utero & 0.020^{ * } & 0.012^{ ** } & 0.000^{  } & -0.003^{  } & 0.011^{  }\\
 & (0.011) & (0.005) & (0.003) & (0.013) & (0.015)\\
Treated $\times$ Childhood & -0.007^{  } & 0.009^{  } & -0.001^{  } & -0.001^{  } & 0.007^{  }\\
 & (0.012) & (0.006) & (0.003) & (0.014) & (0.019)\\
In utero & -0.005^{  } & -0.006^{  } & 0.000^{  } & -0.005^{  } & -0.004^{  }\\
 & (0.006) & (0.004) & (0.002) & (0.008) & (0.010)\\
Childhood & -0.008^{  } & -0.002^{  } & -0.001^{  } & -0.009^{  } & -0.015^{  }\\
 & (0.014) & (0.007) & (0.004) & (0.016) & (0.023)\\
\midrule
\multicolumn{1}{l}{Observations} & \multicolumn{1}{c}{64,923} & \multicolumn{1}{c}{64,917} & \multicolumn{1}{c}{65,060} & \multicolumn{1}{c}{64,926} & \multicolumn{1}{c}{64,933}\\
\multicolumn{1}{l}{Mean dep. var.} & \multicolumn{1}{c}{0.092} & \multicolumn{1}{c}{0.027} & \multicolumn{1}{c}{0.007} & \multicolumn{1}{c}{0.142} & \multicolumn{1}{c}{0.192}\\
\multicolumn{1}{l}{$R^2$} & \multicolumn{1}{c}{0.018} & \multicolumn{1}{c}{0.016} & \multicolumn{1}{c}{0.013} & \multicolumn{1}{c}{0.016} & \multicolumn{1}{c}{0.03}\\
\bottomrule
\end{tabular}
\begin{tablenotes}
\item Columns: (1) ever hospitalised due to respiratory disease, (2) ever hospitalised due to influenza/pneumonia, (3) hospitalisation or death due to COVID-19, (4) ever hospitalised due to cardio-vascular disease, (5) ever hospitalised due to cancer. Includes fixed-effects for district, month of birth, and year of birth. Also controls for gender, weather variation, and year-month linear time trends by administrative county. Urban England and Wales are defined as districts that had a population density above 400 individuals per km$^2$ in 1951. Standard errors are clustered by district. (*): $p < 0.1$, (**): $p<0.05$, (***): $p<0.01$.
\end{tablenotes}
\end{threeparttable}
\end{table}

Next, we examine the impact of the timing of exposure relative to individuals' gestational age. For brevity, we further only report the estimates for education, fluid intelligence, and respiratory conditions (including influenza/pneumonia), since further analysis did not show significant long-term effects on COVID-19, cancer and cardiovascular disease. \autoref{tab:lon_vs_urb-trimester} shows the estimates for \autoref{eq:E}, where we replace the indicator for \textit{in utero} exposure, $E_i^{IU}$, with three indicators for the relevant trimesters. For years of education, this suggests that second trimester exposure is the most important and reduces education by 0.2 years on average, with third trimester and childhood exposure also showing negative effects, though these are not significantly different from zero. For fluid intelligence, we find the largest effects for first trimester exposure, reducing slightly with gestational age. 
The intrauterine effect on respiratory conditions, shown in column~(3), is mainly driven by exposure in the first and, to a lesser extent, the second trimester. However, for influenza/pneumonia in column~(4), we find similar estimates across childhood and trimesters, suggesting that the effect does not depend on the timing of exposure.

\begin{table}[!h]

\caption{\label{tab:lon_vs_urb-trimester}Trimester effects. Difference-in-Difference estimates comparing treated to control districts defined as urban England and Wales.}
\centering
\begin{threeparttable}
\begin{tabular}[t]{ldddd}
\toprule
\multicolumn{1}{c}{\em{}} & \multicolumn{4}{c}{\em{Dependent variable:}} \\
\cmidrule(l{3pt}r{3pt}){2-5}
\multicolumn{1}{c}{} & \multicolumn{1}{c}{(1)} & \multicolumn{1}{c}{(2)} & \multicolumn{1}{c}{(3)} & \multicolumn{1}{c}{(4)} \\
\multicolumn{1}{c}{ } & \multicolumn{1}{c}{{\specialcell[b]{Educational \\ attainment}}} & \multicolumn{1}{c}{{\specialcell[b]{Fluid \\ intelligence}}} & \multicolumn{1}{c}{{\specialcell[b]{Respiratory, \\ any}}} & \multicolumn{1}{c}{{\specialcell[b]{Influenza/ \\ pneumonia}}}\\
\midrule
Treated $\times$ In utero, 1. tri. & 0.064^{  } & -0.147^{ ** } & 0.032^{ * } & 0.011^{  }\\
 & (0.134) & (0.073) & (0.018) & (0.008)\\
Treated $\times$ In utero, 2. tri. & -0.220^{ * } & -0.120^{ * } & 0.019^{  } & 0.012^{  }\\
 & (0.118) & (0.068) & (0.016) & (0.008)\\
Treated $\times$ In utero, 3. tri. & -0.144^{  } & -0.068^{  } & 0.009^{  } & 0.013^{ * }\\
 & (0.121) & (0.089) & (0.015) & (0.008)\\
Treated $\times$ Childhood & -0.143^{  } & -0.155^{ ** } & -0.008^{  } & 0.009^{ * }\\
 & (0.088) & (0.069) & (0.012) & (0.006)\\
In utero 1. tri. & -0.007^{  } & -0.010^{  } & -0.004^{  } & -0.008^{  }\\
 & (0.063) & (0.047) & (0.008) & (0.006)\\
In utero 2. tri. & -0.049^{  } & 0.006^{  } & 0.003^{  } & -0.004^{  }\\
 & (0.061) & (0.040) & (0.009) & (0.005)\\
In utero 3. tri. & 0.084^{  } & 0.042^{  } & -0.023^{ *** } & -0.008^{  }\\
 & (0.081) & (0.051) & (0.008) & (0.005)\\
Childhood & 0.066^{  } & -0.009^{  } & -0.027^{ * } & -0.003^{  }\\
 & (0.134) & (0.080) & (0.015) & (0.008)\\
\midrule
\multicolumn{1}{l}{Observations} & \multicolumn{1}{c}{64,681} & \multicolumn{1}{c}{26,877} & \multicolumn{1}{c}{64,923} & \multicolumn{1}{c}{64,917}\\
\multicolumn{1}{l}{Mean dep. var.} & \multicolumn{1}{c}{13.317} & \multicolumn{1}{c}{0} & \multicolumn{1}{c}{0.092} & \multicolumn{1}{c}{0.027}\\
\multicolumn{1}{l}{$R^2$} & \multicolumn{1}{c}{0.08} & \multicolumn{1}{c}{0.067} & \multicolumn{1}{c}{0.018} & \multicolumn{1}{c}{0.016}\\
\bottomrule
\end{tabular}
\begin{tablenotes}
\item Columns: (1) educational attainment in years, (2) standardised fluid intelligence score, (3) ever hospitalised due to respiratory disease, (4) ever hospitalised due to influenza/pneumonia. Includes fixed-effects for district, month of birth, and year of birth. Also controls for gender, weather variation, and year-month linear time trends by administrative county. Urban England and Wales are defined as districts that had a population density above 400 individuals per km$^2$ in 1951. Standard errors are clustered by district. (*): $p < 0.1$, (**): $p<0.05$, (***): $p<0.01$.
\end{tablenotes}
\end{threeparttable}
\end{table}

\FloatBarrier

\subsection{Treatment effect heterogeneity}
We next explore potential heterogeneity of treatment effects. For this, we investigate three sources of variation: heterogeneity with respect to gender, socio-economic status and individuals' genetic `predisposition'. We discuss each in turn.

\subsubsection{Heterogeneity by gender}
First, we investigate whether the effect of exposure to the smog is similar for men and women. \autoref{tab:lon_vs_urb-heterogeneity_sex} presents the estimates for our main outcomes of years of education, fluid intelligence and respiratory disease, with Panel (a) and (b) presenting the estimates for women and men, respectively. This shows that the negative effect of smog exposure on years of education in the full sample is largely driven by women, with much smaller effect estimates for men. Having said that, however, we cannot reject the null that the two estimates are the same. 
For fluid intelligence and respiratory disease, including influenza/pneumonia, we do not see large differences between men and women, though the estimates are not always significantly different from zero due to the reduced sample sizes and with that, larger standard errors.\footnote{In \autoref{tab:lon_vs_urb-sex_outcome}, we model the gender ratio as the outcome of interest to explore whether the smog caused differential mortality by gender. We find no differences, suggesting that the differential effects in \autoref{tab:lon_vs_urb-heterogeneity_sex} are driven by scarring rather than selection.}

\begin{table}

\caption{\label{tab:lon_vs_urb-heterogeneity_sex}Heterogeneity across sex. Difference-in-Difference estimates comparing treated to control districts defined as urban England and Wales.}
\centering
\begin{threeparttable}
\begin{tabular}[t]{ldddd}
\toprule
\multicolumn{1}{c}{\em{}} & \multicolumn{4}{c}{\em{Dependent variable:}} \\
\cmidrule(l{3pt}r{3pt}){2-5}
\multicolumn{1}{c}{} & \multicolumn{1}{c}{(1)} & \multicolumn{1}{c}{(2)} & \multicolumn{1}{c}{(3)} & \multicolumn{1}{c}{(4)} \\
\multicolumn{1}{c}{ } & \multicolumn{1}{c}{\specialcell[b]{Educational \\ attainment}} & \multicolumn{1}{c}{\specialcell[b]{Fluid \\ intelligence}} & \multicolumn{1}{c}{\specialcell[b]{Respiratory, \\ any}} & \multicolumn{1}{c}{\specialcell[b]{Influenza/ \\ pneumonia}}\\
\midrule
\addlinespace[0.3em]
\multicolumn{5}{l}{\textbf{Panel (a) -- Female}}\\
\hspace{1em}Treated $\times$ In utero & -0.118^{  } & -0.074^{  } & 0.020^{  } & 0.014^{ ** }\\
\hspace{1em} & (0.128) & (0.072) & (0.013) & (0.007)\\
\hspace{1em}Treated $\times$ Childhood & -0.238^{ * } & -0.152^{ * } & -0.010^{  } & 0.006^{  }\\
\hspace{1em} & (0.130) & (0.083) & (0.015) & (0.008)\\
\hspace{1em}In utero & 0.077^{  } & -0.028^{  } & -0.004^{  } & -0.008^{  }\\
\hspace{1em} & (0.074) & (0.048) & (0.008) & (0.005)\\
\hspace{1em}Childhood & 0.127^{  } & -0.064^{  } & -0.012^{  } & -0.002^{  }\\
\hspace{1em} & (0.149) & (0.114) & (0.018) & (0.010)\\
\addlinespace[0.75em]
\hspace{1em}Observations & \multicolumn{1}{D{,}{,}{-3}}{35,972} & \multicolumn{1}{D{,}{,}{-3}}{14,909} & \multicolumn{1}{D{,}{,}{-3}}{36,110} & \multicolumn{1}{D{,}{,}{-3}}{36,105}\\
\hspace{1em}Mean dep. var. & 13.273 & 0 & 0.084 & 0.024\\
\hspace{1em}$R^2$ & 0.089 & 0.084 & 0.028 & 0.027\\
\addlinespace[0.3em]
\multicolumn{5}{l}{\textbf{Panel (b) -- Male}}\\
\hspace{1em}Treated $\times$ In utero & -0.084^{  } & -0.162^{ * } & 0.022^{  } & 0.013^{  }\\
\hspace{1em} & (0.113) & (0.085) & (0.017) & (0.010)\\
\hspace{1em}Treated $\times$ Childhood & -0.022^{  } & -0.184^{ * } & -0.003^{  } & 0.017^{ * }\\
\hspace{1em} & (0.135) & (0.105) & (0.019) & (0.010)\\
\hspace{1em}In utero & -0.080^{  } & 0.055^{  } & -0.007^{  } & -0.004^{  }\\
\hspace{1em} & (0.077) & (0.051) & (0.010) & (0.006)\\
\hspace{1em}Childhood & -0.159^{  } & -0.026^{  } & -0.004^{  } & -0.000^{  }\\
\hspace{1em} & (0.175) & (0.103) & (0.023) & (0.013)\\
\addlinespace[0.75em]
\hspace{1em}Observations & \multicolumn{1}{D{,}{,}{-3}}{28,645} & \multicolumn{1}{D{,}{,}{-3}}{11,830} & \multicolumn{1}{D{,}{,}{-3}}{28,748} & \multicolumn{1}{D{,}{,}{-3}}{28,747}\\
\hspace{1em}Mean dep. var. & 13.373 & 0 & 0.103 & 0.03\\
\hspace{1em}$R^2$ & 0.098 & 0.095 & 0.032 & 0.029\\
\bottomrule
\end{tabular}
\begin{tablenotes}
\item Columns: (1) educational attainment in years, (2) standardised fluid intelligence score, (3) ever experienced a (primary) respiratory hospitalisation, (4) same as (3) but looking only at influenza/pneumonia. Panels: (a) female subsample, (b) male subsample.  Includes fixed-effects for district, month of birth, and year of birth. Also controls for gender, weather variation, and year-month linear time trends by administrative county. Urban England and Wales are defined as districts that had a population density above 400 individuals per km$^2$ in 1951. Standard errors are clustered by district. (*): $p < 0.1$, (**): $p<0.05$, (***): $p<0.01$.
\end{tablenotes}
\end{threeparttable}
\end{table}

\subsubsection{Heterogeneity by social class}
We next explore potential treatment effect heterogeneity with respect to social class. Although the UK Biobank does not include data on individuals' (or parental) socio-economic position at birth, and because individuals' socio-economic position in adulthood is endogenous to the smog exposure, we merge the 1951 UK Census to the UK Biobank, allowing us to characterise the local area of birth in terms of its socio-economic composition relative to other areas in England and Wales. As such, we create a measure of social class based on individuals' occupation at the district level, capturing the share of its population in professional, managerial and technical occupations. We define districts to be of `high social class' if they are above the median of this distribution; we define `low social class' districts as those below the median. 

We estimate the main specification for these `high class' and `low class' subsamples and report the results in \autoref{tab:lon_vs_urb-heterogeneity_ses}. Overall, we find slightly larger adverse effects of smog exposure among the lower social classes, but with the reduced sample sizes, the standard errors are larger and the estimates are not always significantly different from zero. The larger estimates among the low social classes suggest that the impact is disproportionally felt among those born in districts characterised by a lower socio-economic status. Note that this is unlikely to be driven by differential exposure to pollution across districts with different socio-economic compositions, since both poor and affluent districts were similarly affected by the smog \citep{wilkins1954air, Ball2018}.

\begin{table}

\caption{\label{tab:lon_vs_urb-heterogeneity_ses}Heterogeneity across social class. Difference-in-Difference estimates comparing treated to control districts defined as urban England and Wales.}
\centering
\begin{threeparttable}
\begin{tabular}[t]{ldddd}
\toprule
\multicolumn{1}{c}{\em{}} & \multicolumn{4}{c}{\em{Dependent variable:}} \\
\cmidrule(l{3pt}r{3pt}){2-5}
\multicolumn{1}{c}{} & \multicolumn{1}{c}{(1)} & \multicolumn{1}{c}{(2)} & \multicolumn{1}{c}{(3)} & \multicolumn{1}{c}{(4)} \\
\multicolumn{1}{c}{ } & \multicolumn{1}{c}{\specialcell[b]{Educational \\ attainment}} & \multicolumn{1}{c}{\specialcell[b]{Fluid \\ intelligence}} & \multicolumn{1}{c}{\specialcell[b]{Respiratory, \\ any}} & \multicolumn{1}{c}{\specialcell[b]{Influenza/ \\ pneumonia}}\\
\midrule
\addlinespace[0.3em]
\multicolumn{5}{l}{\textbf{Panel (a) -- High class}}\\
\hspace{1em}Treated $\times$ In utero & -0.038^{  } & -0.099^{  } & 0.020^{  } & 0.008^{  }\\
\hspace{1em} & (0.132) & (0.097) & (0.015) & (0.008)\\
\hspace{1em}Treated $\times$ Childhood & 0.022^{  } & -0.122^{  } & -0.009^{  } & 0.015^{ * }\\
\hspace{1em} & (0.129) & (0.117) & (0.016) & (0.008)\\
\hspace{1em}In utero & -0.104^{  } & -0.126^{  } & 0.005^{  } & 0.002^{  }\\
\hspace{1em} & (0.128) & (0.084) & (0.014) & (0.009)\\
\hspace{1em}Childhood & 0.023^{  } & -0.150^{  } & 0.044^{  } & 0.019^{  }\\
\hspace{1em} & (0.238) & (0.163) & (0.031) & (0.015)\\
\addlinespace[0.75em]
\hspace{1em}Observations & \multicolumn{1}{D{,}{,}{-3}}{14,470} & \multicolumn{1}{D{,}{,}{-3}}{6,637} & \multicolumn{1}{D{,}{,}{-3}}{14,488} & \multicolumn{1}{D{,}{,}{-3}}{14,486}\\
\hspace{1em}Mean dep. var. & 13.91 & 0 & 0.078 & 0.022\\
\hspace{1em}$R^2$ & 0.080 & 0.075 & 0.028 & 0.026\\
\addlinespace[0.3em]
\multicolumn{5}{l}{\textbf{Panel (b) -- Low class}}\\
\hspace{1em}Treated $\times$ In utero & -0.086^{  } & -0.038^{  } & 0.028^{  } & 0.014^{  }\\
\hspace{1em} & (0.140) & (0.083) & (0.021) & (0.010)\\
\hspace{1em}Treated $\times$ Childhood & -0.321^{ ** } & -0.188^{ * } & -0.016^{  } & -0.002^{  }\\
\hspace{1em} & (0.145) & (0.111) & (0.023) & (0.010)\\
\hspace{1em}In utero & 0.019^{  } & 0.048^{  } & -0.009^{  } & -0.009^{ * }\\
\hspace{1em} & (0.060) & (0.039) & (0.007) & (0.005)\\
\hspace{1em}Childhood & -0.007^{  } & 0.002^{  } & -0.021^{  } & -0.008^{  }\\
\hspace{1em} & (0.140) & (0.082) & (0.016) & (0.008)\\
\addlinespace[0.75em]
\hspace{1em}Observations & \multicolumn{1}{D{,}{,}{-3}}{49,656} & \multicolumn{1}{D{,}{,}{-3}}{19,923} & \multicolumn{1}{D{,}{,}{-3}}{49,878} & \multicolumn{1}{D{,}{,}{-3}}{49,874}\\
\hspace{1em}Mean dep. var. & 13.143 & 0 & 0.097 & 0.028\\
\hspace{1em}$R^2$ & 0.058 & 0.061 & 0.016 & 0.015\\
\bottomrule
\end{tabular}
\begin{tablenotes}
\item Columns: (1) educational attainment in years, (2) standardised fluid intelligence score, (3) ever experienced a (primary) respiratory hospitalisation, (4) same as (3) but only looks at influenza/pneumonia. Panels: (a) subsample with individuals born in `high class' districts. (b) subsample with individuals born in `low class' districts. Includes fixed-effects for district, month of birth, and year of birth. Also controls for gender, weather variation, and year-month linear time trends by administrative county. Urban England and Wales are defined as districts that had a population density above 400 individuals per km$^2$ in 1951. Standard errors are clustered by district. (*): $p < 0.1$, (**): $p<0.05$, (***): $p<0.01$.
\end{tablenotes}
\end{threeparttable}
\end{table}

For educational attainment in Column~(1), we find negative estimates among the lower social classes, and significantly so for childhood exposure. The estimates for childhood exposure in high and low social class districts are marginally significantly different from each other (p=0.08). For fluid intelligence in Column~(2), we find negative estimates of pre- and postnatal exposure in both panels, though with variation in magnitude and statistical significance. Similar to the findings for educational attainment, these results suggest larger cognitive impacts of pollution exposure in childhood for individuals born in districts characterised by lower social classes. Turning to respiratory disease and influenza/pneumonia in Columns~(3) and (4), we again find slightly larger estimates for intrauterine exposure among low social class districts, but we cannot statistically distinguish between them.

\subsubsection{Genetic heterogeneity}
Finally, to directly incorporate the genetic component into the analysis, we construct variables measuring individuals' `genetic predisposition' to the outcomes of interest. We do this by running our own tailor-made Genome-Wide Association Study (GWAS) for each of the main outcomes on UK Biobank unrelated participants born in the years \textit{outside} our analysis sample (i.e., 1934--1949 and 1957--1970), as well as those born in districts that are not defined as either treated or control during the study years 1950-1956. 
We use the summary statistics from this GWAS to construct so-called polygenic scores (also known as polygenic indices) for those in the (independent) analysis sample covering the birth cohorts 1950--1956.\footnote{This is the current state-of-the-art approach to constructing polygenic sores. The independence between the discovery and analysis sample avoids over-prediction in the latter. See Appendix~\ref{sec:AppendixA} for an introduction to genetics, an explanation of the genetic terms used here, as well as more detail on the construction of the polygenic scores. Note that because genetic variation in fixed at conception, we do not have the problem of reverse causality.} We do the latter using LDpred2, a Bayesian genetic risk prediction method \citep{vilhjalmsson2015modeling, prive2020ldpred2}. All polygenic scores are standardised to have mean zero and unit variance in the analysis sample.

Polygenic scores can be interpreted as the best linear genetic predictor of the outcome of interest \citep{mills2020introduction}. However, this does not mean that it is an immutable, biological relationship. Indeed, in addition to any potential biological effect, the association may capture gene-environment \textit{correlation} (also referred to as $rGE$), either via genetic variation \textit{invoking} certain environmental responses (so-called \textit{evocative} $rGE$), or via individuals \textit{selecting} into certain environments based on their genetic variation (\textit{active} $rGE$). Furthermore it captures `genetic nurture' effects: the fact that parental genetic variation can shape the environment that the child experiences, which in turn can affect their outcome (an example of \textit{passive} $rGE$; see also e.g., \citet{kong2018nature}). In short, this implies that the predictive power of the polygenic score, as shown in \autoref{tab:lon_vs_urb-pgs_test} in Appendix~\ref{sec:AppendixA}, may capture both genetic as well as environmental components. Nevertheless, it allows us to investigate previously unobserved heterogeneity in the effect estimates and with that, quantify the extent to which one's genetic variation may protect or exacerbate the effects of early life adverse circumstances. 

\autoref{tab:lon_vs_urb-heterogeneity_genetics} presents the main difference-in-difference estimates distinguishing between individuals with a high versus low genetic `predisposition' to the outcome, defined as having a polygenic score above or below the median, shown in Panel (a) and (b) respectively. Note that the polygenic score is specific to the outcome of interest to avoid any potential data mining. For example, the polygenic score in Columns (1) and (2) of \autoref{tab:lon_vs_urb-heterogeneity_genetics} is the best linear genetic predictor for education and fluid intelligence, respectively. This shows that the zero effect of the smog on educational attainment conceals substantial genetic heterogeneity. More precisely, the negative effect is substantially larger for those with a high polygenic score for education, for both prenatal and childhood exposure (though we only reject the null that the estimates are the same for prenatal exposure; $p=0.06$), with the estimates being close to zero for those with a polygenic score below the median. 

\begin{table}

\caption{\label{tab:lon_vs_urb-heterogeneity_genetics}Heterogeneity across genetics. Difference-in-Difference estimates comparing treated to control districts defined as urban England and Wales.}
\centering
\begin{threeparttable}
\fontsize{11}{13}\selectfont
\begin{tabular}[t]{ldddd}
\toprule
\multicolumn{1}{c}{\em{}} & \multicolumn{4}{c}{\em{Dependent variable:}} \\
\cmidrule(l{3pt}r{3pt}){2-5}
\multicolumn{1}{c}{} & \multicolumn{1}{c}{(1)} & \multicolumn{1}{c}{(2)} & \multicolumn{1}{c}{(3)} & \multicolumn{1}{c}{(4)} \\
\multicolumn{1}{c}{ } & \multicolumn{1}{c}{\specialcell[b]{Educational \\ attainment}} & \multicolumn{1}{c}{\specialcell[b]{Fluid \\ intelligence}} & \multicolumn{1}{c}{\specialcell[b]{Respiratory, \\ any}} & \multicolumn{1}{c}{\specialcell[b]{Influenza/ \\ pneumonia}}\\
\midrule
\addlinespace[0.3em]
\multicolumn{5}{l}{\textbf{Panel (a) -- High polygenic score}}\\
\hspace{1em}Treated $\times$ In utero & -0.249^{ ** } & -0.091^{  } & 0.019^{  } & 0.018^{ ** }\\
\hspace{1em} & (0.107) & (0.073) & (0.017) & (0.008)\\
\hspace{1em}Treated $\times$ Childhood & -0.197^{ * } & -0.154^{  } & -0.015^{  } & 0.017^{ * }\\
\hspace{1em} & (0.118) & (0.095) & (0.019) & (0.010)\\
\hspace{1em}In utero & -0.010^{  } & 0.020^{  } & -0.016^{ * } & -0.008^{  }\\
\hspace{1em} & (0.075) & (0.050) & (0.010) & (0.006)\\
\hspace{1em}Childhood & -0.168^{  } & -0.052^{  } & -0.007^{  } & -0.004^{  }\\
\hspace{1em} & (0.164) & (0.096) & (0.020) & (0.011)\\
\addlinespace[0.75em]
\hspace{1em}Observations & \multicolumn{1}{D{,}{,}{-3}}{32,376} & \multicolumn{1}{D{,}{,}{-3}}{13,571} & \multicolumn{1}{D{,}{,}{-3}}{32,423} & \multicolumn{1}{D{,}{,}{-3}}{32,420}\\
\hspace{1em}Mean dep. var. & 13.911 & 0 & 0.104 & 0.031\\
\hspace{1em}$R^2$ & 0.083 & 0.087 & 0.030 & 0.027\\
\addlinespace[0.3em]
\multicolumn{5}{l}{\textbf{Panel (b) -- Low polygenic score}}\\
\hspace{1em}Treated $\times$ In utero & 0.096^{  } & -0.086^{  } & 0.022^{  } & 0.007^{  }\\
\hspace{1em} & (0.152) & (0.084) & (0.014) & (0.008)\\
\hspace{1em}Treated $\times$ Childhood & -0.010^{  } & -0.113^{  } & 0.007^{  } & 0.005^{  }\\
\hspace{1em} & (0.161) & (0.104) & (0.016) & (0.008)\\
\hspace{1em}In utero & -0.016^{  } & -0.028^{  } & 0.007^{  } & -0.005^{  }\\
\hspace{1em} & (0.068) & (0.047) & (0.008) & (0.005)\\
\hspace{1em}Childhood & 0.111^{  } & -0.038^{  } & -0.006^{  } & 0.000^{  }\\
\hspace{1em} & (0.152) & (0.113) & (0.018) & (0.009)\\
\addlinespace[0.75em]
\hspace{1em}Observations & \multicolumn{1}{D{,}{,}{-3}}{32,233} & \multicolumn{1}{D{,}{,}{-3}}{13,162} & \multicolumn{1}{D{,}{,}{-3}}{32,442} & \multicolumn{1}{D{,}{,}{-3}}{32,439}\\
\hspace{1em}Mean dep. var. & 12.721 & 0 & 0.081 & 0.023\\
\hspace{1em}$R^2$ & 0.075 & 0.084 & 0.029 & 0.028\\
\bottomrule
\end{tabular}
\begin{tablenotes}
\item Columns: (1) educational attainment in years, (2) standardised fluid intelligence score, (3) ever experienced a (primary) respiratory hospitalisation, (4) same as (3) but only looks at influenza/pneumonia. Panels: (a) subsample with above-median polygenic score. (b) subsample with below-median polygenic score. Includes fixed-effects for district, month of birth, and year of birth. Also controls for year-month linear time trends by administrative county. Urban England and Wales are defined as districts that had a population density above 400 individuals per km$^2$ in 1951. Standard errors are clustered by district. (*): $p < 0.1$, (**): $p<0.05$, (***): $p<0.01$.
\end{tablenotes}
\end{threeparttable}
\end{table}

To explore what may be driving the negative effect on educational attainment for those with a high polygenic score, Table~\ref{tab:qualifications} in Appendix~\ref{sec:AppendixB} examines the effects of pollution exposure on the probability of reaching different levels of qualifications. This shows that the negative effect of pre- and post-natal pollution exposure for those with a high polygenic score is driven by a reduction in the probability of obtaining an degree or upper secondary qualification and -- correspondingly -- a higher probability of lower secondary or no qualifications. This suggests that pollution exposure reduces one's human capital potential, in particular among those with a high \textit{genetic} potential. 

There is little difference between the estimates for individuals with high and low polygenic scores for fluid intelligence, with both showing a negative effect of smog exposure, though with the smaller sample sizes, they are not significantly different from zero. Furthermore, while the effects of prenatal smog exposure on respiratory disease are similar across subgroups, we find that for influenza/pneumonia the effects of both prenatal and childhood exposure are somewhat larger for those with a high polygenic score, suggesting that the respiratory health of individuals who are genetically `predisposed' is more vulnerable to severe pollution events, though the differences between those with high and low polygenic scores are not significant at conventional levels.\footnote{Our genetic heterogeneity analysis is robust to the use of polygenic scores constructed from an alternative GWAS, obtained from the polygenic index repository \citep{becker2021resource}.} 

\FloatBarrier

\section{Robustness analysis}\label{sec:robustness}

We next present a range of sensitivity analyses to explore the robustness of our main findings. First, we investigate the sensitivity of our estimates to the definition of treatment and control groups. Second, we explore whether our estimates are robust to different definitions of the reference group (i.e., those conceived after the smog). Third, we examine whether the exposure effects differ for exposure in infancy versus early childhood. 
In all robustness checks, we run our analyses only on the three main outcomes: educational attainment, fluid intelligence and respiratory disease.

\subsection{Definition of treated and control districts}
We start by exploring the sensitivity of our estimates to (1) alternative definitions of treated districts, and (2) alternative definitions of control districts. First, we consider different definitions of the exposed districts. The results are reported in \autoref{tab:lon_vs_urb-exposure_def_robustness}. In Panel~(a), we start by dropping the districts classified as low exposure (defined in Section~\ref{sec:data}). Assuming these districts were less exposed compared to Central London, excluding them may increase our effect estimates. Panel~(a) indeed shows that dropping low exposure districts results in estimates that are similar or slightly larger relative to those in the main analysis. 

To reduce measurement error in the exposure classifications, Panel~(b) exploits the actual residential locations at birth of individuals (at a 1 km$^2$ resolution) and assigns exposure based on the individual's eastings and northings of birth relative to the pollution boundaries (as opposed to assigning treatment based on whether individuals' district of birth is in the smog-affected area, as in the main specification). 
\autoref{tab:lon_vs_urb-exposure_def_robustness} shows that excluding individuals born outside the high exposure boundaries, but in districts that are (at least partially) exposed does not affect our estimates.

\begin{table}

\caption{\label{tab:lon_vs_urb-exposure_def_robustness}Definition of exposure. Difference-in-Difference estimates comparing treated to control districts defined as urban England and Wales.}
\centering
\begin{threeparttable}
\begin{tabular}[t]{ldddd}
\toprule
\multicolumn{1}{c}{\em{}} & \multicolumn{4}{c}{\em{Sample:}} \\
\cmidrule(l{3pt}r{3pt}){2-5}
\multicolumn{1}{c}{} & \multicolumn{1}{c}{(1)} & \multicolumn{1}{c}{(2)} & \multicolumn{1}{c}{(3)} & \multicolumn{1}{c}{(4)} \\
\multicolumn{1}{c}{ } & \multicolumn{1}{c}{\specialcell[b]{Educational \\ attainment}} & \multicolumn{1}{c}{\specialcell[b]{Fluid \\ intelligence}} & \multicolumn{1}{c}{\specialcell[b]{Respiratory, \\ any}} & \multicolumn{1}{c}{\specialcell[b]{Influenza/ \\ pneumonia}}\\
\midrule
\addlinespace[0.3em]
\multicolumn{5}{l}{\textbf{Panel (a) -- Districts, low exposure dropped}}\\
\hspace{1em}Treated $\times$ In utero & -0.121^{  } & -0.105^{ ** } & 0.018^{  } & 0.013^{ ** }\\
\hspace{1em} & (0.095) & (0.052) & (0.012) & (0.006)\\
\hspace{1em}Treated $\times$ Childhood & -0.147^{  } & -0.185^{ *** } & -0.015^{  } & 0.007^{  }\\
\hspace{1em} & (0.090) & (0.068) & (0.013) & (0.006)\\
\hspace{1em}In utero & -0.001^{  } & 0.012^{  } & -0.005^{  } & -0.006^{  }\\
\hspace{1em} & (0.054) & (0.036) & (0.006) & (0.004)\\
\hspace{1em}Childhood & -0.023^{  } & -0.033^{  } & -0.007^{  } & -0.000^{  }\\
\hspace{1em} & (0.121) & (0.074) & (0.014) & (0.007)\\
\addlinespace[0.75em]
\hspace{1em}Observations & \multicolumn{1}{D{,}{,}{-3}}{63,394} & \multicolumn{1}{D{,}{,}{-3}}{26,294} & \multicolumn{1}{D{,}{,}{-3}}{63,633} & \multicolumn{1}{D{,}{,}{-3}}{63,627}\\
\hspace{1em}Mean dep. var. & 13.301 & 0 & 0.093 & 0.027\\
\hspace{1em}$R^2$ & 0.077 & 0.066 & 0.018 & 0.016\\
\addlinespace[0.3em]
\multicolumn{5}{l}{\textbf{Panel (b) -- Birth location}}\\
\hspace{1em}Treated $\times$ In utero & -0.099^{  } & -0.100^{ ** } & 0.020^{ * } & 0.011^{ * }\\
\hspace{1em} & (0.091) & (0.051) & (0.011) & (0.006)\\
\hspace{1em}Treated $\times$ Childhood & -0.130^{  } & -0.136^{ ** } & -0.005^{  } & 0.007^{  }\\
\hspace{1em} & (0.089) & (0.068) & (0.013) & (0.006)\\
\hspace{1em}In utero & -0.004^{  } & 0.005^{  } & -0.005^{  } & -0.006^{  }\\
\hspace{1em} & (0.053) & (0.035) & (0.006) & (0.004)\\
\hspace{1em}Childhood & -0.016^{  } & -0.057^{  } & -0.008^{  } & -0.001^{  }\\
\hspace{1em} & (0.120) & (0.074) & (0.014) & (0.007)\\
\addlinespace[0.75em]
\hspace{1em}Observations & \multicolumn{1}{D{,}{,}{-3}}{64,681} & \multicolumn{1}{D{,}{,}{-3}}{26,877} & \multicolumn{1}{D{,}{,}{-3}}{64,923} & \multicolumn{1}{D{,}{,}{-3}}{64,917}\\
\hspace{1em}Mean dep. var. & 13.317 & 0 & 0.092 & 0.027\\
\hspace{1em}$R^2$ & 0.080 & 0.067 & 0.018 & 0.016\\
\bottomrule
\end{tabular}
\begin{tablenotes}
\item Columns: (1) educational attainment in years, (2) standardised fluid intelligence score, (3) ever hospitalised due to respiratory disease, (4) ever hospitalised due to influenza/pneumonia. Panels: (a) exposed districts defined as districts overlapping with any pollution boundary but districts with low exposure have been dropped, (b) exposed individuals defined as individuals with birth location inside any pollution boundary. The `control' cities are: Bristol, Cardiff, Leicester, Liverpool, Manchester, Newcastle, Nottingham, Sheffield, and Birmingham (defined according to 1951 districts). Includes fixed-effects for district, month of birth, and year of birth. Also controls for gender, weather variation and linear time trends. Standard errors are clustered by district. (*): $p < 0.1$, (**): $p<0.05$, (***): $p<0.01$.
\end{tablenotes}
\end{threeparttable}
\end{table}

Second, our main analysis defines control districts as those with a population density of at least 400 individuals per km$^2$. We here investigate the sensitivity of our results to this definition. 
\autoref{tab:lon_vs_urb-robustness_density} shows the estimates from \autoref{eq:E} for the main outcomes under different choices of the population density threshold used to define the control districts. From Panel (a) to (c) the control districts become more densely populated. 
For educational attainment in Column~(1), we find insignificant negative estimates for exposure to the smog across the three panels. When using more densely populated control districts, the estimates for prenatal exposure move closer to zero relative to those in the main analysis, while those for postnatal exposure increase in (absolute) magnitude, and the standard errors become larger due to the reduction in sample size. The estimates for fluid intelligence in Column~(2) increase in (absolute) magnitude as the control districts become more densely populated, particularly for childhood exposure. This suggests that the estimates vary somewhat depending on the definition of the control district. However, they are always negative, large, and significantly different from zero. In Columns~(3) and (4), the effect of prenatal exposure to the smog on the likelihood of being diagnosed with respiratory disease is relatively robust to the use of more densely populated control districts. 

\begin{table}

\caption{\label{tab:lon_vs_urb-robustness_density}Sensitivity of main difference-in-difference estimates with respect to the urban density cutoff (population per km$^2$).}
\centering
\begin{threeparttable}
\fontsize{10}{12}\selectfont
\begin{tabular}[t]{ldddd}
\toprule
\multicolumn{1}{c}{\em{}} & \multicolumn{4}{c}{\em{Dependent variable:}} \\
\cmidrule(l{3pt}r{3pt}){2-5}
\multicolumn{1}{c}{} & \multicolumn{1}{c}{(1)} & \multicolumn{1}{c}{(2)} & \multicolumn{1}{c}{(3)} & \multicolumn{1}{c}{(4)} \\
\multicolumn{1}{c}{ } & \multicolumn{1}{c}{\specialcell[b]{Educational \\ attainment}} & \multicolumn{1}{c}{\specialcell[b]{Fluid \\ intelligence}} & \multicolumn{1}{c}{\specialcell[b]{Respiratory, \\ any}} & \multicolumn{1}{c}{\specialcell[b]{Influenza/ \\ pneumonia}}\\
\midrule
\addlinespace[0.3em]
\multicolumn{5}{l}{\textbf{Panel (a) -- 2,000 population per km$^2$}}\\
\hspace{1em}Treated $\times$ In utero & 0.003^{  } & -0.154^{ *** } & 0.021^{  } & 0.015^{ ** }\\
\hspace{1em} & (0.103) & (0.055) & (0.013) & (0.006)\\
\hspace{1em}Treated $\times$ Childhood & -0.104^{  } & -0.221^{ *** } & -0.017^{  } & 0.009^{  }\\
\hspace{1em} & (0.108) & (0.072) & (0.017) & (0.008)\\
\hspace{1em}In utero & -0.059^{  } & 0.042^{  } & -0.009^{  } & -0.010^{ ** }\\
\hspace{1em} & (0.063) & (0.039) & (0.007) & (0.005)\\
\hspace{1em}Childhood & -0.069^{  } & -0.022^{  } & -0.009^{  } & -0.001^{  }\\
\hspace{1em} & (0.150) & (0.086) & (0.017) & (0.008)\\
\hspace{1em}Observations & \multicolumn{1}{D{,}{,}{-3}}{46,908} & \multicolumn{1}{D{,}{,}{-3}}{20,210} & \multicolumn{1}{D{,}{,}{-3}}{47,104} & \multicolumn{1}{D{,}{,}{-3}}{47,099}\\
\hspace{1em}Mean dep. var. & 13.246 & 0 & 0.096 & 0.028\\
\hspace{1em}$R^2$ & 0.068 & 0.054 & 0.012 & 0.011\\
\addlinespace[0.3em]
\multicolumn{5}{l}{\textbf{Panel (b) -- 3,000 population per km$^2$}}\\
\hspace{1em}Treated $\times$ In utero & -0.051^{  } & -0.203^{ *** } & 0.027^{ * } & 0.018^{ *** }\\
\hspace{1em} & (0.111) & (0.058) & (0.015) & (0.007)\\
\hspace{1em}Treated $\times$ Childhood & -0.171^{  } & -0.302^{ *** } & -0.010^{  } & 0.010^{  }\\
\hspace{1em} & (0.116) & (0.073) & (0.020) & (0.008)\\
\hspace{1em}In utero & -0.070^{  } & 0.059^{  } & -0.010^{  } & -0.012^{ ** }\\
\hspace{1em} & (0.070) & (0.044) & (0.008) & (0.005)\\
\hspace{1em}Childhood & -0.044^{  } & -0.022^{  } & -0.016^{  } & -0.004^{  }\\
\hspace{1em} & (0.170) & (0.097) & (0.019) & (0.009)\\
\hspace{1em}Observations & \multicolumn{1}{D{,}{,}{-3}}{38,057} & \multicolumn{1}{D{,}{,}{-3}}{16,650} & \multicolumn{1}{D{,}{,}{-3}}{38,228} & \multicolumn{1}{D{,}{,}{-3}}{38,224}\\
\hspace{1em}Mean dep. var. & 13.186 & 0 & 0.097 & 0.029\\
\hspace{1em}$R^2$ & 0.061 & 0.052 & 0.011 & 0.008\\
\addlinespace[0.3em]
\multicolumn{5}{l}{\textbf{Panel (c) -- 4,000 population per km$^2$}}\\
\hspace{1em}Treated $\times$ In utero & -0.042^{  } & -0.165^{ ** } & 0.021^{  } & 0.014^{ * }\\
\hspace{1em} & (0.129) & (0.066) & (0.018) & (0.008)\\
\hspace{1em}Treated $\times$ Childhood & -0.212^{  } & -0.360^{ *** } & -0.017^{  } & 0.002^{  }\\
\hspace{1em} & (0.133) & (0.082) & (0.022) & (0.009)\\
\hspace{1em}In utero & -0.129^{  } & 0.076^{  } & 0.000^{  } & -0.006^{  }\\
\hspace{1em} & (0.088) & (0.054) & (0.008) & (0.006)\\
\hspace{1em}Childhood & -0.056^{  } & 0.103^{  } & -0.007^{  } & 0.009^{  }\\
\hspace{1em} & (0.181) & (0.110) & (0.026) & (0.011)\\
\hspace{1em}Observations & \multicolumn{1}{D{,}{,}{-3}}{23,108} & \multicolumn{1}{D{,}{,}{-3}}{10,799} & \multicolumn{1}{D{,}{,}{-3}}{23,207} & \multicolumn{1}{D{,}{,}{-3}}{23,203}\\
\hspace{1em}Mean dep. var. & 13.218 & 0 & 0.098 & 0.029\\
\hspace{1em}$R^2$ & 0.057 & 0.049 & 0.012 & 0.008\\
\bottomrule
\end{tabular}
\begin{tablenotes}
\item Columns: (1) educational attainment in years, (2) standardised fluid intelligence score, (3) ever hospitalised due to respiratory disease, (4) ever hospitalised due to influenza/pneumonia. Panels show the estimates for different choices of the urban density cut off as measured in 1951. Includes fixed-effects for district, month of birth, and year of birth. Also controls for gender, weather variation, and year-month linear time trends by administrative county. Standard errors are clustered by district. (*): $p < 0.1$, (**): $p<0.05$, (***): $p<0.01$.
\end{tablenotes}
\end{threeparttable}
\end{table}

Finally, instead of using population density to define the control districts, we use the main major cities in England and Wales: Birmingham, Bristol, Cardiff, Leicester, Liverpool, Manchester, Newcastle, Nottingham and Sheffield.   
Indeed, one interpretation of our results is that the treated not only have a pollution shock, but also face an accumulation of pollution throughout their childhood, which may affect their later-life health. By using specific major cities \textit{only} in the control group, we ensure that both the treated and control groups experience heightened pollution throughout their early lives. \autoref{tab:lon_vs_cit-main_outcomes} shows that this reduces the sample size substantially and with that, increases the standard errors. Despite that, the magnitude of the estimates are very similar to those reported above. This therefore suggests that our findings are not very sensitive to the definition of treatment and control districts.

\begin{table}[!h]

\caption{\label{tab:lon_vs_cit-main_outcomes}Difference-in-Difference estimates comparing treated to control cities.}
\centering
\begin{threeparttable}
\begin{tabular}[t]{ldddd}
\toprule
\multicolumn{1}{c}{\em{}} & \multicolumn{4}{c}{\em{Dependent variable:}} \\
\cmidrule(l{3pt}r{3pt}){2-5}
\multicolumn{1}{c}{} & \multicolumn{1}{c}{(1)} & \multicolumn{1}{c}{(2)} & \multicolumn{1}{c}{(3)} & \multicolumn{1}{c}{(4)} \\
\multicolumn{1}{c}{ } & \multicolumn{1}{c}{{\specialcell[b]{Educational \\ attainment}}} & \multicolumn{1}{c}{{\specialcell[b]{Fluid \\ intelligence}}} & \multicolumn{1}{c}{{\specialcell[b]{Respiratory, \\ any}}} & \multicolumn{1}{c}{{\specialcell[b]{Influenza/ \\ pneumonia}}}\\
\midrule
Treated $\times$ In utero & -0.099^{  } & -0.177^{ *** } & 0.020^{ * } & 0.016^{ *** }\\
 & (0.105) & (0.050) & (0.012) & (0.006)\\
Treated $\times$ Childhood & -0.133^{  } & -0.269^{ *** } & 0.000^{  } & 0.016^{ ** }\\
 & (0.099) & (0.077) & (0.017) & (0.008)\\
In utero & -0.040^{  } & 0.071^{  } & -0.007^{  } & -0.012^{ * }\\
 & (0.083) & (0.053) & (0.008) & (0.006)\\
Childhood & -0.076^{  } & 0.005^{  } & -0.016^{  } & -0.009^{  }\\
 & (0.212) & (0.130) & (0.021) & (0.011)\\
\midrule
\multicolumn{1}{l}{Observations} & \multicolumn{1}{c}{27,279} & \multicolumn{1}{c}{12,509} & \multicolumn{1}{c}{27,386} & \multicolumn{1}{c}{27,384}\\
\multicolumn{1}{l}{Mean dep. var.} & \multicolumn{1}{c}{13.276} & \multicolumn{1}{c}{0} & \multicolumn{1}{c}{0.093} & \multicolumn{1}{c}{0.027}\\
\multicolumn{1}{l}{$R^2$} & \multicolumn{1}{c}{0.072} & \multicolumn{1}{c}{0.042} & \multicolumn{1}{c}{0.012} & \multicolumn{1}{c}{0.009}\\
\bottomrule
\end{tabular}
\begin{tablenotes}
\item Columns: (1) educational attainment in years, (2) standardised fluid intelligence score, (3) ever hospitalised due to respiratory disease, (4) ever hospitalised due to influenza/pneumonia. The `control' cities are: Bristol, Cardiff, Leicester, Liverpool, Manchester, Newcastle, Nottingham, Sheffield, and Birmingham (defined according to 1951 districts). Includes fixed-effects for district, month of birth, and year of birth. Also controls for gender, weather variation and linear time trends. Standard errors are clustered by district. (*): $p < 0.1$, (**): $p<0.05$, (***): $p<0.01$.
\end{tablenotes}
\end{threeparttable}
\end{table}

\FloatBarrier

\subsection{Definition of the reference group}
Our main analysis compares those exposed to the smog \textit{in utero} or in early childhood to those conceived \textit{after} the smog. The latter (reference, or unexposed) group includes those born between September 1953 and December 1956. We next explore the sensitivity of this definition by restricting the year-months of birth of those in the reference group to be nearer the date of the smog.
Panels (a) and (b) of \autoref{tab:lon_vs_urb-robustness_after} show the estimates of \autoref{eq:E} under two different choices of the end date used to define those who are not exposed to the smog: the end of 1954 and the end of 1955. \autoref{tab:lon_vs_urb-robustness_after} shows that our results are relatively stable across these alternative definitions of the reference (unexposed) group.
For educational attainment in Column~(1), when we restrict the reference group to be closer to the smog event, as shown in Panel~(a) and (b), the effect of childhood exposure to the smog increases in absolute terms, but with a smaller sample size, the standard errors are larger and the confidence intervals always overlap with zero. The estimates for prenatal exposure remains around -0.1 across the two panels, but it is insignificantly different from zero throughout.

For fluid intelligence in Column~(2), the estimates for prenatal and childhood exposure to the smog are negative, and very similar across the panels. The estimate for the effect of childhood exposure on the probability of being diagnosed with respiratory disease (Column~(3)) increases in (absolute) magnitude and the effect of prenatal exposure reduces when we restrict the reference group. Finally, the intrauterine effect on the risk of hospitalisation with influenza/pneumonia is similar across panels (Column~(4)), whilst there is a slight reduction in the estimates for childhood exposure as the reference group is further restricted.

\begin{table}

\caption{\label{tab:lon_vs_urb-robustness_after}Sensitivity of main difference-in-difference estimates with respect to the after cutoff.}
\centering
\begin{threeparttable}
\begin{tabular}[t]{ldddd}
\toprule
\multicolumn{1}{c}{\em{}} & \multicolumn{4}{c}{\em{Dependent variable:}} \\
\cmidrule(l{3pt}r{3pt}){2-5}
\multicolumn{1}{c}{} & \multicolumn{1}{c}{(1)} & \multicolumn{1}{c}{(2)} & \multicolumn{1}{c}{(3)} & \multicolumn{1}{c}{(4)} \\
\multicolumn{1}{c}{ } & \multicolumn{1}{c}{\specialcell[b]{Educational \\ attainment}} & \multicolumn{1}{c}{\specialcell[b]{Fluid \\ intelligence}} & \multicolumn{1}{c}{\specialcell[b]{Respiratory, \\ any}} & \multicolumn{1}{c}{\specialcell[b]{Influenza/ \\ pneumonia}}\\
\midrule
\addlinespace[0.3em]
\multicolumn{5}{l}{\textbf{Panel (a) -- Controls from September 1953 to December 1954}}\\
\hspace{1em}Treated $\times$ In utero & -0.103^{  } & -0.110^{ ** } & 0.011^{  } & 0.010^{ * }\\
\hspace{1em} & (0.098) & (0.050) & (0.012) & (0.006)\\
\hspace{1em}Treated $\times$ Childhood & -0.181^{  } & -0.178^{ ** } & -0.031^{ ** } & 0.002^{  }\\
\hspace{1em} & (0.119) & (0.082) & (0.016) & (0.008)\\
\hspace{1em}In utero & -0.003^{  } & 0.040^{  } & 0.001^{  } & -0.004^{  }\\
\hspace{1em} & (0.040) & (0.028) & (0.005) & (0.003)\\
\hspace{1em}Childhood & -0.027^{  } & 0.047^{  } & 0.009^{  } & -0.004^{  }\\
\hspace{1em} & (0.063) & (0.048) & (0.007) & (0.005)\\
\hspace{1em}Observations & \multicolumn{1}{D{,}{,}{-3}}{45,232} & \multicolumn{1}{D{,}{,}{-3}}{18,770} & \multicolumn{1}{D{,}{,}{-3}}{45,395} & \multicolumn{1}{D{,}{,}{-3}}{45,390}\\
\hspace{1em}Mean dep. var. & 13.292 & 0 & 0.095 & 0.028\\
\hspace{1em}$R^2$ & 0.084 & 0.074 & 0.023 & 0.022\\
\addlinespace[0.3em]
\multicolumn{5}{l}{\textbf{Panel (b) -- Controls from September 1953 to December 1955}}\\
\hspace{1em}Treated $\times$ In utero & -0.096^{  } & -0.094^{ ** } & 0.018^{  } & 0.011^{ * }\\
\hspace{1em} & (0.094) & (0.047) & (0.011) & (0.006)\\
\hspace{1em}Treated $\times$ Childhood & -0.127^{  } & -0.098^{  } & -0.014^{  } & 0.005^{  }\\
\hspace{1em} & (0.106) & (0.068) & (0.014) & (0.007)\\
\hspace{1em}In utero & 0.001^{  } & 0.026^{  } & -0.000^{  } & -0.004^{  }\\
\hspace{1em} & (0.041) & (0.025) & (0.004) & (0.003)\\
\hspace{1em}Childhood & -0.010^{  } & -0.001^{  } & 0.007^{  } & -0.003^{  }\\
\hspace{1em} & (0.053) & (0.036) & (0.007) & (0.004)\\
\hspace{1em}Observations & \multicolumn{1}{D{,}{,}{-3}}{55,196} & \multicolumn{1}{D{,}{,}{-3}}{22,946} & \multicolumn{1}{D{,}{,}{-3}}{55,402} & \multicolumn{1}{D{,}{,}{-3}}{55,396}\\
\hspace{1em}Mean dep. var. & 13.308 & 0 & 0.093 & 0.028\\
\hspace{1em}$R^2$ & 0.081 & 0.072 & 0.020 & 0.018\\
\bottomrule
\end{tabular}
\begin{tablenotes}
\item Columns: (1) educational attainment in years, (2) standardised fluid intelligence score, (3) ever hospitalised due to respiratory disease, (4) ever hospitalised due to influenza/pneumonia. Panels show estimates under different choices of cutoff for defining the control cohorts born after the smog. Includes fixed-effects for district, month of birth, and year of birth. Also controls for gender, weather variation, and year-month linear time trends by administrative county. Urban England and Wales are defined as districts that had a population density above 400 individuals per km$^2$ in 1951. Standard errors are clustered by district. (*): $p < 0.1$, (**): $p<0.05$, (***): $p<0.01$.
\end{tablenotes}
\end{threeparttable}
\end{table}

\FloatBarrier

\subsection{Childhood exposure}
We next explore whether the childhood exposure effect differs for exposure in infancy (age 0) or later (age 1). \autoref{tab:lon_vs_urb-childhood-main_outcomes} presents the estimates that distinguish between the two ages in early childhood, showing that the effect on fluid intelligence is driven mainly by exposure in infancy. We also find larger effects on years of education for exposure in infancy. Although the effect of exposure at age 1 remains negative on education and intelligence, it is insignificantly different from zero for both outcomes. 

\begin{table}[!h]

\caption{\label{tab:lon_vs_urb-childhood-main_outcomes}Childhood effects at age 0 and 1. Difference-in-Difference estimates comparing treated to control districts defined as urban England and Wales.}
\centering
\begin{threeparttable}
\begin{tabular}[t]{ldddd}
\toprule
\multicolumn{1}{c}{\em{}} & \multicolumn{4}{c}{\em{Dependent variable:}} \\
\cmidrule(l{3pt}r{3pt}){2-5}
\multicolumn{1}{c}{} & \multicolumn{1}{c}{(1)} & \multicolumn{1}{c}{(2)} & \multicolumn{1}{c}{(3)} & \multicolumn{1}{c}{(4)} \\
\multicolumn{1}{c}{ } & \multicolumn{1}{c}{{\specialcell[b]{Educational \\ attainment}}} & \multicolumn{1}{c}{{\specialcell[b]{Fluid \\ intelligence}}} & \multicolumn{1}{c}{{\specialcell[b]{Respiratory, \\ any}}} & \multicolumn{1}{c}{{\specialcell[b]{Influenza/ \\ pneumonia}}}\\
\midrule
Treated $\times$ In utero & -0.093^{  } & -0.094^{ * } & 0.022^{ ** } & 0.012^{ ** }\\
 & (0.090) & (0.051) & (0.011) & (0.005)\\
Treated $\times$ Childhood, age 0 & -0.142^{  } & -0.172^{ ** } & -0.010^{  } & 0.009^{  }\\
 & (0.088) & (0.067) & (0.012) & (0.006)\\
Treated $\times$ Childhood, age 1 & -0.107^{  } & -0.089^{  } & 0.005^{  } & 0.010^{  }\\
 & (0.114) & (0.082) & (0.015) & (0.008)\\
In utero & -0.012^{  } & 0.005^{  } & -0.005^{  } & -0.006^{  }\\
 & (0.055) & (0.036) & (0.006) & (0.004)\\
Childhood, age 0 & -0.050^{  } & -0.043^{  } & -0.005^{  } & -0.001^{  }\\
 & (0.126) & (0.079) & (0.015) & (0.007)\\
Childhood, age 1 & -0.181^{  } & -0.045^{  } & 0.001^{  } & -0.002^{  }\\
 & (0.181) & (0.122) & (0.022) & (0.011)\\
\midrule
\multicolumn{1}{l}{Observations} & \multicolumn{1}{c}{64,681} & \multicolumn{1}{c}{26,877} & \multicolumn{1}{c}{64,923} & \multicolumn{1}{c}{64,917}\\
\multicolumn{1}{l}{Mean dep. var.} & \multicolumn{1}{c}{13.317} & \multicolumn{1}{c}{0} & \multicolumn{1}{c}{0.092} & \multicolumn{1}{c}{0.027}\\
\multicolumn{1}{l}{$R^2$} & \multicolumn{1}{c}{0.08} & \multicolumn{1}{c}{0.067} & \multicolumn{1}{c}{0.018} & \multicolumn{1}{c}{0.016}\\
\bottomrule
\end{tabular}
\begin{tablenotes}
\item Columns: (1) educational attainment in years, (2) standardised fluid intelligence score, (3) ever experienced a (primary) respiratory hospitalisation, (4) same as (3) but only looks at influenza/pneumonia. Includes fixed-effects for district, month of birth, and year of birth. Also controls for gender, weather variation, and year-month linear time trends by administrative county. Urban England and Wales are defined as districts that had a population density above 400 individuals per km$^2$ in 1951. Standard errors are clustered by district. (*): $p < 0.1$, (**): $p<0.05$, (***): $p<0.01$.
\end{tablenotes}
\end{threeparttable}
\end{table}

\FloatBarrier

\section{Conclusions}\label{sec:conclusion}

There is a substantial literature documenting the \textit{contemporaneous} effects of exposure to pollution on individuals' human capital and health outcomes. Much less is known, however, about the potential longer-term effects of early-life pollution exposure, despite the fact that the intrauterine and early childhood environment are crucial for shaping individuals' outcomes in older age. Indeed, most of the literature that investigates the effects of early-life pollution focuses on short-term effects, such as outcomes at birth, finding largely negative impacts. A lack of historical pollution data with good coverage of geographical locations means it is often not possible to use actual pollution measurements and relate those to later-life outcomes. To shed light on the longer-term effects, research instead has to rely on reduced form analysis and natural experiments. That is exactly what we do in this paper, estimating the very long-term effects of being exposed to a severe pollution event on a range of outcomes measured at age $\sim$60. 

We present new evidence of the very long-term effects of an early-life pollution shock. The London smog affected Londoners between 5--9 December 1952, when a thermal inversion trapped pollution over London, which -- due to weather conditions at the time -- was not dispersed. We focus on the long-term human capital and health effects of exposure to the smog. We compare individuals exposed to the smog in London in either the intrauterine or infancy period to those born in other urban areas, as well as to those conceived after the smog. Our difference-in-difference analysis shows that those exposed to the smog have lower fluid intelligence scores, with some suggestive evidence that they also have fewer years of education. We find that exposure in infancy has slightly larger effects compared to exposure \textit{in utero}. Investigating the long-term \textit{health} effects, we find a large increase in the probability of being diagnosed with respiratory disease due to intrauterine exposure, which is driven by hospitalisations for influenza/pneumonia, with no impacts on cardiovascular disease and cancer diagnoses. 

We next study potential differential effects of \textit{in utero} exposure, distinguishing between exposure in the first, second, and third trimester. We find some evidence for differential gestational effects for fluid intelligence, with larger effects for exposure at early gestational ages. Similarly, we see that the increase in respiratory conditions is larger for first trimester exposure, but there is no evidence of trimester-specific effects for influenza/pneumonia.

We then model the heterogeneity of our effect estimates with respect to three sources of (predetermined) variation: gender, socio-economic status at birth, and individuals' genetic predisposition.
Although we often cannot reject the null that the estimates are the same across subgroups, our findings are suggestive of larger negative effects of pollution on schooling for women, larger impacts on human capital and respiratory health for those in low social class environments, as well as larger impacts on education and respiratory health among those genetically more susceptible to the outcome of interest. 

Our estimates are quantitatively and qualitatively important, but there are four key points regarding their interpretation. First, they estimate the effects of exposure to a \textit{severe} pollution event. Indeed, London nowadays experiences pollution levels that are still high, but nowhere near those observed in 1952. Hence, our results cannot be extrapolated to smog events occurring in London or most other cities in Europe nowadays. Despite that, they \textit{do} compare to smog events that are happening each year in industrialising economies such as India and China. Hence, our findings are relevant to those settings, suggesting that such extreme pollution events do not only affect contemporaneous outcomes, but also have longer term adverse effects. 

Second, our analysis compares pollution during the smog to `standard' pollution levels in control districts as well as in the years immediately after the smog. Although these levels of pollution are indeed lower than those in inner London, they are not comparable to current levels of pollution. Hence, our estimates capture the effect of being exposed to a large pollution shock, relative to already high levels throughout early childhood. Again, the results can therefore better be extrapolated to industrialising economies with higher pollution levels in general, as well as larger pollution shocks.

Third, although we argue that the main cause of the reduction in human capital and health is the exposure to the smog, other potential channels are worth mentioning. For example, if the smog caused an increase in the probability of the mother losing one of her parents, this may have also affected her children through the longer-term effects of exposure to prenatal stress \citep[see e.g.,][]{persson2018family}.

Fourth, there is at least one reason to believe that our estimates are an upper bound of the `true' effect of the smog, but also three reasons to believe they are a lower bound. If the smog differentially induced individuals to engage in pollution avoidance behaviour, their subsequent reduced exposure to pollution may lead us to over-estimate the effect of the smog. Although we cannot address this directly, the fact that we cannot reject the common trend assumption post-smog, that we find no differential trends in the socio-economic composition of treated and control districts, and no evidence of a differential probability of moving for sibling pairs across treated and control districts puts some confidence in our estimates.

Having said that, there are three reasons to believe we underestimate the long-term effects of the smog. First, the evidence suggests that the smog led to a small increase in infant mortality and foetal loss, and potentially a reduction in access to pre- and post-natal care. Assuming that those who were affected by this were more vulnerable and those who survived were stronger, this suggests that our estimates are likely to be a lower bound. 
Second, and relatedly, since individuals in the UK Biobank were invited to participate in 2006-2010, we implicitly condition on survival until this time. In the presence of frailty selection, where fragile individuals are more likely to die prior to assessment leaving stronger survivors in the sample, our estimates are likely to be attenuated. The third reason why our estimates are likely to be downward biased is due to measurement error. For one, since we do not observe gestational age, we assume all individuals were \textit{in utero} for nine months prior to their year-month of birth, and we assume individuals were born on the first of the month. In reality, some individuals would have had a shorter gestational period, potentially misclassifying them as being exposed to the smog. Two, we are reliant on publications from the 1950s, showing the extent of the smog as well as its variability across London. We therefore define individuals as either exposed or unexposed, but in reality, pollution would have shown more regional variability that we are unable to capture in our analyses. Three, related to this, we observe individuals' location of birth (eastings and northings) with a 1km resolution. Given that pollution changes across space, individuals who are born on the boundary of our exposed and unexposed districts may be misclassified, leading to additional measurement error.
 
As with any research, our analysis comes with its limitations. First, we cannot identify \textit{which} pollutants matter more for individuals' human capital and health outcomes. Indeed, the literature suggests that multiple pollutants increased during the smog, and we cannot determine whether one or more of these are driving the deterioration in later-life outcomes. Second, the UK Biobank is a large database of those aged 45--69 in the United Kingdom, but it is not representative, with its participants being healthier and wealthier than the UK population \citep{fry2017comparison}. This has implications for our research and the interpretation of our findings. To the extent that air pollution disproportionately affects groups of lower socio-economic status, our results may not accurately reflect the impacts of pollution for these groups, potentially underestimating the overall impacts of pollution exposure. Having said that, the data is unique in combining information from questionnaires, objective measurements, genetic information, and administrative data on a very large sample of UK residents, with information on year-month and location of birth, allowing us to identify which participants were likely to be exposed and which were not. With that, our paper highlights the very long-run pollution effects, identifying reductions in individuals' human capital and health outcomes up to $\sim$60 years after the actual exposure. 

Finally, our findings have clear policy implications. They suggest that reducing pollution has large, long-term benefits for the population. From the population perspective, given the ease and low cost of pollution forecasting, the benefits of avoiding pollution are substantial. From a policy maker perspective, this should encourage the implementation of incentives or regulation that reduces pollution from e.g., residential homes, industry, or transportation. Indeed, in addition to any \textit{contemporaneous} benefits of a reduction in pollution, we show that there are substantial \textit{long-term} benefits of pollution reductions. Not taking these into account would lead to a large underestimation of the welfare effects that can be achieved by policies that cut emissions.

\newpage 
\singlespacing
\printbibliography
\newpage 
\doublespacing

\clearpage
\section*{Online Appendix}

\setcounter{table}{0} 
\setcounter{figure}{0} 
\setcounter{equation}{0} 
\renewcommand{\thetable}{\Alph{subsection}.\arabic{table}}
\renewcommand{\thefigure}{\Alph{subsection}.\arabic{figure}} %
\renewcommand{\theequation}{\Alph{subsection}.\arabic{equation}}
\renewcommand{\thesubsection}{\Alph{subsection}}
\renewcommand{\thesubsubsection}{\Alph{subsection}.\arabic{subsubsection}}

\subsection{Additional Tables and Figures} \label{sec:AppendixB}

\begin{figure}[!h]\caption{\label{fig:dementia_by_year_of_birth}Frequency of hospitalisations with dementia in the UK Biobank by birth year.}\centering\includegraphics[width=0.7\textwidth]{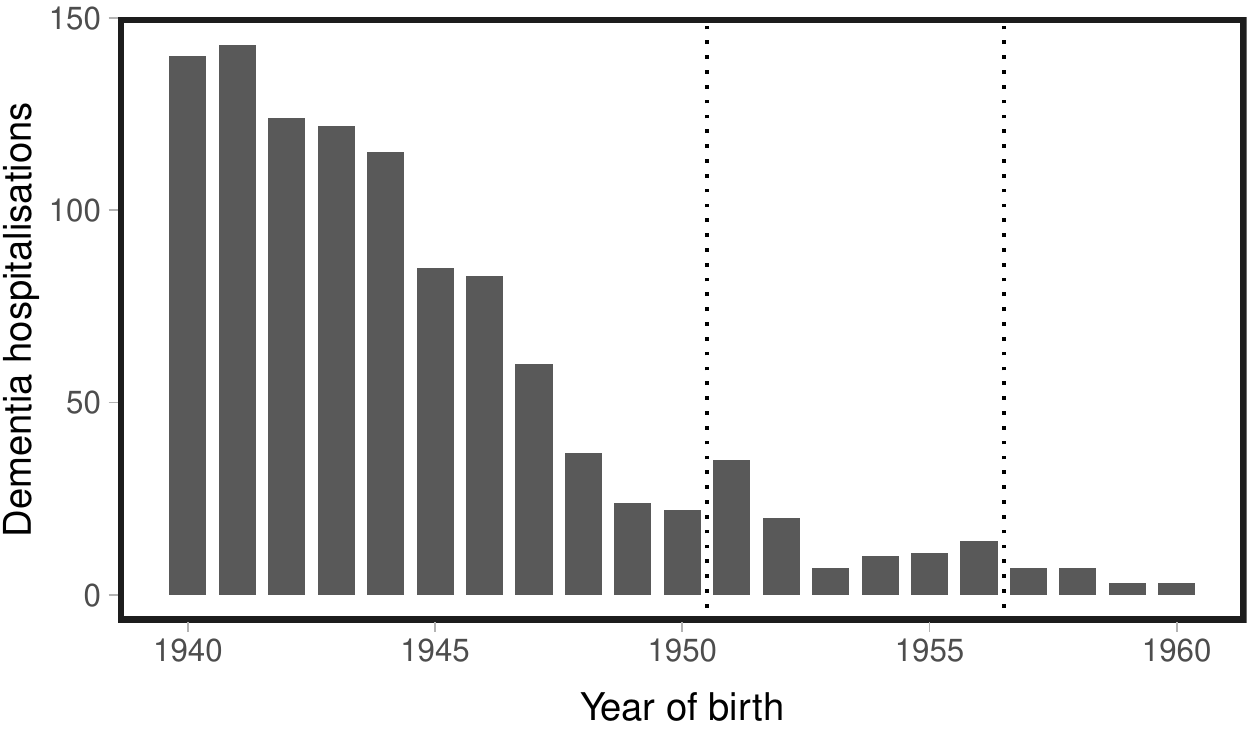}\caption*{The vertical dotted lines show our approximate sample window (1950-1956).}\end{figure}

\begin{table}[h]
\caption{\label{tab:years_educ_definition}Mapping between qualifications and years of education.}
\centering
\begin{threeparttable}
\begin{tabular}{lc}
\toprule
Qualifications & Years of education \\ \midrule
College or university degree & 16 \\
A/AS levels $+$ NVQ/HND/HNC & 14 \\
A/AS levels $+$ Other professional qualifications & 15 \\
NVQ/HND/HNC & 13 \\
Other professional qualifications & 12 \\
A/AS levels & 13 \\
CSEs, GCSEs, or O levels & 11 \\
No qualifications & 10 \\ \bottomrule
\end{tabular}
\begin{tablenotes}
\item Columns: (1) the qualifications recorded in the UK Biobank, (2) the assigned years of education. A plus indicates that the individual must hold both of the specified qualifications simultaneously.
\end{tablenotes}
\end{threeparttable}
\end{table}

\begin{figure}[h!]
\caption{Population density (in population per $\text{km}^2$) at the district level in England and Wales.}
\centering\includegraphics[width=\textwidth]{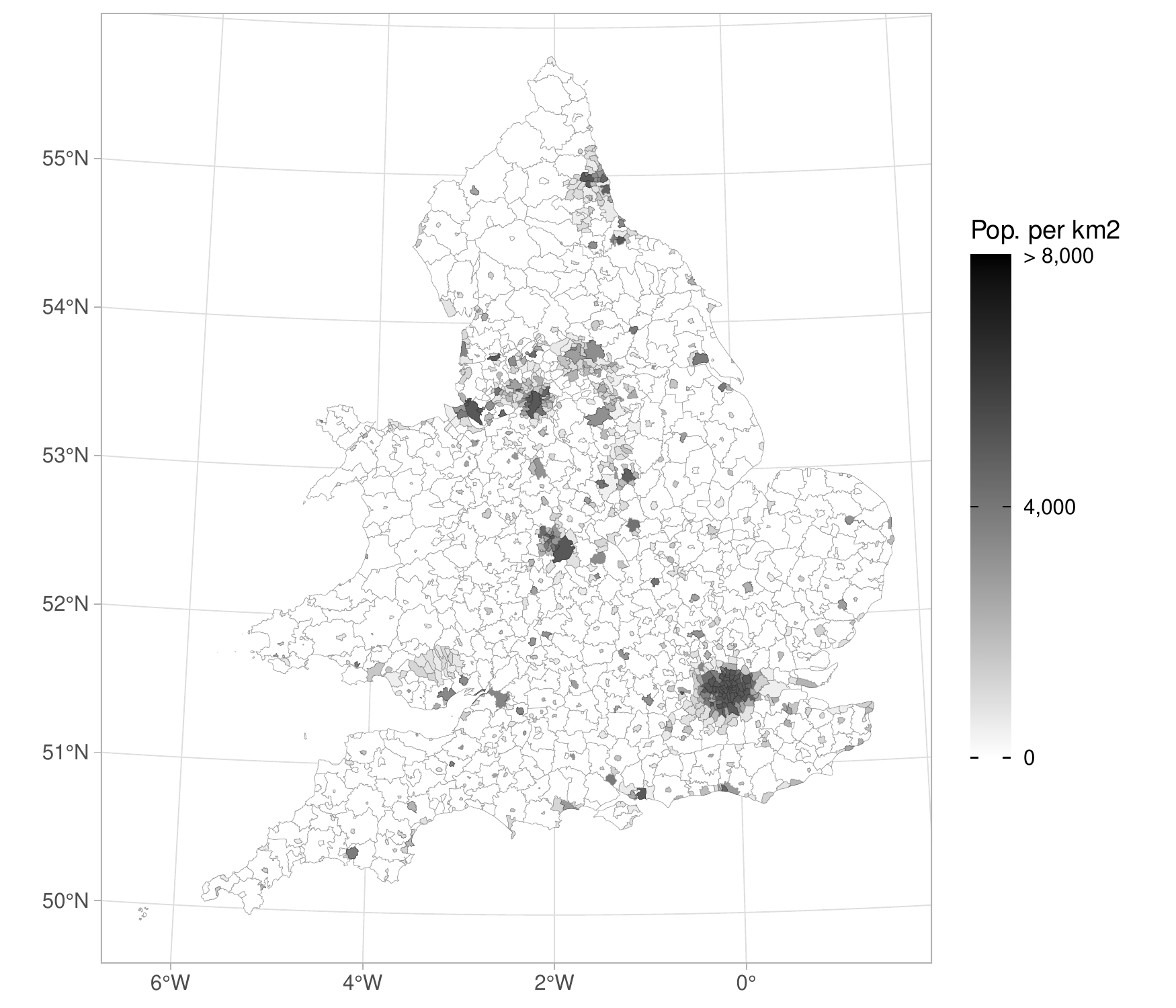}
\caption*{This map uses data provided through \url{www.visionofbritain.org.uk} and uses historical material which is copyright of the Great Britain Historical GIS Project and the University of Portsmouth \citep{southall2011rebuilding, GBHDGIS2011}.}
\label{fig:districts_density}
\end{figure}

\begin{figure}[!h]\caption{\label{fig:weather}Time series for minimum temperature, sunshine, rainfall, and wind speed in control and treated districts. }\centering\includegraphics[width=0.85\textwidth]{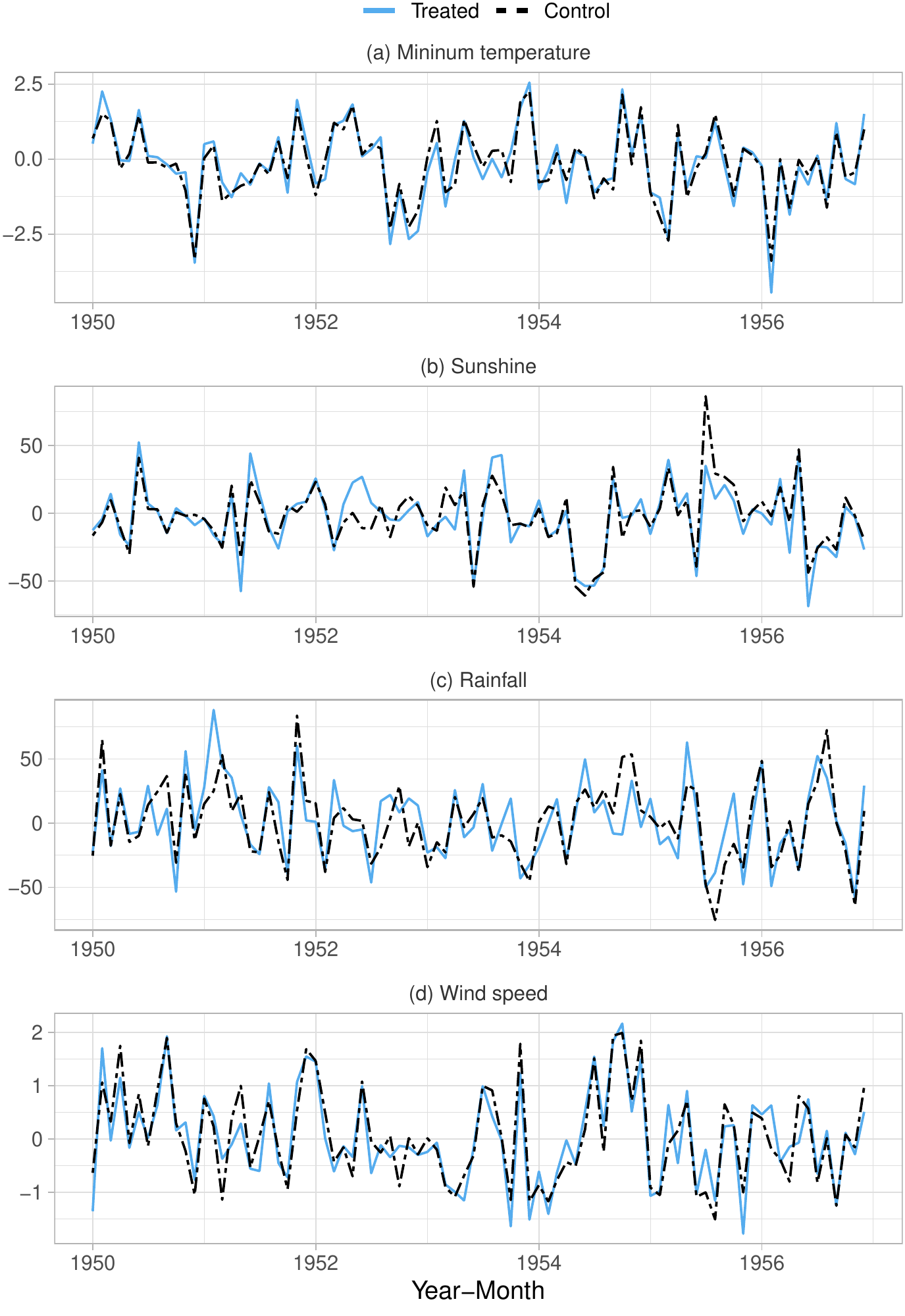}\caption*{We take the measurements at the birth locations of all individuals in our sample and average these by year-month and treatment status. Before averaging we remove seasonality using a set of month dummies.}\end{figure}

\FloatBarrier

\begin{figure}[!h]\caption{\label{fig:census_ses_shares}Socioeconomic composition across the 1951, 1961, and 1971 censuses in control and treated districts.}\centering\includegraphics[width=0.8\textwidth]{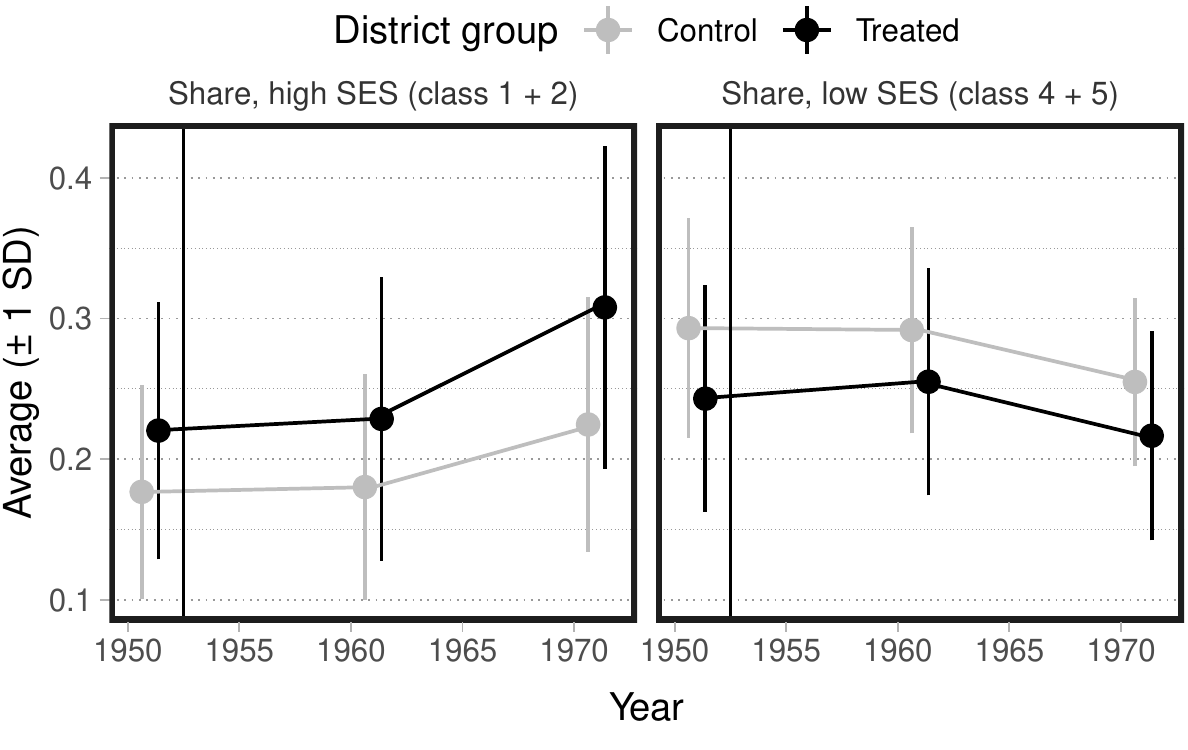}\caption*{Shows shares of high SES (left panel) and low SES (right panel) residents in respectively control districts (gray) and treated districts (black). Vertical black line marks 1952; the year of the London smog.}\end{figure}

\begin{table}[!h]

\caption{\label{tab:moving_probability}Impact of smog exposure on the birth locations of sibling pairs.}
\centering
\begin{threeparttable}
\fontsize{10}{12}\selectfont
\begin{tabular}[t]{ldd}
\toprule
\multicolumn{1}{c}{\em{}} & \multicolumn{2}{c}{\em{Dependent variable:}} \\
\cmidrule(l{3pt}r{3pt}){2-3}
\multicolumn{1}{c}{} & \multicolumn{1}{c}{(1)} & \multicolumn{1}{c}{(2)} \\
\multicolumn{1}{c}{ } & \multicolumn{1}{c}{{Moved}} & \multicolumn{1}{c}{{Moved }}\\
\midrule
Post & -0.001^{  } & -0.003^{  }\\
 & (0.018) & (0.034)\\
Treated & 0.249^{ *** } & {}\\
 & (0.049) & {}\\
Treated $\times$ Post & 0.020^{  } & 0.043^{  }\\
 & (0.052) & (0.056)\\
Age & 0.002^{ *** } & 0.002^{ *** }\\
 & (0.001) & (0.001)\\
\midrule
\multicolumn{1}{l}{Observations} & \multicolumn{1}{c}{3,079} & \multicolumn{1}{c}{3,079}\\
\multicolumn{1}{l}{$R^2$} & \multicolumn{1}{c}{0.045} & \multicolumn{1}{c}{0.318}\\
\bottomrule
\end{tabular}
\begin{tablenotes}
\item Columns: (1) without fixed effects, (2) with district, year-of-birth, and month-of-birth fixed effects. (*): $p < 0.1$, (**): $p<0.05$, (***): $p<0.01$.
\end{tablenotes}
\end{threeparttable}
\end{table}

\begin{figure}[!h]\caption{\label{fig:impacts_on_rates}Estimates from event-study specifications comparing births, mortality, mortality under age 1, and mortality under 4 weeks in treated and control districts over time.}\centering\includegraphics[width=0.75\textwidth]{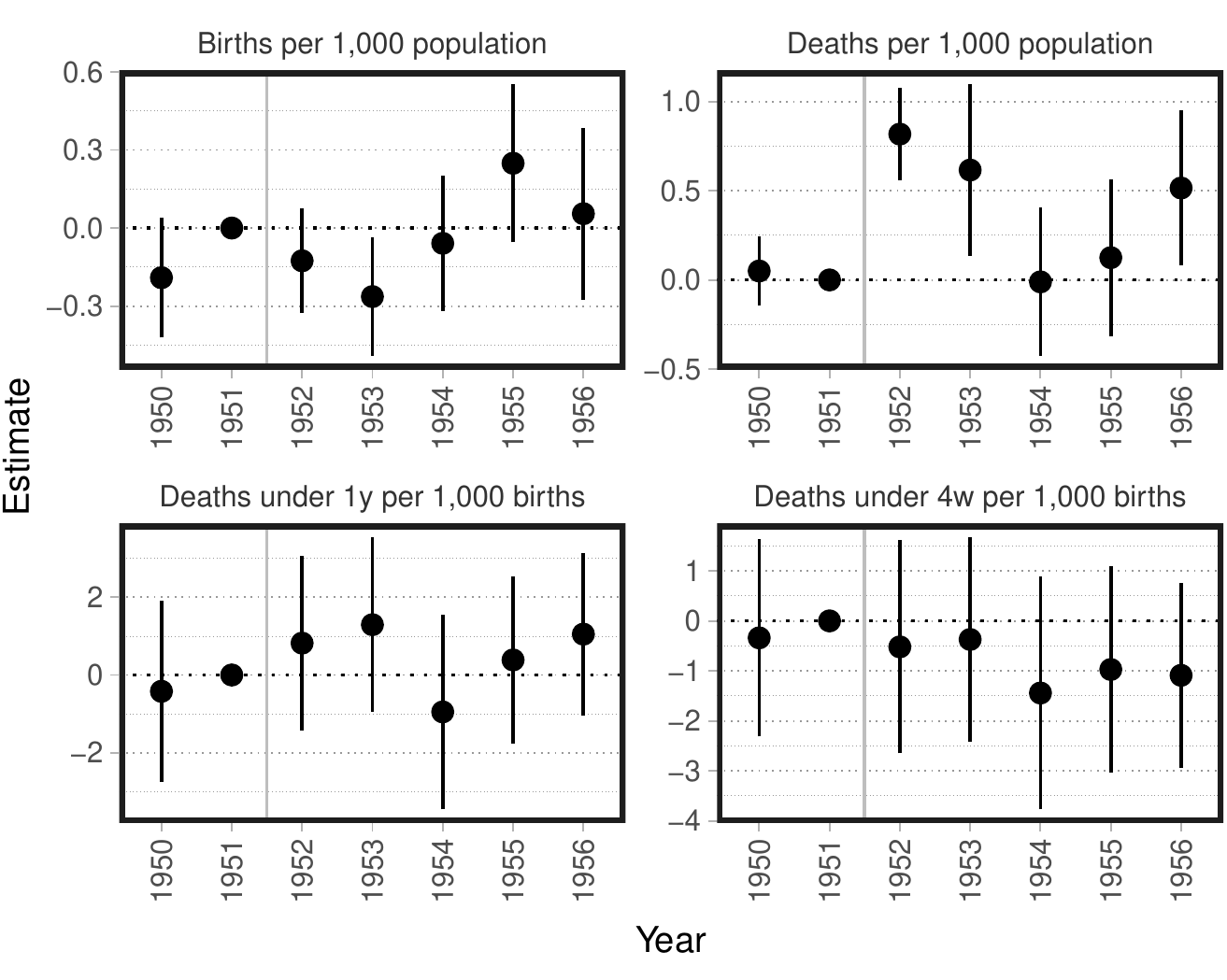}\caption*{Years to the right of the vertical line are post smog. 1951 is the reference year. The vertical lines on top of the point estimates show 0.95 confidence intervals.}\end{figure}

\begin{table}[!h]

\caption{\label{tab:lon_vs_urb-qualifications}Difference-in-Difference estimates comparing treated to control districts defined as urban England and Wales.}
\centering
\begin{threeparttable}
\begin{tabular}[t]{ldddd}
\toprule
\multicolumn{1}{c}{\em{}} & \multicolumn{4}{c}{\em{Exits with highest qualification:}} \\
\cmidrule(l{3pt}r{3pt}){2-5}
\multicolumn{1}{c}{} & \multicolumn{1}{c}{(1)} & \multicolumn{1}{c}{(2)} & \multicolumn{1}{c}{(3)} & \multicolumn{1}{c}{(4)} \\
\multicolumn{1}{c}{ } & \multicolumn{1}{c}{{\specialcell[b]{Degree}}} & \multicolumn{1}{c}{{\specialcell[b]{Upper secondary}}} & \multicolumn{1}{c}{{\specialcell[b]{Lower secondary}}} & \multicolumn{1}{c}{{\specialcell[b]{None}}}\\
\midrule
Treated $\times$ In utero & -0.001^{  } & -0.022^{  } & 0.011^{  } & 0.012^{  }\\
 & (0.019) & (0.017) & (0.012) & (0.011)\\
Treated $\times$ Childhood & -0.010^{  } & -0.018^{  } & 0.028^{ * } & 0.001^{  }\\
 & (0.019) & (0.018) & (0.015) & (0.011)\\
In utero & -0.016^{  } & 0.023^{ ** } & 0.000^{  } & -0.008^{  }\\
 & (0.011) & (0.012) & (0.009) & (0.009)\\
Childhood & -0.017^{  } & 0.037^{ * } & -0.007^{  } & -0.013^{  }\\
 & (0.023) & (0.022) & (0.017) & (0.017)\\
\midrule
\multicolumn{1}{l}{Observations} & \multicolumn{1}{c}{64,681} & \multicolumn{1}{c}{64,681} & \multicolumn{1}{c}{64,681} & \multicolumn{1}{c}{64,681}\\
\multicolumn{1}{l}{Mean dep. var.} & \multicolumn{1}{c}{0.347} & \multicolumn{1}{c}{0.356} & \multicolumn{1}{c}{0.17} & \multicolumn{1}{c}{0.126}\\
\multicolumn{1}{l}{$R^2$} & \multicolumn{1}{c}{0.07} & \multicolumn{1}{c}{0.021} & \multicolumn{1}{c}{0.029} & \multicolumn{1}{c}{0.036}\\
\bottomrule
\end{tabular}
\begin{tablenotes}
\item Columns: (1) exits with university/college, (2) exits at upper secondary level (A/AS-levels, professional/vocational training), (3) exits at lower secondary level (CSEs, GCSEs, O-levels), (4) exits with no qualifications. Includes fixed-effects for district, month of birth, and year of birth. Also controls for gender, weather variation, and year-month linear time trends by administrative county. Urban England and Wales are defined as districts that had a population density above 400 individuals per km$^2$ in 1951. Standard errors are clustered by district. (*): $p < 0.1$, (**): $p<0.05$, (***): $p<0.01$.
\end{tablenotes}
\end{threeparttable}
\end{table}

\begin{landscape}
\begin{table}[!h]

\caption{\label{tab:lon_vs_urb-icd_j_subgroups}Difference-in-Difference estimates comparing treated to control districts defined as urban England and Wales.}
\centering
\begin{threeparttable}
\fontsize{10}{12}\selectfont
\begin{tabular}[t]{ldddddddddd}
\toprule
\multicolumn{1}{c}{\em{}} & \multicolumn{10}{c}{\em{Dependent variable:}} \\
\cmidrule(l{3pt}r{3pt}){2-11}
\multicolumn{1}{c}{} & \multicolumn{1}{c}{(1)} & \multicolumn{1}{c}{(2)} & \multicolumn{1}{c}{(3)} & \multicolumn{1}{c}{(4)} & \multicolumn{1}{c}{(5)} & \multicolumn{1}{c}{(6)} & \multicolumn{1}{c}{(7)} & \multicolumn{1}{c}{(8)} & \multicolumn{1}{c}{(9)} & \multicolumn{1}{c}{(10)} \\
\multicolumn{1}{c}{ } & \multicolumn{1}{c}{{J00-J06}} & \multicolumn{1}{c}{{J09-J18}} & \multicolumn{1}{c}{{J20-J22}} & \multicolumn{1}{c}{{J30-J39}} & \multicolumn{1}{c}{{J40-J47}} & \multicolumn{1}{c}{{J60-J70}} & \multicolumn{1}{c}{{J80-J84}} & \multicolumn{1}{c}{{J85-J86}} & \multicolumn{1}{c}{{J90-J94}} & \multicolumn{1}{c}{{J95-J99}}\\
\midrule
Treated $\times$ In utero & 0.000^{  } & 0.012^{ ** } & -0.002^{  } & 0.008^{  } & -0.002^{  } & 0.000^{  } & 0.003^{  } & -0.001^{  } & -0.002^{  } & 0.004^{  }\\
 & (0.003) & (0.005) & (0.004) & (0.007) & (0.004) & (0.001) & (0.002) & (0.001) & (0.002) & (0.003)\\
Treated $\times$ Childhood & 0.003^{  } & 0.009^{  } & -0.003^{  } & -0.012^{  } & -0.005^{  } & 0.001^{  } & 0.000^{  } & 0.001^{  } & -0.002^{  } & 0.002^{  }\\
 & (0.003) & (0.006) & (0.005) & (0.007) & (0.005) & (0.001) & (0.002) & (0.002) & (0.003) & (0.003)\\
In utero & 0.003^{ * } & -0.006^{  } & -0.006^{ * } & -0.003^{  } & 0.004^{  } & 0.001^{  } & -0.001^{  } & 0.000^{  } & 0.002^{  } & -0.003^{ * }\\
 & (0.002) & (0.004) & (0.003) & (0.004) & (0.003) & (0.001) & (0.001) & (0.001) & (0.002) & (0.002)\\
Childhood & 0.004^{  } & -0.002^{  } & -0.012^{ * } & -0.012^{  } & 0.008^{  } & -0.002^{  } & 0.001^{  } & -0.001^{  } & 0.006^{ ** } & -0.001^{  }\\
 & (0.004) & (0.007) & (0.006) & (0.009) & (0.006) & (0.002) & (0.002) & (0.002) & (0.003) & (0.003)\\
\midrule
\multicolumn{1}{l}{Observations} & \multicolumn{1}{c}{64,917} & \multicolumn{1}{c}{64,917} & \multicolumn{1}{c}{64,919} & \multicolumn{1}{c}{64,919} & \multicolumn{1}{c}{64,920} & \multicolumn{1}{c}{64,917} & \multicolumn{1}{c}{64,917} & \multicolumn{1}{c}{64,917} & \multicolumn{1}{c}{64,918} & \multicolumn{1}{c}{64,917}\\
\multicolumn{1}{l}{Mean dep. var.} & \multicolumn{1}{c}{0.006} & \multicolumn{1}{c}{0.027} & \multicolumn{1}{c}{0.015} & \multicolumn{1}{c}{0.036} & \multicolumn{1}{c}{0.015} & \multicolumn{1}{c}{0.001} & \multicolumn{1}{c}{0.002} & \multicolumn{1}{c}{0.001} & \multicolumn{1}{c}{0.005} & \multicolumn{1}{c}{0.004}\\
\multicolumn{1}{l}{$R^2$} & \multicolumn{1}{c}{0.011} & \multicolumn{1}{c}{0.016} & \multicolumn{1}{c}{0.014} & \multicolumn{1}{c}{0.016} & \multicolumn{1}{c}{0.013} & \multicolumn{1}{c}{0.011} & \multicolumn{1}{c}{0.014} & \multicolumn{1}{c}{0.014} & \multicolumn{1}{c}{0.014} & \multicolumn{1}{c}{0.015}\\
\bottomrule
\end{tabular}
\begin{tablenotes}
\item Columns: Ever hospitalisatised due to (1) J00-J06 Acute upper respiratory infections, (2) J09-J18 Influenza and pneumonia, (3) J20-J22 Other acute lower respiratory infections, (4) J30-J39 Other diseases of upper respiratory tract, (5) J40-J47 Chronic lower respiratory diseases, (6) J60-J70 Lung diseases due to external agents, (7) J80-J84 Other respiratory diseases principally affecting the interstitium, (8) J85-J86 Suppurative and necrotic conditions of lower respiratory tract, (9) J90-J94 Other diseases of pleura, (10) J95-J99 Other diseases of the respiratory system. Includes fixed-effects for district, month of birth, and year of birth. Also controls for gender, weather variation, and year-month linear time trends by administrative county. Urban England and Wales are defined as districts that had a population density above 400 individuals per km$^2$ in 1951. Standard errors are clustered by district. (*): $p < 0.1$, (**): $p<0.05$, (***): $p<0.01$.
\end{tablenotes}
\end{threeparttable}
\end{table}

\end{landscape}

\begin{table}
\setlength{\tabcolsep}{1pt}
\caption{\label{tab:lon_vs_urb-robustness_add_regressors}Sensitivity of main difference-in-difference estimates to different sets of regressors).}
\centering
\begin{threeparttable}
\fontsize{7}{9}\selectfont
\begin{tabular}[t]{lddddddd}
\toprule
\multicolumn{1}{c}{\em{}} & \multicolumn{7}{c}{\em{Dependent variable:}} \\
\cmidrule(l{3pt}r{3pt}){2-8}
\multicolumn{1}{c}{} & \multicolumn{1}{c}{(1)} & \multicolumn{1}{c}{(2)} & \multicolumn{1}{c}{(3)} & \multicolumn{1}{c}{(4)} & \multicolumn{1}{c}{(5)} & \multicolumn{1}{c}{(6)} & \multicolumn{1}{c}{(7)} \\
\multicolumn{1}{c}{ } & \multicolumn{1}{c}{\specialcell[b]{Educational \\ attainment}} & \multicolumn{1}{c}{\specialcell[b]{Fluid \\ intelligence}} & \multicolumn{1}{c}{\specialcell[b]{Respiratory, \\ any}} & \multicolumn{1}{c}{\specialcell[b]{Influenza/ \\ pneumonia}} & \multicolumn{1}{c}{\specialcell[b]{COVID-19}} & \multicolumn{1}{c}{\specialcell[b]{Cancer}} & \multicolumn{1}{c}{\specialcell[b]{Cardio- \\ vascular}}\\
\midrule
\addlinespace[0.3em]
\multicolumn{8}{l}{\textbf{Panel (a) -- Base}}\\
\hspace{1em}Treated $\times$ In utero & -0.031^{  } & -0.070^{  } & 0.024^{ *** } & 0.007^{  } & 0.000^{  } & -0.000^{  } & 0.005^{  }\\
\hspace{1em} & (0.077) & (0.044) & (0.008) & (0.005) & (0.003) & (0.011) & (0.010)\\
\hspace{1em}Treated $\times$ Childhood & -0.060^{  } & -0.092^{ *** } & -0.000^{  } & 0.003^{  } & -0.001^{  } & 0.006^{  } & -0.003^{  }\\
\hspace{1em} & (0.047) & (0.033) & (0.008) & (0.004) & (0.001) & (0.007) & (0.008)\\
\hspace{1em}In utero & -0.047^{  } & 0.021^{  } & -0.001^{  } & 0.001^{  } & 0.001^{  } & 0.020^{ *** } & 0.017^{ *** }\\
\hspace{1em} & (0.031) & (0.020) & (0.003) & (0.002) & (0.001) & (0.005) & (0.005)\\
\hspace{1em}Childhood & -0.083^{ *** } & -0.003^{  } & 0.008^{ *** } & 0.005^{ *** } & 0.001^{  } & 0.028^{ *** } & 0.026^{ *** }\\
\hspace{1em} & (0.022) & (0.018) & (0.003) & (0.002) & (0.001) & (0.004) & (0.004)\\
\hspace{1em}Observations & \multicolumn{1}{D{,}{,}{-3}}{64,702} & \multicolumn{1}{D{,}{,}{-3}}{26,934} & \multicolumn{1}{D{,}{,}{-3}}{64,944} & \multicolumn{1}{D{,}{,}{-3}}{64,938} & \multicolumn{1}{D{,}{,}{-3}}{65,081} & \multicolumn{1}{D{,}{,}{-3}}{64,947} & \multicolumn{1}{D{,}{,}{-3}}{64,954}\\
\hspace{1em}Mean dep. var. & 13.318 & 6.43 & 0.092 & 0.027 & 0.007 & 0.142 & 0.192\\
\hspace{1em}$R^2$ & 0.010 & 0.005 & 0.002 & 0.001 & 0.001 & 0.002 & 0.014\\
\addlinespace[0.3em]
\multicolumn{8}{l}{\textbf{Panel (b) -- Add district FE}}\\
\hspace{1em}Treated $\times$ In utero & -0.049^{  } & -0.071^{  } & 0.024^{ *** } & 0.007^{  } & 0.000^{  } & 0.000^{  } & 0.005^{  }\\
\hspace{1em} & (0.076) & (0.043) & (0.008) & (0.005) & (0.003) & (0.011) & (0.010)\\
\hspace{1em}Treated $\times$ Childhood & -0.045^{  } & -0.086^{ ** } & -0.001^{  } & 0.002^{  } & -0.001^{  } & 0.006^{  } & -0.002^{  }\\
\hspace{1em} & (0.046) & (0.034) & (0.008) & (0.004) & (0.001) & (0.007) & (0.008)\\
\hspace{1em}In utero & -0.036^{  } & 0.019^{  } & -0.001^{  } & 0.000^{  } & 0.001^{  } & 0.020^{ *** } & 0.018^{ *** }\\
\hspace{1em} & (0.031) & (0.019) & (0.003) & (0.002) & (0.001) & (0.005) & (0.005)\\
\hspace{1em}Childhood & -0.088^{ *** } & -0.011^{  } & 0.009^{ *** } & 0.005^{ *** } & 0.001^{  } & 0.029^{ *** } & 0.026^{ *** }\\
\hspace{1em} & (0.021) & (0.019) & (0.003) & (0.001) & (0.001) & (0.004) & (0.004)\\
\hspace{1em}Observations & \multicolumn{1}{D{,}{,}{-3}}{64,681} & \multicolumn{1}{D{,}{,}{-3}}{26,877} & \multicolumn{1}{D{,}{,}{-3}}{64,923} & \multicolumn{1}{D{,}{,}{-3}}{64,917} & \multicolumn{1}{D{,}{,}{-3}}{65,060} & \multicolumn{1}{D{,}{,}{-3}}{64,926} &  \multicolumn{1}{D{,}{,}{-3}}{64,933}\\
\hspace{1em}Mean dep. var. & 13.317 & 6.429 & 0.092 & 0.027 & 0.007 & 0.142 &  0.192\\
\hspace{1em}$R^2$ & 0.076 & 0.058 & 0.014 & 0.013 & 0.011 & 0.013 & 0.027\\
\addlinespace[0.3em]
\multicolumn{8}{l}{\textbf{Panel (c) -- Add time (year and month of birth) FE and trends}}\\
\hspace{1em}Treated $\times$ In utero & -0.100^{  } & -0.111^{ ** } & 0.019^{ * } & 0.012^{ ** } & 0.000^{  } & -0.003^{  } & 0.011^{  }\\
\hspace{1em} & (0.089) & (0.051) & (0.011) & (0.005) & (0.003) & (0.013) &  (0.015)\\
\hspace{1em}Treated $\times$ Childhood & -0.135^{  } & -0.158^{ ** } & -0.007^{  } & 0.009^{  } & -0.001^{  } & -0.001^{  } & 0.006^{  }\\
\hspace{1em} & (0.088) & (0.068) & (0.012) & (0.006) & (0.003) & (0.014) &  (0.019)\\
\hspace{1em}In utero & -0.002^{  } & 0.008^{  } & -0.006^{  } & -0.006^{  } & -0.000^{  } & -0.005^{  } & -0.004^{  }\\
\hspace{1em} & (0.054) & (0.036) & (0.006) & (0.004) & (0.002) & (0.008) &  (0.010)\\
\hspace{1em}Childhood & -0.014^{  } & -0.052^{  } & -0.008^{  } & -0.002^{  } & -0.001^{  } & -0.009^{  } & -0.016^{  }\\
\hspace{1em} & (0.120) & (0.074) & (0.014) & (0.007) & (0.004) & (0.016) &  (0.023)\\
\hspace{1em}Observations & \multicolumn{1}{D{,}{,}{-3}}{64,681} & \multicolumn{1}{D{,}{,}{-3}}{26,877} & \multicolumn{1}{D{,}{,}{-3}}{64,923} & \multicolumn{1}{D{,}{,}{-3}}{64,917} & \multicolumn{1}{D{,}{,}{-3}}{65,060} & \multicolumn{1}{D{,}{,}{-3}}{64,926} &  \multicolumn{1}{D{,}{,}{-3}}{64,933}\\
\hspace{1em}Mean dep. var. & 13.317 & 6.429 & 0.092 & 0.027 & 0.007 & 0.142 &  0.192\\
\hspace{1em}$R^2$ & 0.080 & 0.067 & 0.018 & 0.016 & 0.013 & 0.016 &  0.030\\
\addlinespace[0.3em]
\multicolumn{8}{l}{\textbf{Panel (d) -- Add weather covariates}}\\
\hspace{1em}Treated $\times$ In utero & -0.100^{  } & -0.112^{ ** } & 0.020^{ * } & 0.012^{ ** } & 0.000^{  } & -0.003^{  } & 0.011^{  }\\
\hspace{1em} & (0.089) & (0.051) & (0.011) & (0.005) & (0.003) & (0.013) & (0.015)\\
\hspace{1em}Treated $\times$ Childhood & -0.135^{  } & -0.158^{ ** } & -0.007^{  } & 0.009^{  } & -0.001^{  } & -0.001^{  } & 0.007^{  }\\
\hspace{1em} & (0.088) & (0.068) & (0.012) & (0.006) & (0.003) & (0.014) & (0.019)\\
\hspace{1em}In utero & -0.003^{  } & 0.008^{  } & -0.005^{  } & -0.006^{  } & -0.000^{  } & -0.005^{  } & -0.004^{  }\\
\hspace{1em} & (0.054) & (0.036) & (0.006) & (0.004) & (0.002) & (0.008) & (0.010)\\
\hspace{1em}Childhood & -0.014^{  } & -0.051^{  } & -0.008^{  } & -0.002^{  } & -0.001^{  } & -0.009^{  } & -0.015^{  }\\
\hspace{1em} & (0.120) & (0.074) & (0.014) & (0.007) & (0.004) & (0.016) & (0.023)\\
\hspace{1em}Observations & \multicolumn{1}{D{,}{,}{-3}}{64,681} & \multicolumn{1}{D{,}{,}{-3}}{26,877} & \multicolumn{1}{D{,}{,}{-3}}{64,923} & \multicolumn{1}{D{,}{,}{-3}}{64,917} & \multicolumn{1}{D{,}{,}{-3}}{65,060} & \multicolumn{1}{D{,}{,}{-3}}{64,926} & \multicolumn{1}{D{,}{,}{-3}}{64,933}\\
\hspace{1em}Mean dep. var. & 13.317 & 6.429 & 0.092 & 0.027 & 0.007 & 0.142 & 0.192\\
\hspace{1em}$R^2$ & 0.080 & 0.067 & 0.018 & 0.016 & 0.013 & 0.016 & 0.030\\
\bottomrule
\end{tabular}
\begin{tablenotes}
\item Columns: (1) educational attainment in years, (2) standardised fluid intelligence score, (3) ever hospitalised due to respiratory disease, (4) ever hospitalised due to influenza/pneumonia, (5) hospitalisation or death due to COVID-19, (6) ever hospitalised due to cardio-vascular disease, (7) ever hospitalised due to cancer. Panels show the estimates for increasingly complex sets of regressors. The base specification in Panel (a) only contains indicators for whether the individual was born in a treatment district and whether the individual belonged to the childhood, in utero, or after birth cohorts. Panels (b)-(d) add additional regressors to this specification. Standard errors are clustered by district. (*): $p < 0.1$, (**): $p<0.05$, (***): $p<0.01$.
\end{tablenotes}
\end{threeparttable}
\end{table}

\begin{landscape}
\begin{table}

\caption{\label{tab:lon_vs_urb-borusyak}Robust estimates of average treatment effects using estimator from Borusyak et al. (2021). Compares treated to control districts defined as urban England and Wales.}
\centering
\begin{threeparttable}
\fontsize{10}{12}\selectfont
\begin{tabular}[t]{lddddddd}
\toprule
\multicolumn{1}{c}{\em{}} & \multicolumn{7}{c}{\em{Dependent variable:}} \\
\cmidrule(l{3pt}r{3pt}){2-8}
\multicolumn{1}{c}{} & \multicolumn{1}{c}{(1)} & \multicolumn{1}{c}{(2)} & \multicolumn{1}{c}{(3)} & \multicolumn{1}{c}{(4)} & \multicolumn{1}{c}{(5)} & \multicolumn{1}{c}{(6)} & \multicolumn{1}{c}{(7)} \\
\multicolumn{1}{c}{ } & \multicolumn{1}{c}{\specialcell[b]{Educational \\ attainment}} & \multicolumn{1}{c}{\specialcell[b]{Fluid \\ intelligence}} & \multicolumn{1}{c}{\specialcell[b]{Respiratory, \\ any}} & \multicolumn{1}{c}{\specialcell[b]{Influenza/ \\ pneumonia}} & \multicolumn{1}{c}{\specialcell[b]{COVID-19}} & \multicolumn{1}{c}{\specialcell[b]{Cancer}} & \multicolumn{1}{c}{\specialcell[b]{Cardio- \\ vascular}}\\
\midrule
ATT, In utero & -0.049^{  } & -0.137^{  } & 0.025^{ *** } & 0.008^{  } & 0.000^{  } & 0.000^{  } & 0.004^{  }\\
 & (0.075) & (0.089) & (0.009) & (0.005) & (0.003) & (0.011) & (0.011)\\
ATT, Childhood & -0.043^{  } & -0.161^{ ** } & -0.001^{  } & 0.002^{  } & -0.000^{  } & 0.006^{  } & -0.002^{  }\\
 & (0.047) & (0.069) & (0.008) & (0.004) & (0.001) & (0.007) & (0.008)\\
\bottomrule
\end{tabular}
\begin{tablenotes}
\item Columns: (1) educational attainment in years, (2) standardised fluid intelligence score, (3) ever hospitalised due to respiratory disease, (4) ever hospitalised due to influenza/pneumonia, (5) hospitalisation or death due to COVID-19, (6) hospitalised due to cardio-vascular disease, (7) hospitalised due to cancer. Reports the robust estimates of the ATTs for the in utero and childhood cohorts when we assign equal weights across units in these cohorts. Includes fixed-effects for district, month of birth, and year of birth. Also controls for year-month linear time trends by administrative county. Urban England and Wales are defined as districts that had a population density above 400 individuals per km$^2$ in 1951. Standard errors are clustered by district. (*): $p < 0.1$, (**): $p<0.05$, (***): $p<0.01$.
\end{tablenotes}
\end{threeparttable}
\end{table}

\end{landscape}

\begin{table}[!h]

\caption{\label{tab:lon_vs_urb-economic_outcomes_no-ancestry}Difference-in-Difference estimates comparing treated to control districts defined as urban England and Wales.}
\centering
\begin{threeparttable}
\begin{tabular}[t]{ldd}
\toprule
\multicolumn{1}{c}{\em{}} & \multicolumn{2}{c}{\em{Dependent variable:}} \\
\cmidrule(l{3pt}r{3pt}){2-3}
\multicolumn{1}{c}{} & \multicolumn{1}{c}{(1)} & \multicolumn{1}{c}{(2)} \\
\multicolumn{1}{c}{ } & \multicolumn{1}{c}{{\specialcell[b]{Educational \\ attainment}}} & \multicolumn{1}{c}{{\specialcell[b]{Fluid \\ intelligence}}}\\
\midrule
Treated $\times$ In utero & -0.078^{  } & -0.096^{ * }\\
 & (0.085) & (0.050)\\
Treated $\times$ Childhood & -0.120^{  } & -0.163^{ ** }\\
 & (0.084) & (0.066)\\
In utero & -0.019^{  } & 0.003^{  }\\
 & (0.053) & (0.035)\\
Childhood & -0.037^{  } & -0.057^{  }\\
 & (0.118) & (0.072)\\
\midrule
\multicolumn{1}{l}{Observations} & \multicolumn{1}{c}{65,724} & \multicolumn{1}{c}{27,333}\\
\multicolumn{1}{l}{Mean dep. var.} & \multicolumn{1}{c}{13.32} & \multicolumn{1}{c}{0}\\
\multicolumn{1}{l}{$R^2$} & \multicolumn{1}{c}{0.079} & \multicolumn{1}{c}{0.066}\\
\bottomrule
\end{tabular}
\begin{tablenotes}
\item Columns: (1) educational attainment in years, (2) standardised fluid intelligence score. Includes fixed-effects for district, month of birth, and year of birth. Also controls for gender, weather variation, and year-month linear time trends by administrative county. Urban England and Wales are defined as districts that had a population density above 400 individuals per km$^2$ in 1951. Standard errors are clustered by district. (*): $p < 0.1$, (**): $p<0.05$, (***): $p<0.01$.
\end{tablenotes}
\end{threeparttable}
\end{table}

\begin{table}[!h]

\caption{\label{tab:lon_vs_urb-health_outcomes_no-ancestry}Difference-in-Difference estimates comparing treated to control districts defined as urban England and Wales.}
\centering
\begin{threeparttable}
\fontsize{10}{12}\selectfont
\begin{tabular}[t]{lddddd}
\toprule
\multicolumn{1}{c}{\em{}} & \multicolumn{5}{c}{\em{Dependent variable:}} \\
\cmidrule(l{3pt}r{3pt}){2-6}
\multicolumn{1}{c}{} & \multicolumn{1}{c}{(1)} & \multicolumn{1}{c}{(2)} & \multicolumn{1}{c}{(3)} & \multicolumn{1}{c}{(4)} & \multicolumn{1}{c}{(5)} \\
\multicolumn{1}{c}{ } & \multicolumn{1}{c}{{\specialcell[b]{Respiratory, \\ any}}} & \multicolumn{1}{c}{{\specialcell[b]{Influenza/ \\ pneumonia}}} & \multicolumn{1}{c}{{\specialcell[b]{COVID-19}}} & \multicolumn{1}{c}{{\specialcell[b]{Cancer}}} & \multicolumn{1}{c}{{\specialcell[b]{Cardio- \\ vascular}}}\\
\midrule
Treated $\times$ In utero & 0.018^{ * } & 0.011^{ ** } & 0.000^{  } & -0.005^{  } & 0.012^{  }\\
 & (0.010) & (0.005) & (0.003) & (0.013) & (0.015)\\
Treated $\times$ Childhood & -0.009^{  } & 0.009^{  } & -0.001^{  } & -0.005^{  } & 0.005^{  }\\
 & (0.012) & (0.006) & (0.003) & (0.014) & (0.019)\\
In utero & -0.006^{  } & -0.006^{  } & 0.000^{  } & -0.004^{  } & -0.004^{  }\\
 & (0.006) & (0.004) & (0.002) & (0.008) & (0.010)\\
Childhood & -0.007^{  } & -0.002^{  } & -0.001^{  } & -0.007^{  } & -0.013^{  }\\
 & (0.014) & (0.007) & (0.004) & (0.016) & (0.023)\\
\midrule
\multicolumn{1}{l}{Observations} & \multicolumn{1}{c}{65,977} & \multicolumn{1}{c}{65,971} & \multicolumn{1}{c}{66,117} & \multicolumn{1}{c}{65,980} & \multicolumn{1}{c}{65,987}\\
\multicolumn{1}{l}{Mean dep. var.} & \multicolumn{1}{c}{0.093} & \multicolumn{1}{c}{0.027} & \multicolumn{1}{c}{0.007} & \multicolumn{1}{c}{0.142} & \multicolumn{1}{c}{0.193}\\
\multicolumn{1}{l}{$R^2$} & \multicolumn{1}{c}{0.018} & \multicolumn{1}{c}{0.016} & \multicolumn{1}{c}{0.013} & \multicolumn{1}{c}{0.016} & \multicolumn{1}{c}{0.03}\\
\bottomrule
\end{tabular}
\begin{tablenotes}
\item Columns: (1) ever hospitalised due to respiratory disease, (2) ever hospitalised due to influenza/pneumonia, (3) hospitalisation or death due to COVID-19, (4) ever hospitalised due to cardio-vascular disease, (5) ever hospitalised due to cancer. Includes fixed-effects for district, month of birth, and year of birth. Also controls for gender, weather variation, and year-month linear time trends by administrative county. Urban England and Wales are defined as districts that had a population density above 400 individuals per km$^2$ in 1951. Standard errors are clustered by district. (*): $p < 0.1$, (**): $p<0.05$, (***): $p<0.01$.
\end{tablenotes}
\end{threeparttable}
\end{table}

\begin{table}[!h]

\caption{\label{tab:lon_vs_urb-sex_outcome}Difference-in-Difference estimates comparing treated to control districts defined as urban England and Wales.}
\centering
\begin{threeparttable}
\begin{tabular}[t]{ld}
\toprule
\multicolumn{1}{c}{\em{}} & \multicolumn{1}{c}{\em{Dependent variable:}} \\
\cmidrule(l{3pt}r{3pt}){2-2}
\multicolumn{1}{c}{} & \multicolumn{1}{c}{(1)} \\
\multicolumn{1}{c}{ } & \multicolumn{1}{c}{{\specialcell[b]{Male}}}\\
\midrule
Treated $\times$ In utero & -0.007^{  }\\
 & (0.021)\\
Treated $\times$ Childhood & 0.011^{  }\\
 & (0.020)\\
In utero & 0.001^{  }\\
 & (0.011)\\
Childhood & -0.044^{ ** }\\
 & (0.022)\\
\midrule
\multicolumn{1}{l}{Observations} & \multicolumn{1}{c}{65,060}\\
\multicolumn{1}{l}{Mean dep. var.} & \multicolumn{1}{c}{0.443}\\
\multicolumn{1}{l}{$R^2$} & \multicolumn{1}{c}{0.014}\\
\bottomrule
\end{tabular}
\begin{tablenotes}
\item Columns: (1) being born as male. Includes fixed-effects for district, month of birth, and year of birth. Also controls for gender, weather variation, and year-month linear time trends by administrative county. Urban England and Wales are defined as districts that had a population density above 400 individuals per km$^2$ in 1951. Standard errors are clustered by district. (*): $p < 0.1$, (**): $p<0.05$, (***): $p<0.01$.
\end{tablenotes}
\end{threeparttable}
\end{table}

\begin{table}

\caption{\label{tab:qualifications}Heterogeneity across genetics -- Qualifications. Difference-in-Difference estimates comparing treated to control districts defined as urban England and Wales.}
\centering
\begin{threeparttable}
\begin{tabular}[t]{ldddd}
\toprule
\multicolumn{1}{c}{\em{}} & \multicolumn{4}{c}{\em{Exits with highest qualification:}} \\
\cmidrule(l{3pt}r{3pt}){2-5}
\multicolumn{1}{c}{} & \multicolumn{1}{c}{(1)} & \multicolumn{1}{c}{(2)} & \multicolumn{1}{c}{(3)} & \multicolumn{1}{c}{(4)} \\
\multicolumn{1}{c}{ } & \multicolumn{1}{c}{\specialcell[b]{Degree}} & \multicolumn{1}{c}{\specialcell[b]{Upper secondary}} & \multicolumn{1}{c}{\specialcell[b]{Lower secondary}} & \multicolumn{1}{c}{\specialcell[b]{None}}\\
\midrule
\addlinespace[0.3em]
\multicolumn{5}{l}{\textbf{Panel (a) -- High polygenic score}}\\
\hspace{1em}Treated $\times$ In utero & -0.036^{  } & -0.015^{  } & 0.047^{ *** } & 0.005^{  }\\
\hspace{1em} & (0.024) & (0.025) & (0.017) & (0.009)\\
\hspace{1em}Treated $\times$ Childhood & -0.017^{  } & -0.025^{  } & 0.023^{  } & 0.020^{ ** }\\
\hspace{1em} & (0.029) & (0.027) & (0.019) & (0.010)\\
\hspace{1em}In utero & -0.014^{  } & 0.027^{  } & -0.015^{  } & 0.002^{  }\\
\hspace{1em} & (0.016) & (0.018) & (0.012) & (0.008)\\
\hspace{1em}Childhood & -0.054^{  } & 0.045^{  } & 0.011^{  } & -0.002^{  }\\
\hspace{1em} & (0.033) & (0.033) & (0.024) & (0.019)\\
\addlinespace[0.75em]
\hspace{1em}Observations & \multicolumn{1}{D{,}{,}{-3}}{32,376} & \multicolumn{1}{D{,}{,}{-3}}{32,376} & \multicolumn{1}{D{,}{,}{-3}}{32,376} & \multicolumn{1}{D{,}{,}{-3}}{32,376}\\
\hspace{1em}Mean dep. var. & 0.454 & 0.337 & 0.139 & 0.07\\
\hspace{1em}$R^2$ & 0.077 & 0.039 & 0.041 & 0.039\\
\addlinespace[0.3em]
\multicolumn{5}{l}{\textbf{Panel (b) -- Low polygenic score}}\\
\hspace{1em}Treated $\times$ In utero & 0.045^{  } & -0.029^{  } & -0.028^{  } & 0.012^{  }\\
\hspace{1em} & (0.031) & (0.031) & (0.022) & (0.021)\\
\hspace{1em}Treated $\times$ Childhood & 0.004^{  } & -0.008^{  } & 0.040^{  } & -0.036^{  }\\
\hspace{1em} & (0.031) & (0.027) & (0.027) & (0.022)\\
\hspace{1em}In utero & -0.023^{  } & 0.023^{  } & 0.015^{  } & -0.016^{  }\\
\hspace{1em} & (0.014) & (0.016) & (0.014) & (0.014)\\
\hspace{1em}Childhood & 0.012^{  } & 0.034^{  } & -0.023^{  } & -0.023^{  }\\
\hspace{1em} & (0.029) & (0.032) & (0.026) & (0.028)\\
\addlinespace[0.75em]
\hspace{1em}Observations & \multicolumn{1}{D{,}{,}{-3}}{32,233} & \multicolumn{1}{D{,}{,}{-3}}{32,233} & \multicolumn{1}{D{,}{,}{-3}}{32,233} & \multicolumn{1}{D{,}{,}{-3}}{32,233}\\
\hspace{1em}Mean dep. var. & 0.24 & 0.376 & 0.202 & 0.183\\
\hspace{1em}$R^2$ & 0.069 & 0.032 & 0.040 & 0.044\\
\bottomrule
\end{tabular}
\begin{tablenotes}
\item Columns: (1) exits with university/college degree, (2) exits at upper secondary level (A/AS-levels, professional/vocational training), (2) exits at lower secondary level (CSEs, GCSEs, O-levels), (3) exits with no qualifications. Panels: (a) high PGS subsample, (b) low PGS subsample. Includes fixed-effects for district, month of birth, and year of birth. Also controls for gender, weather variation, and year-month linear time trends by administrative county. Urban England and Wales are defined as districts that had a population density above 400 individuals per km$^2$ in 1951. Standard errors are clustered by district. (*): $p < 0.1$, (**): $p<0.05$, (***): $p<0.01$.
\end{tablenotes}
\end{threeparttable}
\end{table}

\FloatBarrier

\setcounter{table}{0} 
\setcounter{figure}{0} 
\setcounter{equation}{0} 
\renewcommand{\thetable}{\Alph{subsection}.\arabic{table}}
\renewcommand{\thefigure}{\Alph{subsection}.\arabic{figure}} %
\renewcommand{\theequation}{\Alph{subsection}.\arabic{equation}}
\renewcommand{\thesubsection}{\Alph{subsection}}
\renewcommand{\thesubsubsection}{\Alph{subsection}.\arabic{subsubsection}}
\subsection{Choice of time trends}\label{sec:trend_robustness}
The main analysis controls for administrative-county-specific (year-month) trends to allow the outcome of interest to trend differently in each administrative county. We here explore the sensitivity of the trend-specifications. Panel~(a) of \autoref{tab:lon_vs_urb-robustness-trends}, includes administrative county-specific \textit{annual} (as opposed to year-month) trends. Panel~(b) specifies \textit{year-month} trends for each of the over 1400 districts observed in our data, and Panel~(c) allows for district-specific \textit{annual} trends. Finally, Panel~(d) does not include any trends and only accounts for year of birth dummies.

\begin{table}

\caption{\label{tab:lon_vs_urb-robustness-trends}Specification of trends. Difference‐in‐Difference estimates comparing treated to control districts defined as urban England and Wales.}
\centering
\begin{threeparttable}
\begin{tabular}[t]{ldddd}
\toprule
\multicolumn{1}{c}{\em{}} & \multicolumn{4}{c}{\em{Dependent variable:}} \\
\cmidrule(l{3pt}r{3pt}){2-5}
\multicolumn{1}{c}{} & \multicolumn{1}{c}{(1)} & \multicolumn{1}{c}{(2)} & \multicolumn{1}{c}{(3)} & \multicolumn{1}{c}{(4)} \\
\multicolumn{1}{c}{ } & \multicolumn{1}{c}{\specialcell[b]{Educational \\ attainment}} & \multicolumn{1}{c}{\specialcell[b]{Fluid \\ intelligence}} & \multicolumn{1}{c}{\specialcell[b]{Respiratory, \\ any}} & \multicolumn{1}{c}{\specialcell[b]{Influenza/ \\ pneumonia}}\\
\midrule
\addlinespace[0.3em]
\multicolumn{5}{l}{\textbf{Panel (a) -- Year by administrative county}}\\
\hspace{1em}Treated $\times$ In utero & -0.087^{  } & -0.101^{ ** } & 0.016^{  } & 0.010^{ * }\\
\hspace{1em} & (0.089) & (0.050) & (0.010) & (0.005)\\
\hspace{1em}Treated $\times$ Childhood & -0.119^{  } & -0.142^{ ** } & -0.014^{  } & 0.007^{  }\\
\hspace{1em} & (0.086) & (0.066) & (0.012) & (0.006)\\
\addlinespace[0.3em]
\multicolumn{5}{l}{\textbf{Panel (b) -- Year-month by district}}\\
\hspace{1em}Treated $\times$ In utero & -0.118^{  } & -0.164^{ *** } & 0.022^{ * } & 0.017^{ *** }\\
\hspace{1em} & (0.091) & (0.048) & (0.012) & (0.006)\\
\hspace{1em}Treated $\times$ Childhood & -0.173^{ * } & -0.248^{ *** } & -0.003^{  } & 0.017^{ ** }\\
\hspace{1em} & (0.098) & (0.064) & (0.016) & (0.007)\\
\addlinespace[0.3em]
\multicolumn{5}{l}{\textbf{Panel (c) -- Year by district}}\\
\hspace{1em}Treated $\times$ In utero & -0.111^{  } & -0.150^{ *** } & 0.017^{  } & 0.013^{ ** }\\
\hspace{1em} & (0.090) & (0.047) & (0.012) & (0.006)\\
\hspace{1em}Treated $\times$ Childhood & -0.165^{ * } & -0.230^{ *** } & -0.013^{  } & 0.012^{ * }\\
\hspace{1em} & (0.093) & (0.062) & (0.015) & (0.007)\\
\addlinespace[0.3em]
\multicolumn{5}{l}{\textbf{Panel (d) -- No trend}}\\
\hspace{1em}Treated $\times$ In utero & -0.049^{  } & -0.072^{ * } & 0.024^{ *** } & 0.008^{  }\\
\hspace{1em} & (0.075) & (0.043) & (0.008) & (0.005)\\
\hspace{1em}Treated $\times$ Childhood & -0.046^{  } & -0.086^{ ** } & -0.001^{  } & 0.002^{  }\\
\hspace{1em} & (0.046) & (0.034) & (0.008) & (0.004)\\
\bottomrule
\end{tabular}
\begin{tablenotes}
\item Columns: (1) educational attainment in years, (2) standardised fluid intelligence score, (3) ever hospitalised due to respiratory disease, (4) ever hospitalised due to influenza/pneumonia. Panels: (a) Year trend at administrative county ($n=174$) level, (b) Year-month trend at district ($n=785$) level, (c) Year trend at district level. (d) No trends.  We always include district FE, year-of-birth FE, and month-of-birth FE, and we always control for gender and weather variation. Standard errors are clustered by district.
\end{tablenotes}
\end{threeparttable}
\end{table}

The estimates are generally consistent across the different specifications. For educational attainment, we find negative estimates throughout for both \textit{in utero} and childhood exposure. For fluid intelligence, we find clear evidence of a negative effect of smog exposure that is slightly larger for exposure in childhood compared to prenatally. The in utero estimates are always positive for respiratory disease, but they are slightly smaller when accounting for annual trends at either the administrative county or district level. For influenza/pneumonia, the estimates always have the same sign, but are somewhat smaller when controlling for annual trends or not including a trend. Nevertheless, for both respiratory disease and influenza/pneumonia, the magnitudes of the estimates remains in the same ballpark, suggesting that smog exposure increases the probability of being diagnosed with respiratory disease.

\clearpage

\setcounter{table}{0} 
\setcounter{figure}{0} 
\setcounter{equation}{0} 
\renewcommand{\thetable}{\Alph{subsection}.\arabic{table}}
\renewcommand{\thefigure}{\Alph{subsection}.\arabic{figure}} %
\renewcommand{\theequation}{\Alph{subsection}.\arabic{equation}}
\renewcommand{\thesubsection}{\Alph{subsection}}
\renewcommand{\thesubsubsection}{\Alph{subsection}.\arabic{subsubsection}}
\subsection{Common time trends} \label{sec:AppendixC}
Our specification implicitly assumes that individuals who were born in districts that were affected by the London smog would have had similar trends in their outcomes of interest in the absence of the smog compared to those born in districts that were not affected. To explore this common trend assumption empirically, we compare the trends in the relevant outcomes of interest among those conceived at different points in time throughout our observation period in treated and control districts. Note here, that we are mainly interested in comparing individuals in treated and control districts who are conceived \textit{after} the smog, since those who are conceived before or during the smog were potentially exposed either \textit{in utero} or in childhood. We here focus on our main outcomes of interest: education, fluid intelligence and respiratory disease.

\begin{figure}[!h]\caption{\label{fig:trends}The conditional differences in the means of the relevant outcome for those born in treated versus control districts.}\centering\includegraphics[width=\textwidth]{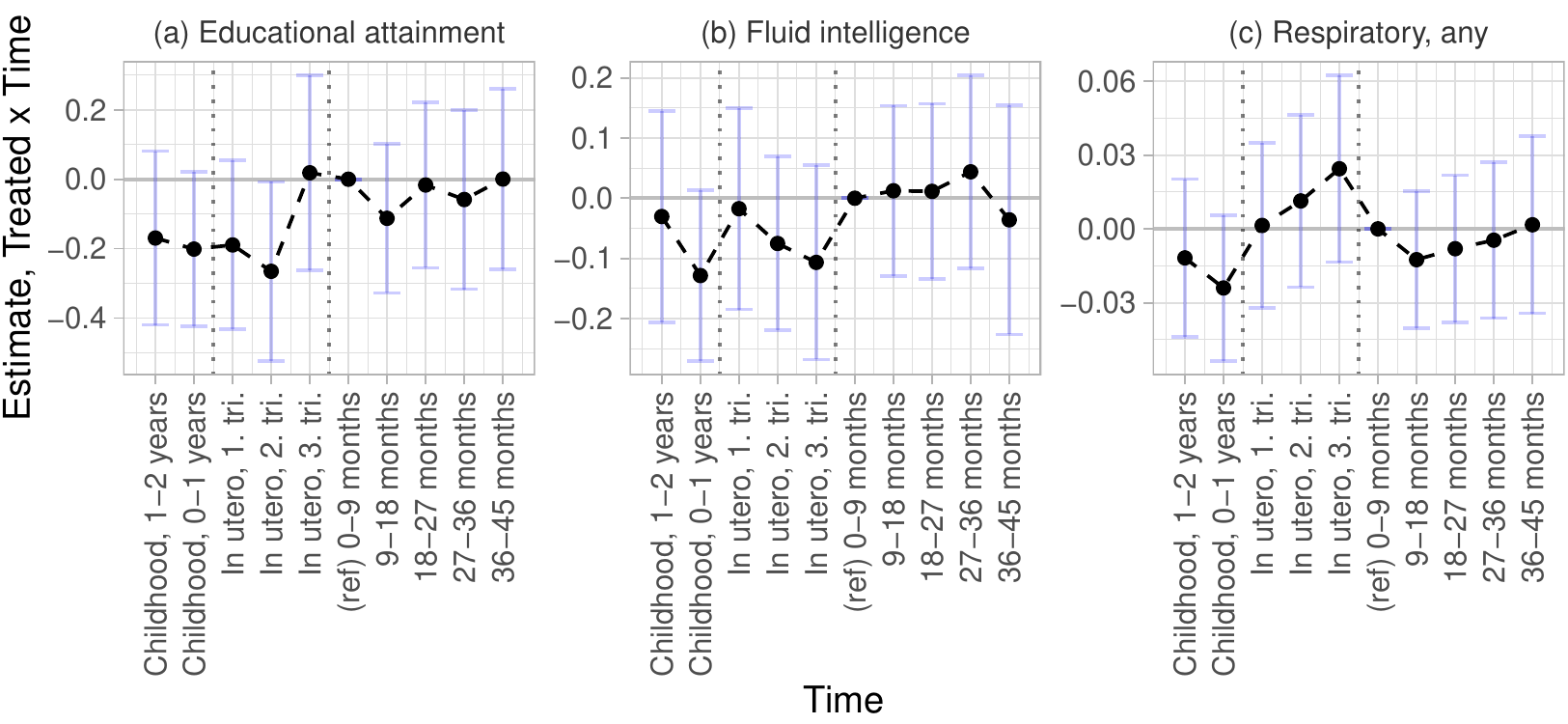}\caption*{Shows the conditional differences across childhood, trimesters in utero, and 9 month intervals throughout our observation period. We condition on the same controls as in the main analysis, and we include fixed-effects for district. We also control for gender, weather variation, and year-month linear time trends by administrative county. The two vertical dotted lines indicate the threshold for potential exposure in childhood and \textit{in utero}. The vertical bars show $90\%$ confidence intervals around the point estimates. Control districts are defined as districts in England and Wales that had a population density above 400 individuals per km$^2$ in 1951. Standard errors are clustered by district.}\end{figure}

\autoref{fig:trends} shows the conditional difference in the mean of the relevant outcome for those born in treated versus control districts across childhood, trimesters \textit{in utero}, and 9 month intervals throughout our observation period. We condition on the same controls and fixed-effects as in the main analysis. The two vertical dotted lines indicate the threshold for potential exposure in childhood and \textit{in utero}. 

\autoref{fig:trends}a and \autoref{fig:trends}b show that those exposed \textit{in utero} or in early childhood have lower education and fluid intelligence compared to those born at the same time, but in control districts. Similarly, \autoref{fig:trends}c shows that those who are exposed to the smog whilst \textit{in utero} have a higher probability of being diagnosed with respiratory disease compared to those born at the same time, but in control districts. For all outcomes, we see no suggestion of differential trends for those conceived after the smog, i.e., those to the right of the second vertical dotted line. In other words, we find no evidence to suggest that those born in treated districts have differential trends in our outcomes of interest compared to those born control districts. 

\FloatBarrier

\clearpage


\setcounter{table}{0} 
\setcounter{figure}{0} 
\setcounter{equation}{0} 
\renewcommand{\thetable}{\Alph{subsection}.\arabic{table}}
\renewcommand{\thefigure}{\Alph{subsection}.\arabic{figure}} %
\renewcommand{\theequation}{\Alph{subsection}.\arabic{equation}}
\renewcommand{\thesubsection}{\Alph{subsection}}
\renewcommand{\thesubsubsection}{\Alph{subsection}.\arabic{subsubsection}}

\subsection{A brief background to genetics} \label{sec:AppendixA}
Humans have 46 chromosomes stored in every cell apart from sex-cells. The chromosomes exist in pairs such that each pair has a maternal and paternal copy. A single chromosome consists of a double-strand of deoxyribonucleic acid (DNA) containing a large number of `base pairs': pairs of nucleotide molecules (referred to as the `letters' A (adenine) that binds with T (thymine), and G (guanine) that binds with C (cytosine)) that together make up the human genome. In a population there will be variation in the base pairs at some locations. Such variation is known as a single nucleotide polymorphism (SNP, pronounced `snip') -- a change in the base pair at one particular locus (location) -- and is the most commonly studied genetic variation. When there are two possible base pairs at a given location (i.e., two alleles), the most frequent base pair is called the major allele, while the less frequent is called the minor allele. As humans have two copies of each chromosome, any given individual can have either zero, one, or two copies of the minor allele. 

To identify specific SNPs that are robustly associated with a particular outcome of interest, so-called Genome-Wide Association Studies (GWAS) relate each SNP to the outcome in a hypothesis-free approach. As there are more SNPs than individuals, the SNP effects cannot be identified in a multivariate regression model. Instead, a GWAS runs a large number of univariate regressions of the outcome on each SNP. These analyses have shown that most outcomes of interest in the social sciences are `polygenic': they are affected by a large number of SNPs, each with a very small effect. To increase the predictive power of the SNPs, it is therefore custom to aggregate the individual SNPs into so-called polygenic scores, as:
$$
G_i = \sum_{j=1}^J \beta_j X_{ij},
$$
where $X_{ij}$ is a count of the number of minor alleles (i.e., 0, 1 or 2) at SNP $j$ for  individual $i$, and $\beta_j$ is its effect size obtained from an independent GWAS. Hence, the polygenic scores are weighted linear combinations of SNPs, where the weights are estimated in an independent GWAS. This is motivated by an additive genetic model where all SNPs contribute additively to the overall genetic predisposition of an individual  \citep[see e.g.,][]{purcell2009prs}. 

We conduct our own tailor-made GWAS for each of the main outcomes in our analysis. To avoid overfitting, we partition the UK Biobank into three non-overlapping samples: (1) a GWAS discovery sample, (2) a reference/tuning sample, and (3) the analysis sample. We use samples (1) and (2) to construct polygenic scores for the individuals in the analysis sample (i.e., sample (3)). We then use these polygenic scores to explore the extent to which one's genetic predisposition can `protect' or `exacerbate' the effect of early-life pollution exposure. The analysis sample is outlined in Section~\ref{sec:data} and contains individuals born in treated or control districts in 1950--1956. The GWAS discovery sample contains individuals born outside the study period, i.e., 1934-1949 and 1957-1970, as well as individuals born outside treated and control districts in the years 1950-1956. From the GWAS discovery sample, we randomly sample 20,000 unrelated individuals of white British ancestry that we exclusively use for the reference sample.\footnote{We use the reference sample to estimate genetic correlations (LD structure) and to select tuning parameters for the LDpred2 method.} 

The GWAS discovery sample sizes and descriptive statistics are reported in \autoref{tab:gwas_descriptives}. The sample size varies depending on the outcome of interest. For our GWAS, we follow the quality control (QC) procedure described by \citet{mitchell2019gwas} to remove genetic outliers and ensure the genotypes are well-measured. 
We follow the literature and include minimal covariates in the GWAS, controlling for gender, genotyping array, birth year, and the first 20 genetic principal components.\footnote{Principal components are commonly used to control for population stratification, see \citet{price2006principal} and \citet{novembre2008interpreting}.} 
To maximise the size of the discovery sample, we use BOLT-LMM \citep{boltlmm} to run the GWAS. Since BOLT-LMM uses a linear mixed model, it allow us to include some related individuals and to relax the restrictions on ancestry (i.e., European ancestry instead of white British individuals only). 

Using the GWAS estimates, we use LDPred2 \citep{prive2020ldpred2} to construct polygenic scores. We construct the polygenic scores under both an infinitesimal model (all SNPs are causal) and a model where the proportion of causal SNPs is estimated as an additional parameter using a grid search \citep[for more information, see][]{prive2020ldpred2}. To avoid overfitting, this grid search is done in the reference sample. We reduce the computational burden by including only SNPs that are in HapMap3 \citep{hapmap3}, resulting in approximately 1.6 million SNPs.\footnote{See also \url{https://www.sanger.ac.uk/resources/downloads/human/hapmap3.html} (accessed 18 October 2021).} We also filter the SNPs using a minor allele frequency threshold of $0.01$ and an info score threshold of $0.97$.

All polygenic scores are standardised to have zero mean and unit standard deviation in the analysis sample. To validate their predictive power, we use a linear regression model to test them against their target outcome in the analysis sample. We control for sex and the first 20 genetic principal components, and we include fixed effects for year-month of birth. We report the incremental $R^2$ defined as the increase in $R^2$ when the polygenic score is included as a covariate.  Table~\ref{tab:lon_vs_urb-pgs_test} reports the results, showing that each polygenic score is highly predictive of its outcome, with the incremental $R^2$ ranging between 0.1\% and 10\%.

\begin{table}[!h]
\caption{\label{tab:gwas_descriptives}Descriptive statistics -- GWAS sample.}
\centering
\begin{threeparttable}
\begin{tabular}[t]{lD{.}{.}{0}D{.}{.}{-3}D{.}{.}{-3}}
\toprule
\multicolumn{1}{l}{} & \multicolumn{1}{c}{Obs.} & \multicolumn{1}{c}{Mean} & \multicolumn{1}{c}{Std. dev.} \\
& \multicolumn{1}{c}{(1)} & \multicolumn{1}{c}{(2)} & \multicolumn{1}{c}{(3)} \\ \midrule
Educational attainment 	& 378,503 & 14.736 & 5.175 \\
Fluid intelligence 		& 138,933 & 0.065 & 0.981 \\
Respiratory disease 		& 377,577 & 0.111 & 0.314 \\
\bottomrule
\end{tabular}
\begin{tablenotes}
\item Columns: (1) Number of observations in the GWAS sample. (2) Sample mean in the GWAS sample. (3) Sample standard deviation in the GWAS sample. Rows: (1) educational attainment in years, (2) standardised fluid intelligence score, (3) ever hospitalised due to respiratory disease. Fluid intelligence is standardised but during the GWAS routine a small number of individuals are discarded from the sample and this causes the mean and variance to differ slightly from zero and unity above. 
\end{tablenotes}
\end{threeparttable}
\end{table}

\begin{table}[!h]

\caption{\label{tab:lon_vs_urb-pgs_test}Predictive power of polygenic scores. Linear regression estimates of the main outcomes regressed on their corresponding polygenic score.}
\centering
\begin{threeparttable}
\begin{tabular}[t]{ldddd}
\toprule
\multicolumn{1}{c}{\em{}} & \multicolumn{4}{c}{\em{Dependent variable:}} \\
\cmidrule(l{3pt}r{3pt}){2-5}
\multicolumn{1}{c}{} & \multicolumn{1}{c}{(1)} & \multicolumn{1}{c}{(2)} & \multicolumn{1}{c}{(3)} & \multicolumn{1}{c}{(4)} \\
\multicolumn{1}{c}{ } & \multicolumn{1}{c}{{\specialcell[b]{Educational \\ attainment}}} & \multicolumn{1}{c}{{\specialcell[b]{Fluid \\ intelligence}}} & \multicolumn{1}{c}{{\specialcell[b]{Respiratory, \\ any}}} & \multicolumn{1}{c}{{\specialcell[b]{Influenza/ \\ pneumonia}}}\\
\midrule
Polygenic score & 0.723^{ *** } & 0.631^{ *** } & 0.014^{ *** } & 0.004^{ *** }\\
 & (0.008) & (0.012) & (0.001) & (0.001)\\
\midrule
\multicolumn{1}{l}{Observations} & \multicolumn{1}{D{,}{,}{-3}}{64,681} & \multicolumn{1}{D{,}{,}{-3}}{26,877} & \multicolumn{1}{D{,}{,}{-3}}{64,923} & \multicolumn{1}{D{,}{,}{-3}}{64,917}\\
\multicolumn{1}{l}{$R^2$} & \multicolumn{1}{d}{0.112} & \multicolumn{1}{d}{0.097} & \multicolumn{1}{d}{0.005} & \multicolumn{1}{d}{0.002}\\
\multicolumn{1}{l}{Incremental $R^2$} & \multicolumn{1}{d}{0.100} & \multicolumn{1}{d}{0.087} & \multicolumn{1}{d}{0.002} & \multicolumn{1}{d}{0.001}\\
\bottomrule
\end{tabular}
\begin{tablenotes}
\item Columns: (1) educational attainment in years, (2) standardised fluid intelligence score, (3) ever hospitalised due to respiratory disease, (4) ever hospitalised due to influenza/pneumonia. We include sex and the first 20 genetic principal components as covariates. All specifications contain year-of-birth and month-of-birth fixed effects. Standard errors are heteroskedasticity robust. The incremental $R^2$ is the increase in $R^2$ relative to a null model excluding the polygenic score. (*): $p < 0.1$, (**): $p<0.05$, (***): $p<0.01$.
\end{tablenotes}
\end{threeparttable}
\end{table}

\FloatBarrier

\end{document}